\documentclass[longauth]{aa}  

\usepackage[dvipsnames]{xcolor}

\usepackage[T1]{fontenc} 
\usepackage{siunitx}
\usepackage{amsmath}
\usepackage{amssymb}
\usepackage{graphicx}
\usepackage{longtable}   
\usepackage{gensymb}
\usepackage{lscape}
\usepackage[export]{adjustbox}
\usepackage{natbib}
\usepackage{subcaption}
\usepackage{longtable}
\usepackage{calrsfs}
\usepackage{array}
\usepackage{multirow}
\usepackage{tabularx}
\usepackage[ruled,vlined]{algorithm2e}
\usepackage{upgreek}
\usepackage[colorlinks=true,allcolors=DarkBlue]{hyperref}

\bibpunct{(}{)}{;}{a}{}{,}

\newcommand{\comment}[1]{}

\newcommand*{\V}[1]{\boldsymbol{#1}}   
\newcommand*{\M}[1]{\mathbf{#1}}       

\usepackage{natbib}
\usepackage[utf8]{inputenc}
\DeclareUnicodeCharacter{2113}{\ensuremath{\ell}} 

\SetCommentSty{mycommfont}
\SetKwComment{tcp}{}{}
\SetArgSty{textnormal}

\usepackage{stmaryrd}
\usepackage[ruled,vlined]{algorithm2e}
\usepackage{amsmath}
\usepackage{amssymb}

\definecolor{DarkBlue}{RGB}{0,69,134}  

\begin{document} 
   \title{Preparing an unsupervised massive analysis of SPHERE \\ high contrast data with the PACO algorithm}
   \subtitle{Optimizations and benchmarking on a sample of 24 solar-type stars}


   \author{A. Chomez\inst{1,2}
    \and
        A.-M. Lagrange\inst{1,2}
        \and
        P. Delorme \inst{2}
        \and 
        M. Langlois \inst{3}
        \and
        G. Chauvin \inst{4}
        \and
        O. Flasseur \inst{3}
        \and
        J. Dallant \inst{3}
        \and 
        F. Philipot \inst{1}
        \and
        S. Bergeon \inst{2}
        \and
        D. Albert \inst{6}
        \and
        N. Meunier \inst{2}
        \and
        P. Rubini \inst{5}
          }

   \institute{
LESIA, Observatoire de Paris, Universit\'{e} PSL, CNRS, 5 Place Jules Janssen, 92190 Meudon, France
\email{antoine.chomez@obspm.fr}
\and
Université Grenoble Alpes, CNRS, IPAG, 38000 Grenoble, France
\and
CRAL, UMR 5574, CNRS, Université de Lyon, ENS, 9 avenue Charles André, 69561 Saint Genis Laval Cedex, France
\and
Université Côte d’Azur, OCA, CNRS, Lagrange, France
\and
Pixyl S.A. La Tronche, France
\and
Université Grenoble Alpes, CNRS, Observatoire des Sciences de l’Univers de Grenoble (OSUG), Grenoble, France
       }
   \date{\today}


  \abstract
  {Despite tremendous progress in the detection and characterisation of extrasolar planetary systems in the last 25 years, we have not found Solar System analogues. In particular, Jupiter-like planets (either mature or old) are barely detectable beyond 5 au with indirect techniques and are still out of reach of direct imaging.}
   {We aim at searching for exoplanets on the whole ESO/VLT-SPHERE archive with improved and unsupervised data analysis algorithm that could allow to detect massive giant planets at 5 au. 
    To prepare, test and optimize our approach, we gathered a sample of twenty four solar-type  stars observed with SPHERE using angular and spectral differential imaging  modes.}
   {We use PACO, a new generation algorithm recently developed, that has been shown to outperform classical methods. We also improve the SPHERE pre-reduction pipeline, and optimize the outputs of PACO to enhance the detection performance. We develop custom built spectral prior libraries to optimize the detection capability of the ASDI mode for both IRDIS and IFS. }
   {Compared to previous works conducted with more classical algorithms than PACO, the contrast limits we derived are more reliable and significantly better, especially at short angular separations where a gain by a factor ten is obtained between 0.2 and 0.5 arcsec. Under good observing conditions, planets down to 5 $\text{M}_{\text{Jup}}$, orbiting at 5 au could be detected around stars within 60 parsec. We identified two exoplanet candidates that require follow-up to test for common proper motion.}
  {In this work, we demonstrated on a small sample the benefits of PACO in terms of achievable contrast and of control of the confidence levels. Besides, we have developed custom tools to take full benefits of this algorithm and to quantity the total error budget on the estimated astrometry and photometry. This work paves the way towards an end-to-end, homogeneous, and unsupervised massive re-reduction of archival direct imaging surveys in the quest of new exoJupiters.}


   \titlerunning{Preparing an unsupervised massive analysis of SPHERE data with the PACO algorithm}
   \maketitle

%

\section{Introduction}

Since the discovery of the first, giant planets around  solar type stars almost 30 years ago, thousand of planets have been discovered, with  masses down to a few Earth-mass. Yet, solar system analogues\footnote{defined as planetary systems around solar-type stars, hosting inner Earth-mass planets with at least one in the habitable zone, and outer sub-Jupiter/Jupiter masses planets.} have not been detected, and we therefore still do not know if our own planetary system is unique or not. Detecting Earth twins is still not possible, and requires major  on-going research to correct for stellar activity with radial velocity (RV) techniques at the appropriate level \citep{2019A&A...629A..42M}. Also, detecting Jupiter twins orbiting at 5 au is very challenging (and not possible at larger separations) with radial velocity because decade(s) of careful monitoring are needed, and long term stellar activity is also responsible for a long term noise. As a consequence, the orbital parameters of the -- still rare -- RV planets announced beyond 5 au are poorly characterized (\cite{2016MNRAS.455.1398W}, \cite{2019ApJ...874...81F}, \cite{RV_FULTON_2021}). These limitations prevent precise comparisons between the radial distribution of giant planets beyond their forming regions and predictions from population synthesis models to constrain formation scenarios (Lagrange et al, 2022, submitted). 

Giant Planets (GPs) played a significant role in the building of the solar system (see e.g., \cite{2003AJ....125.2692L, 2014prpl.conf..595R, 2012AREPS..40..251M}) and of exoplanetary systems (see e.g., \cite{2016AAS...22840405Q, 2019MNRAS.485..541C}). Various mechanisms may be involved, among which dynamical interactions with lighter bodies (e.g., telluric planets, planetesimals) once the proto-planetary disk has dissipated and the dynamics is no longer controlled by
gas. GPs could even play a role in the development of life on Earth analogues \citep{2010A&G....51f..16H}, and could have driven the delivery of water on Earth \citep{2012AREPS..40..251M}. From an observational point of view, remote GPs may also impact the detectability of lighter and closer-in planets with the RV technique and with astrometry because of the more complex RV or astrometric signals in case of multiple systems. Hence, knowing their giant planet population is key to model individual systems.

Detecting giant planets is therefore crucial to understand planetary system formation and evolution. While RV or transit techniques are best suited to detect GPs orbiting within typically 5 au, they are not well adapted to detect and characterize more remote ones.  Absolute astrometry is well adapted for giants in the 5-10 au \citep{2014ApJ...797...14P}, even though such long period planets may be difficult to fully characterize \citep{2018A&A...614A..30R}, especially in the case of multiple systems. Micro-lensing will also be very useful to constrain the giant planet demographics in the 5-10 au range  \citep{2021EPSC...15..298B}. High contrast direct imaging (DI) is probably the most promising technique to detect and characterize analogues of our solar system giants planets in the future. Yet, current high contrast instruments like SPHERE \citep{beuzit2019sphere} or GPI \citep{MacintoshGPI} are sensitive to massive young giants orbiting  typically beyond 10 au. As an example, the SPHERE SHINE GTO survey had a 20\% chance of detecting a 2 $\text{M}_{\text{Jup}}$ at 20 au, and a 10\% chance to detect a 4 $\text{M}_{\text{Jup}}$ planet at 5 au \citep{2021A&A...651A..72V}. These limited performances are due to  1) limitations at the instrumental and data 
processing levels, and 2) a geometrical effect: unless on pole-on orbits, short 
period planets may be missed in a single observation because of a small projected separations at the time of the observation. For instance, due to geometrical effect, a single observation allows us to explore only half of the 5 au region around a star located 30 pc away when the planet orbit is seen edge-on. Fortunately, this geometrical effect can be easily overcome: the region explored is increased by more than 70\% by observing twice the star a few years apart \citep{2017A&A...603A..54L}.

In this paper, we apply PACO (patch covariance), a promising detection algorithm (\cite{2018A&A...618A.138F}; \cite{2020A&A...637A...9F};  \cite{2020A&A...634A...2F} on a small sample of stars representative of SPHERE targets, members of young close associations observed as part of the SPHERE/SHINE survey \citep{2021A&A...651A..70D} and observed under a wide range of atmospheric conditions. This analysis constitutes a test-bed for a forthcoming massive reduction of the SPHERE archive. Our aim is to find the best analysis strategy and to estimate the detection limits achievable on these stars. We moreover define our sample so as to address the following  astrophysical question: how far are we from detecting solar system young giant planets siblings?

Our paper is organized as follows: the sample and the data are described in Sect. \ref{sec:star_sample}. The data reduction and analysis are described in Sects. \ref{sec:data_reduction_frame_centering} and \ref{sec:analysis_pipeline}, respectively. Section \ref{sec:contrast_comparison} describes the achievable performance, and Sect. \ref{sec:results} presents the astrophysical results. A brief summary and a presentation of future work are finally provided in Sect. \ref{sec:conclusion}. 

\section{Star sample}
\label{sec:star_sample}

\begin{figure*}[htp]
\centering

\includegraphics[width=.315\textwidth]{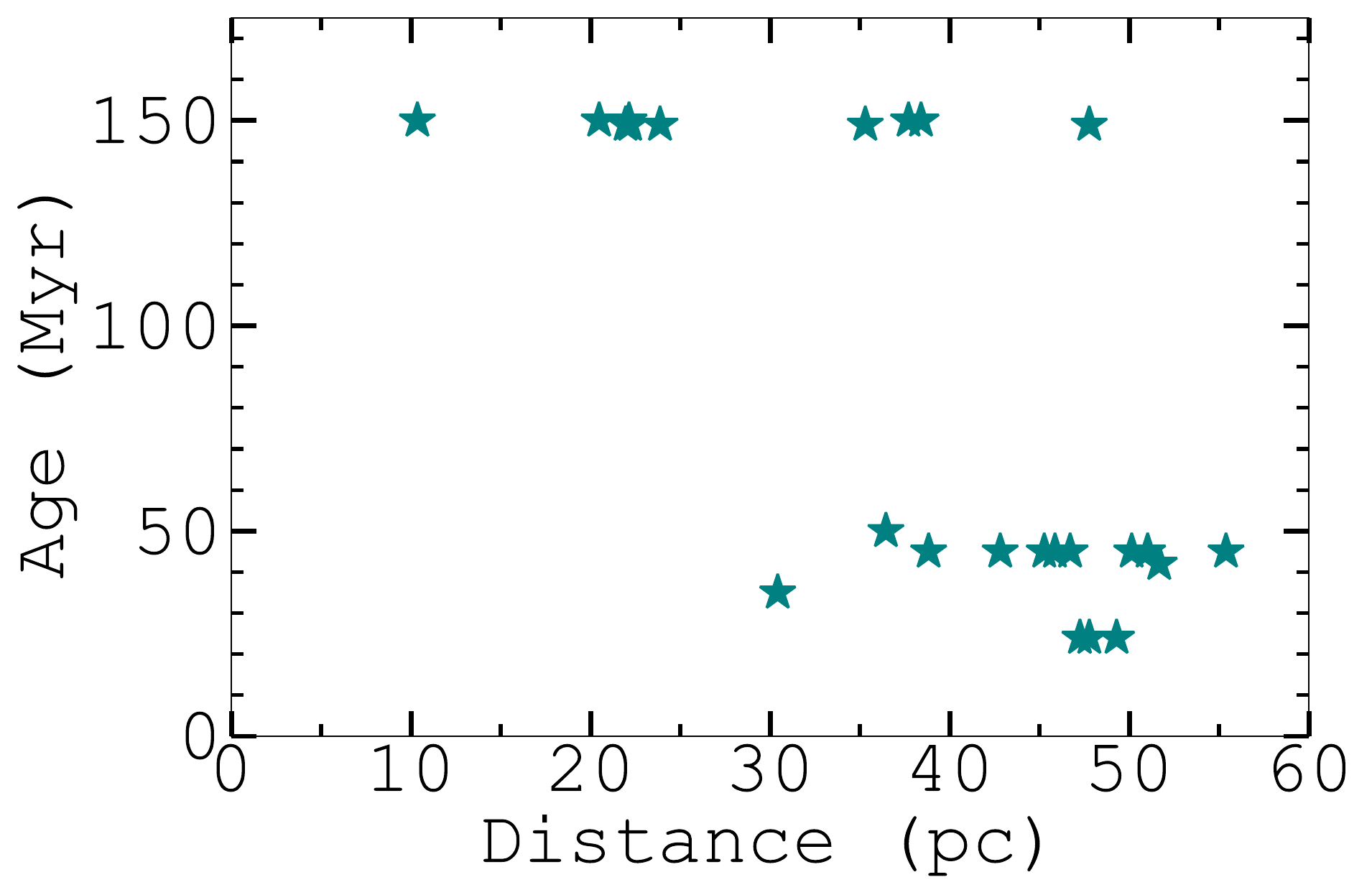}\quad
\includegraphics[width=.305\textwidth]{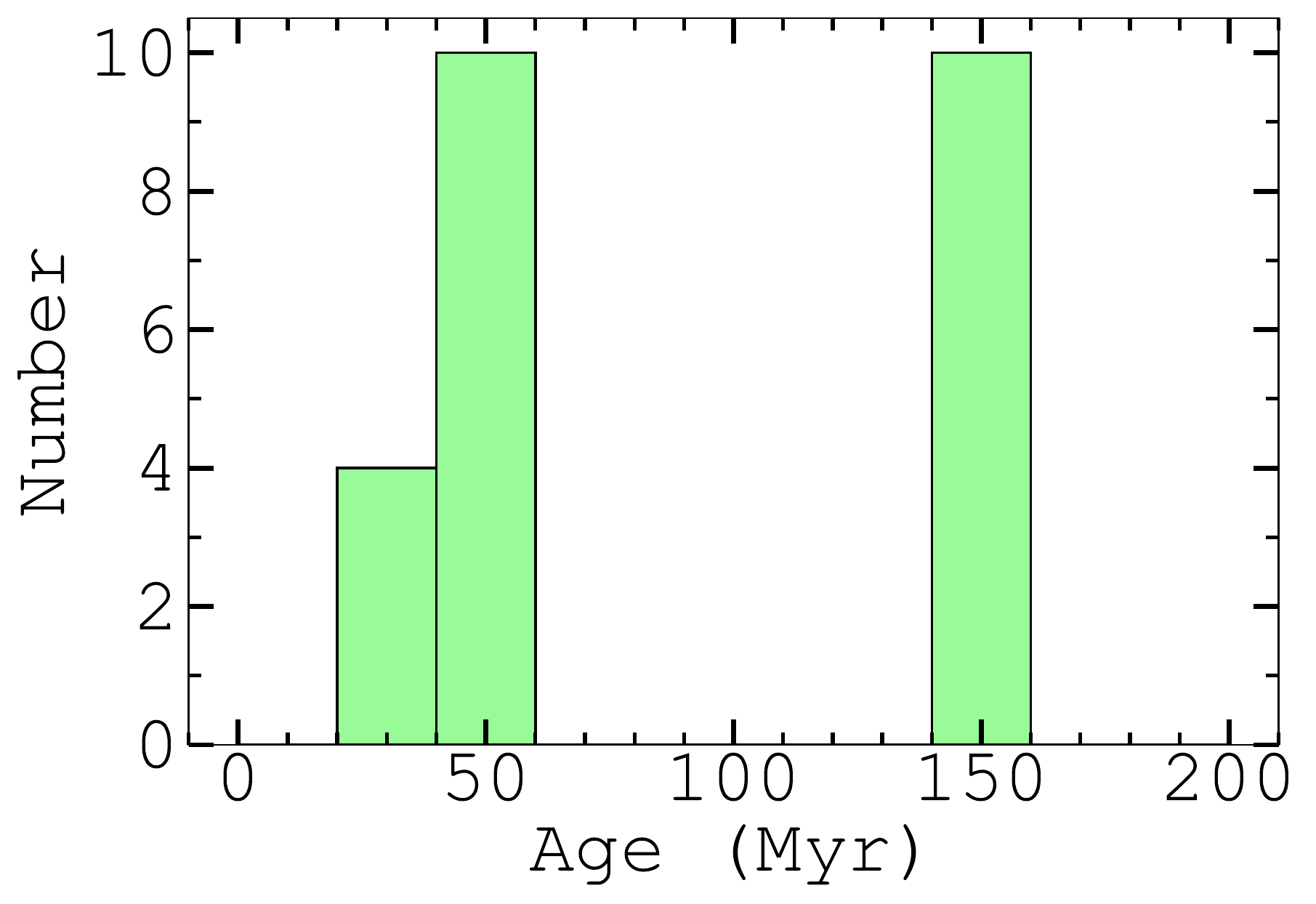}\quad
\includegraphics[width=.3\textwidth]{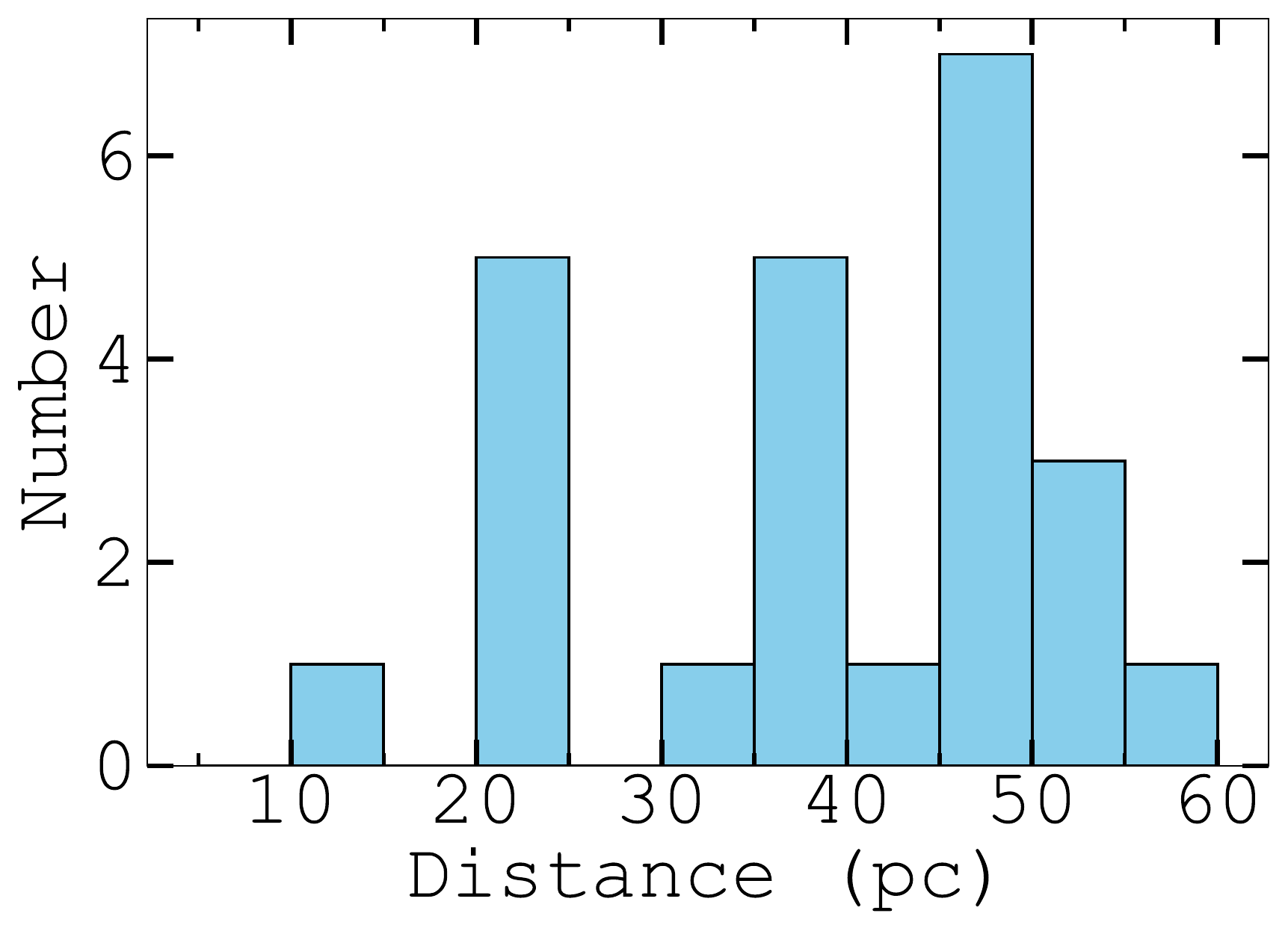}

\medskip

\includegraphics[width=.3\textwidth]{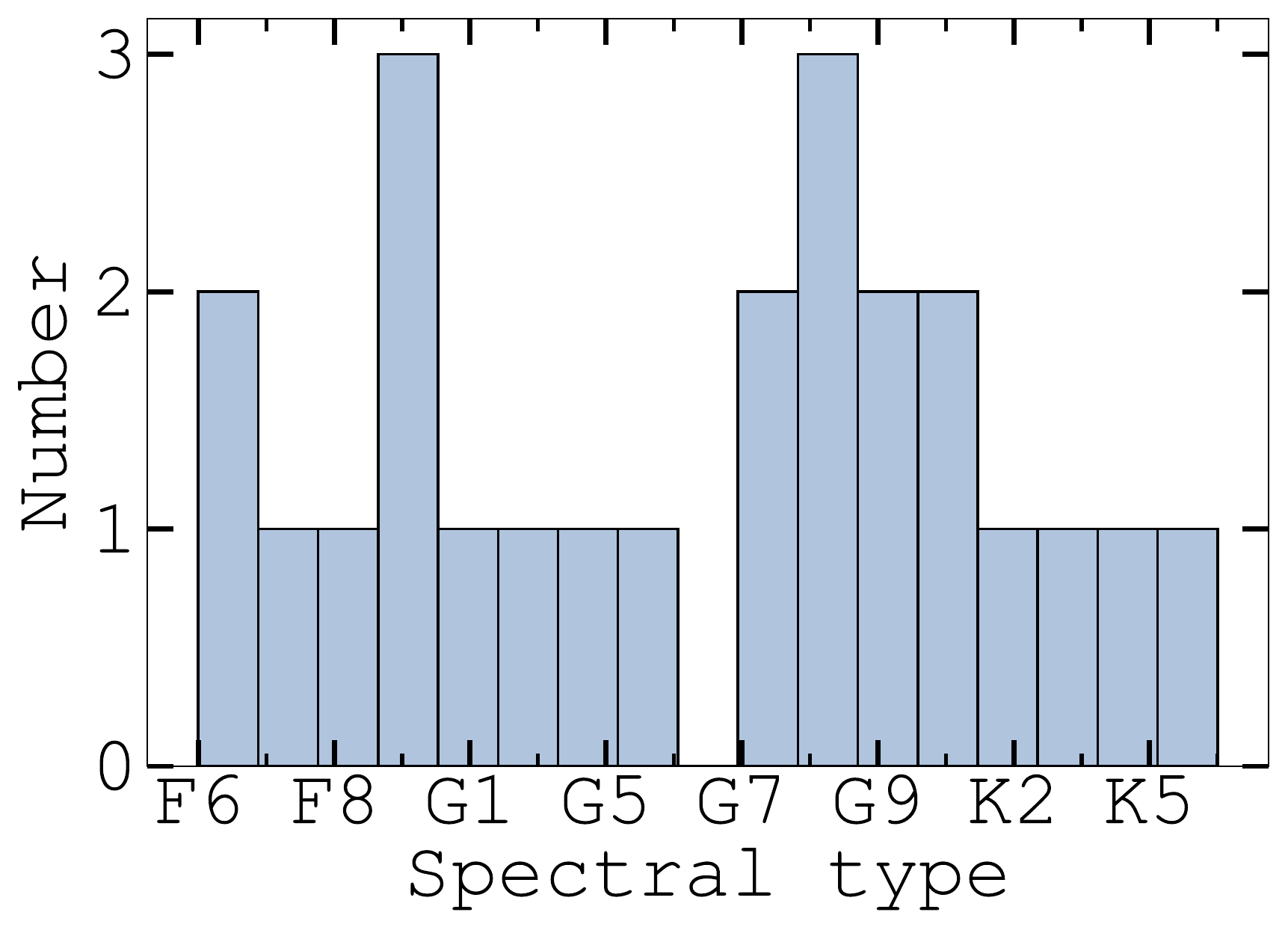}\quad
\includegraphics[width=.3\textwidth]{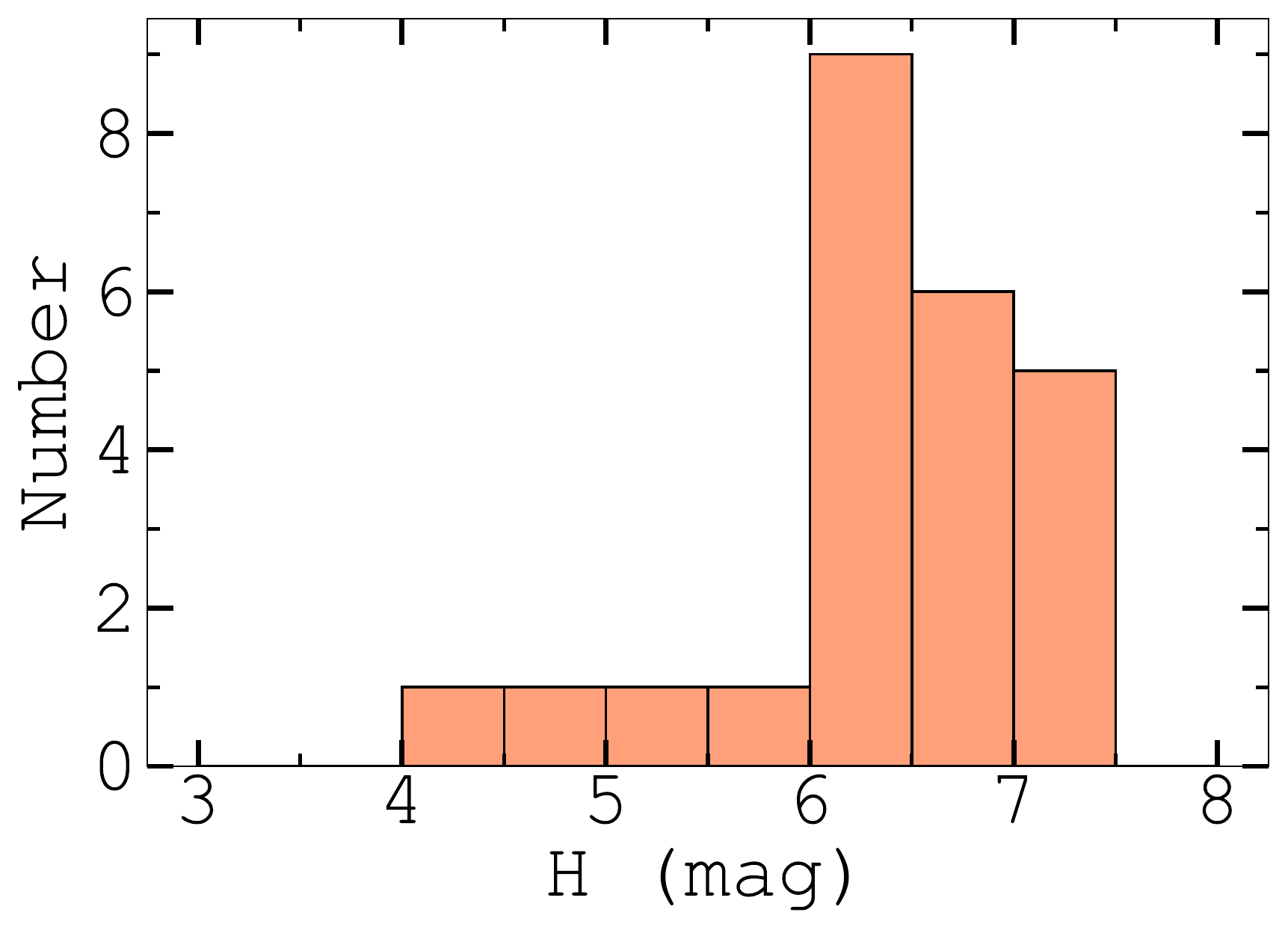}

\caption{Properties of the sample stars in terms of distance, age, spectral type, and magnitude.}
\label{sample}
\end{figure*}

Our sample gathers all (24) young ($\leq$ 150 Myr), close by (< 60 pc), solar-type stars (FGK) observed during the SPHERE/SHINE survey early science release \citep{2021A&A...651A..70D} and already analyzed with conventional post-processing algorithms in \cite{shine2}, hereafter referred to as F150. The thresholds in age and distance were chosen to ensure the best detection limits (5 $\text{M}_{\text{Jup}}$, down to possibly 1 $\text{M}_{\text{Jup}}$) possibly down to 5-10 au from the stars, i.e., at the locations of the solar system giants\footnote{Our sample is therefore biased, and the present study is not meant to have a statistical value. }.  Figure \ref{sample} and Table \ref{star_table} show the properties of the stars in our sample directly extracted from \cite{2021A&A...651A..70D}. It can be noted that the histogram of the ages of the considered stars is bimodal, with one subset aged between 20 and 60 Myr, and the other aged about 150 Myr. This bimodal distribution is caused by some stars belonging to young co-moving groups like AB Doradus (ABDO, 150 Myr),  Tucana-Horologium (TUC, 45 Myr), Carina (CAR, 45 Myr), or $\beta$ Pictoris (BPIC, 24 Myr).

All our targets were observed in angular (and spectral) differential imaging (A(S)DI, \cite{2006ApJ...641..556M}) using the telescope in pupil tracking mode. The standard observing mode of the SHINE survey, namely the IRDIFS mode was used, with IRDIS dual band images in H2 and H3  \citep{2008SPIE.7014E..3LD} and IFS \citep{2008SPIE.7014E..3EC} data covering the YJ bands\footnote{Although no observations using the K12/YJH filters combination are studied in this paper, the methodological developments presented in Sects. \ref{sec:data_reduction_frame_centering}, \ref{sec:analysis_pipeline}, \& \ref{sec:contrast_comparison} are treated without any loss of generality}. Table \ref{obs_table_1} and Table \ref{obs_table_2} provides descriptions of the observations and of the associated atmospheric conditions during these observations. The stars were observed around the time of the meridian crossing, to ensure the largest amplitude of parallactic angle variations (at least 30 degrees). The atmospheric conditions were heterogeneous, with a seeing ranging from 0.5$\arcsec$ to 1.$\arcsec$ most of the time, and the coherence time ranging between 1 to 10 ms.  

\section{Data reduction and frame centering}
\label{sec:data_reduction_frame_centering}


The reduction pipeline from raw to centered datasets is similar to the one built for the SHINE survey and described in \cite{delorme17sphere} and  \cite{shine2}.


To improve the  centering of the IRDIS and IFS frames, a custom built routine that uses the waffle center calibration has been developed (Dallant et al., submitted). As part of the observing sequence, before and after the coronagraphic sequence, two coronagraphic images are recorded with a waffle pattern applied to the deformable mirror to create four replicas of the point spread function (PSF) at a separation of about 14 $\lambda$/D  from the central star. These replicas, called \textit{satellite spots}, are used to determine precisely the position of the central star behind the coronagraphic mask before and after the long coronagraphic sequence. 

To determine the accurate positions of the satellite spots, small circular regions are extracted around their theoretical positions and a bi-dimensional, non-istropic Gaussian fit is performed using a trust region reflective algorithm \citep{branch1999subspace}, particularly suited for large sparse problems with bound constraints. Estimates of the central star positions are then computed via the centroid of the resulting fitted satellite spots and their associated uncertainties. The frames are re-centered using the mean value of the two estimated centroids. This new routine is marginally more precise than the one currently implemented in the SPHERE data center, but its main advantage is a much faster computational time: assembling the 4D datacube takes only a few minutes, i.e.,  more than one order of magnitude faster than the current pipeline, without any loss in precision.

Note that when  precise astrometric measurements of known companions are needed, the satellite spots are generated during the whole coronagraphic sequence. In such a case, all frames are recentered individually using their own satellite spots.

\section{Analysis pipeline}
\label{sec:analysis_pipeline}

The analysis pipeline is based on the PACO A(S)DI pipeline described in \cite{2020A&A...637A...9F} and \cite{2020A&A...634A...2F}. A few important upgrades were made, though: 
\begin{itemize}
    \item Improvement of PACO robustness, i.e., capability to  run PACO on diverse and heterogeneous datasets while consistently providing reliable results, see Sect. \ref{subsec:paco_improvement};
    \item Optimization of spectral priors for PACO ASDI, see Sect. \ref{subsec:prior_asdi};
    \item Automated and improved computation of astrometric and photometric error bars for each characterized source, see Sect. \ref{subsec:astro_photo_err};
    \item Automated classification of the status of any identified candidate companion in case of multi epoch observations, see Sect. \ref{subsec:MEDT}. 
    \end{itemize}
Finally, in view of the forthcoming massive re-reduction of all SPHERE data, a tool was developed to automatically identify any potential companion, and gather associated astrophysical information (astrometry, photometry, spectra, etc) needed for further analysis. These upgrades are described in the following.
Both the centering routines and the analysis pipeline are hosted on the COBREX data center, a modified and improved server based on the SPHERE data center. 

\subsection{Improvements of PACO robustness}
\label{subsec:paco_improvement}
The principle of the PACO algorithm is  described in \cite{2018A&A...618A.138F}. No fundamental modifications were made   concerning the core and the technical elements of the methods. The main updates are:
\begin{itemize}
    \item A refinement of the PSF fitting routine, now implementing a robust strategy based on iteratively re-weighted least-squares \citep{huber2011robust}. This routine improves the robustness of the fit in the case of very low signal-to-noise ratio (S/N) in the measured off-axis PSF (e.g., in absorption bands).
    \item The management of time-variable missing data, with a time-variable mask, to account for possible evolution (during the sequence of acquisition) of the field-of-view with exploitable data.
\end{itemize}
Besides, an engineering effort has been made to validate through numerical experiments the faithfulness of the astrophysical quantities produced by the algorithm, in particular concerning the astrometry and photometry, and their associated error bars. Due to residual noise, the flux estimate in the absence of a source is not completely zero on average. The average level of this effect was estimated for both ADI and ASDI. Thus, flux estimates for which the detection confidence is less than 1$\sigma$ in ASDI and 2.5$\sigma$ in ADI will not be considered or used in this analysis or in the massive reduction. Along the same line, code upgrades (accelerations, automations, and case-specific handlings) have been implemented to allow for massive reductions performed on a computer server.

\subsection{PACO ASDI spectral priors to increase the sensitivity}
\label{subsec:prior_asdi}

\begin{figure}[t!]
\centering
  \includegraphics[width=\linewidth]{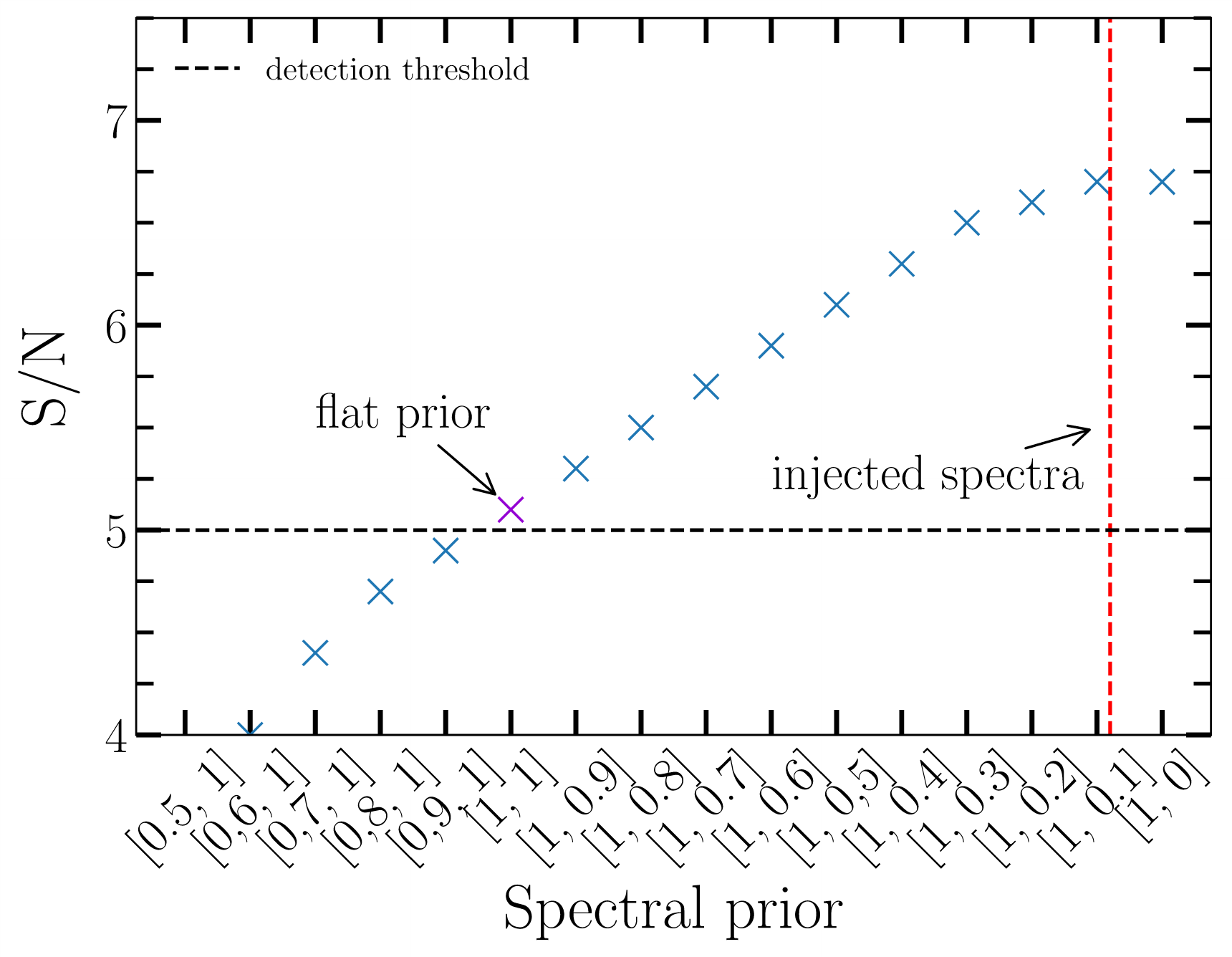}
  \caption{Evolution of the S/N (blue) for various priors for a point source observed with IRDIS, and whose spectrum corresponds to the red mark. The $x$-axis gives various values of the priors. The S/N corresponding to the [0.5,1] prior (significantly different from the normalized contrast of the injected source) is below the $4\sigma$ threshold used for this analysis, while the S/N reached for spectral prior [1, 0.1] (similar to the normalized contrast of the injected source) leads to a clear detection with a significance above the $5\sigma$ detection threshold (dashed black line).}
\label{snr_evo}
\end{figure}

\begin{figure}[t!]
\centering
  \includegraphics[width=\linewidth]{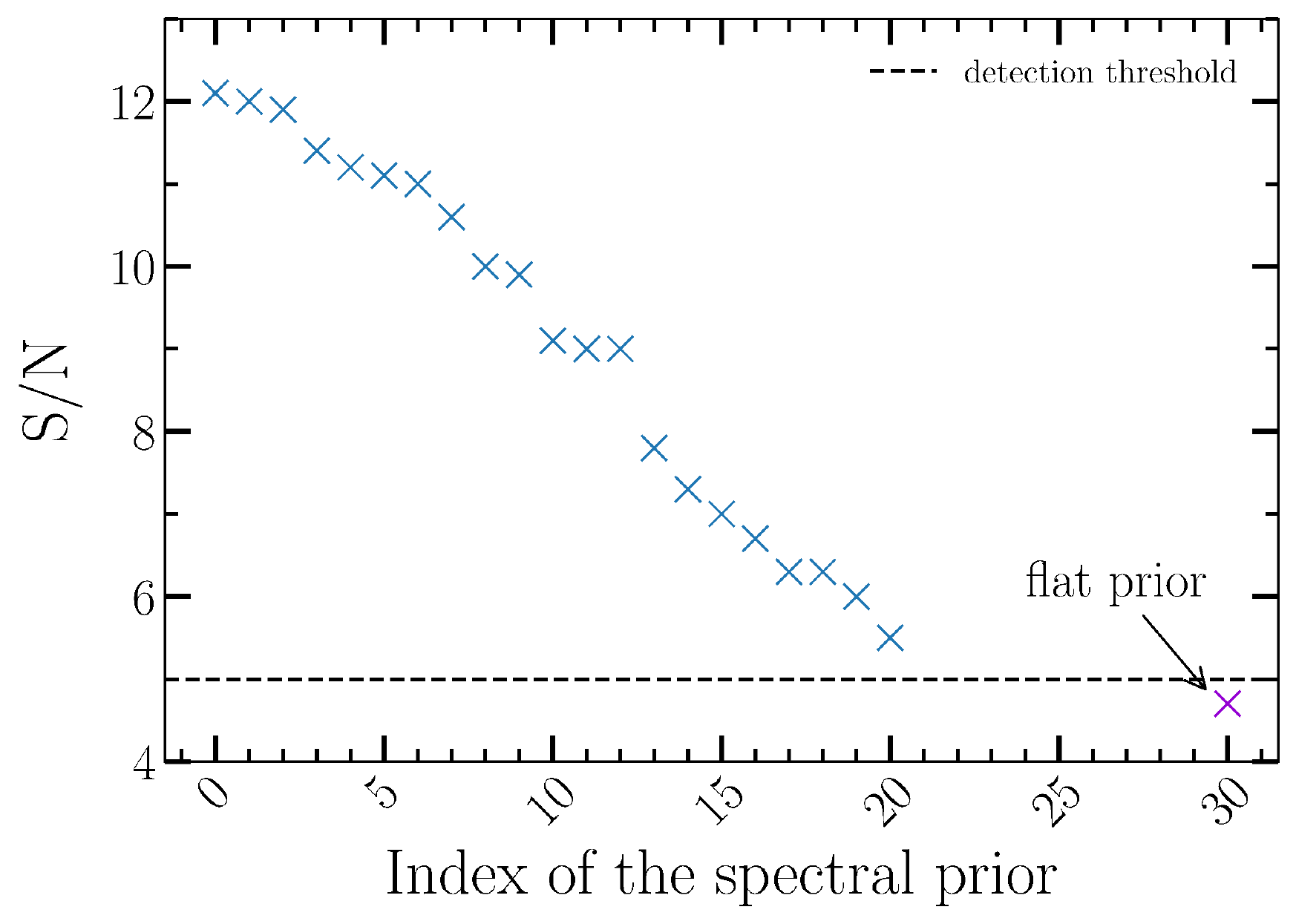}
  \caption{Evolution of the S/N (blue) for various priors for HD 206893b (2017-07-13, IFS, YJ band). The $x$-axis gives the index of the priors. The S/N corresponding to the 10 last priors are below the $5\sigma$ detection threshold. Priors are ordered by decreasing S/N for clarity purposes. The S/N reached for spectral prior close to the spectrum (in contrast) leads to a detection with a significance above the $5\sigma$ detection threshold.}
\label{snr_evo_ifs}
\end{figure}

The ASDI mode of PACO offers the possibility to combine multi-wavelength datasets into a detection map, using specific weights to maximize the detection efficiency. These weights, $\lbrace w_\ell \rbrace_{\ell=1:L}\in \left[ 0 ; 1 \right]$, are referred as spectral priors 
They are represented by vectors with as many components as wavelengths: $L=2$ for IRDIS data, and $L=39$ for IFS. Since all the photometric measurements within PACO are expressed with respect to the target star, the priors should also be expressed as the expected companion contrast relative to the host star (shape-wise). They are then normalized between 0 and 1 (dynamic-wise). When simultaneously using multiple priors, the PACO algorithm computes the S/N of the detected source  for each prior.
As an illustration, we show in  Fig. \ref{snr_evo} the S/N measured on point sources considering various priors for IRDIS data in the case of an injected fake planet, and, in Fig. \ref{snr_evo_ifs} the same kind of plot for an IFS dataset and a real point-like source: HD 206893b \citep{2017A&A...597L...2M}. A more classical ASDI spectral combination approach, as implemented in the PCA and TLOCI versions of the SPHERE data center, is somewhat similar\footnote{In practice, PACO ASDI also accounts for a confidence weight estimated locally for each spectral channel, giving more weight to the spectral channels where the  variance of the estimated flux is the lowest. This information is never accounted for in classical algorithms like TLOCI and PCA.} to the \textit{flat prior} combination (i.e., assuming that the sought exoplanets have the same spectral energy distribution (SED) as their host stars), which is highlighted in Figs. \ref{snr_evo} and \ref{snr_evo_ifs}.

As expected, the S/N is higher when the spectral prior is  similar to the planet spectrum. Thus, we must define sets of spectral priors representative of the variety of the  spectra of the potential exoplanets to optimize the detection capabilities.
Besides, while increasing the number of priors improves the sensitivity to different types of objects\footnote{It has been shown in \cite{2020A&A...634A...2F} that the S/N of detection is only marginally degraded in the case where the prior SEDs differ significantly from the true SED of the sources. In any case, the probability of false alarms remains controlled at the prescribed detection threshold $\tau$.}, it also significantly increases the computational time and (moderately) increases the number of non-redundant false positives at a given detection threshold (see Sect. \ref{subsubsec:prior_selection_irdis}).  A trade-off must then be found.

\subsubsection{Selection of spectral priors using fake planets injections}

\begin{figure}[t!]
\centering
  \includegraphics[width=\linewidth]{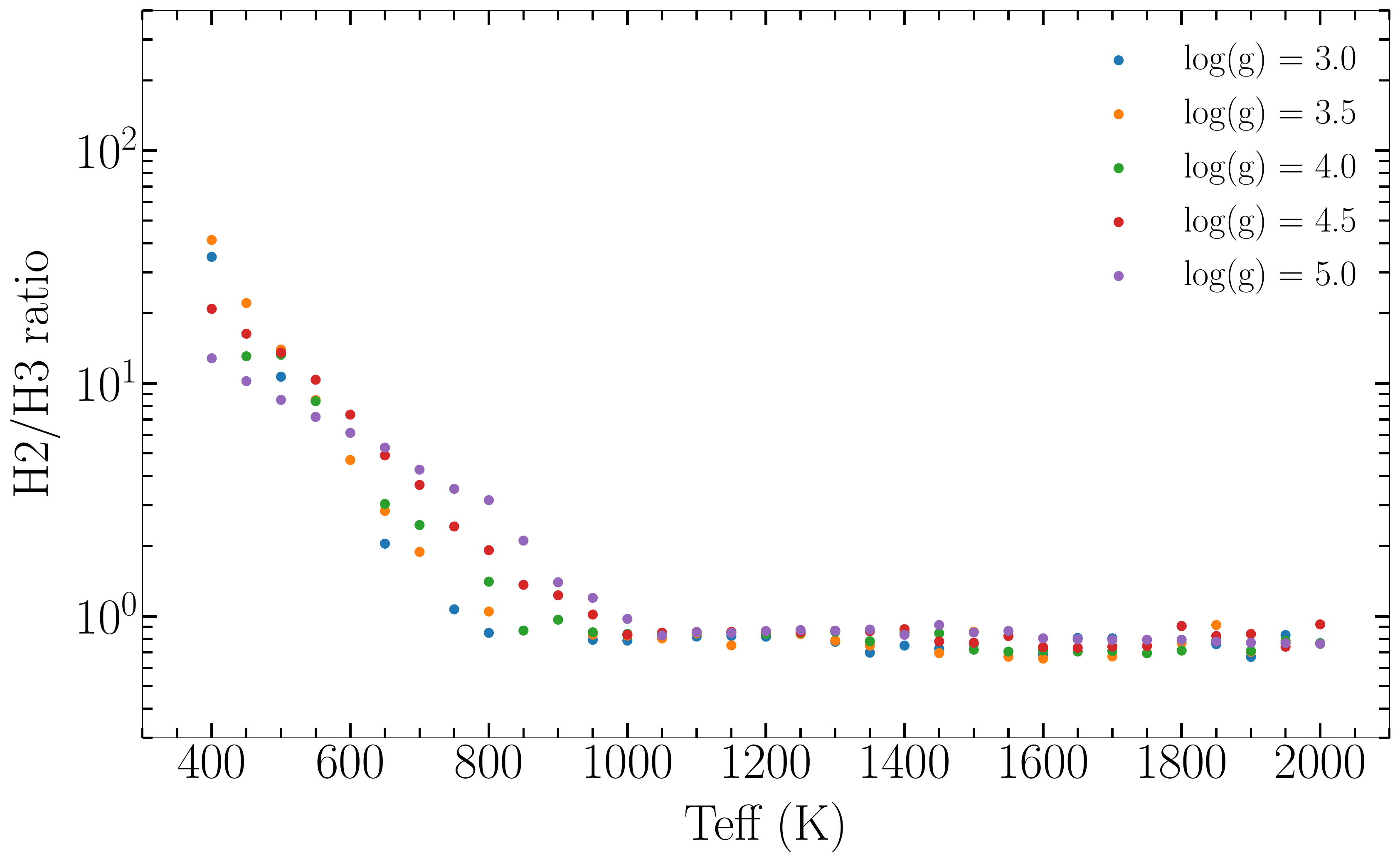}
  \caption{Integrated H2/H3 flux ratio computed for a planet with a solar metallicity, C/O=0.50, and with different values of log(g) and $\text{T}_{\text{eff}}$. ExoREM spectra were used.}
\label{H23_ratio}
\end{figure}

\begin{figure}[t!]
\centering
  \includegraphics[width=\linewidth]{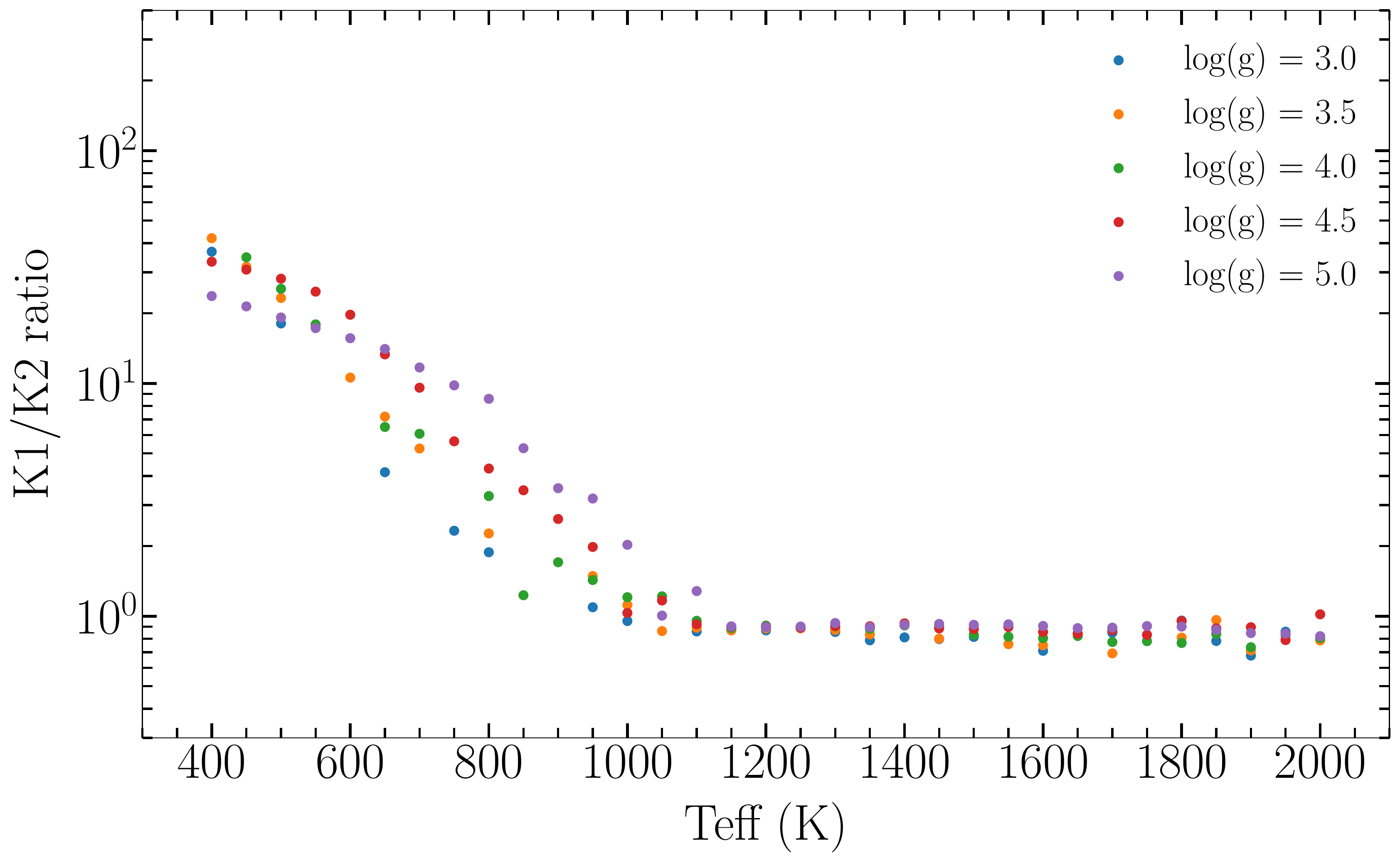}
  \caption{Integrated K1/K2 flux ratio computed for a planet with a solar metallicity, C/O=0.50, and with different values of log(g) and $\text{T}_{\text{eff}}$. ExoREM spectra were used.}
\label{K12_ratio}
\end{figure}

To select the spectral priors, we used a set of four targeted stars that were used in a recent internal blindtest conducted by the SHINE consortium that was aimed at comparing the performance of various detection/characterization algorithms. The targets properties are provided in Table~\ref{star_table_inj}, and the observing and atmospheric conditions are given in Table~\ref{obs_table_inj}. Several hundreds fake planets with various properties were randomly injected between 0.12$\arcsec$ and 5.5$\arcsec$ in the case of IRDIS data, avoiding nonetheless blending between injected sources. For IFS data, about 100 fake point sources (FPSs) were injected. The injected FPS spectra were taken from the BT-Settl grid \citep{2014IAUS..299..271A} and the priors were built using the ExoREM \citep{2019EPSC...13.1450C} spectra. We purposely used different models to inject the fake planets and to build the priors, to avoid biases that could occur when using the same library for both the injection process and the prior definition.

\subsubsection{Priors selection for IRDIS}
\label{subsubsec:prior_selection_irdis}


To find the trade-off between sensitivity and computation time, we considered all possible available spectra, and for each raw spectrum, we computed the ratio $\frac{\text{H2}}{\text{H3}}$ (or  $\frac{\text{K1}}{\text{K2}}$) where H2 and H3 (respectively K1 and K2) represent the integrated fluxes in these spectral bands. Figure~\ref{H23_ratio} and Fig. \ref{K12_ratio} show examples of such ratios for a source with various effective temperatures, a solar metallicity and a C/O ratio of 0.5 for both filter combination. Two regimes can be identified: (i) a \textit{cold} one, corresponding to $\textit{T}_{\text{eff}}$ roughly below 1000 K, where the $\frac{\text{H2}}{\text{H3}}$ (resp $\frac{\text{K1}}{\text{K2}}$) ratio varies from several tens even hundreds down to 1 mostly because of strong $\text{CH}_4$ absorption in H3 and K2 filters, and (ii) a \textit{hot} one, corresponding to $\textit{T}_{\text{eff}}$ above 1000 K, where this ratio is almost constant -- between 1 and 0.6. A similar behavior is observed for all metallicities and C/O ratio.  

\medskip

\noindent The goal is to find the optimal number of priors to explore these features. To do so, we proceeded as follows:

-- We first used all the injected FPSs and measured in each case the S/N considering different priors. These priors correspond to the 15 priors showed in the $x$-axis of Fig.~\ref{snr_evo}. In this example, the maximum S/N is 6.7.
We then repeated the process using instead 8 spectral priors uniformly spread over the considered parameter space (by steps of 0.2), 5 (by steps of 0.3), and 4 (by steps of 0.4). We then considered the evolution of the maximum S/N as a function of the number of priors (for each injected FPS). In the case of relatively faint FPSs (sources that can not be identified on a single frame), using 5 priors (or more) does not significantly degrade the S/N (less than 1$\% $), while using 4 (or less) does.

-- Second, we then studied the impact of the number of priors (between 1 and 8) on the rate of false positives. To do so, we computed the average number of false positives identified by PACO in the IRDIS FoV (limited to 5.5$\arcsec$ to avoid edge effects) on a total of 20 datasets and compared it to the theoretical value.
We can compute the theoretical expected number of false positives as follows :
because the pixel distribution on the S/N maps is Gaussian, the number $\textit{N}_{\text{fp}}$ of false positive  per map  according to the number $\textit{n}_{\text{pixel}}$ of pixels processed  and the probability of false alarm $\textit{P}_{\text{FA}}(\tau)$ at a given detection threshold $\tau$:
\begin{eqnarray}
\textit{N}_{\text{fp}} = \textit{n}_{\text{pixel}} \times \textit{P}_{\text{FA}}(\tau)\,. \label{eq1}
\end{eqnarray}
As an illustration, the number of pixels to process in each IRDIS dataset is approximately equal to one million and, with a $5\sigma$ detection confidence (i.e., $\tau=5$, $P_{\text{FA}}(5) \simeq 2.87\times10^{-7}$), we have:
\begin{eqnarray}
N_{\text{fp}} \simeq 10^6 \times 2.87\times10^{-7} \sim 0.287 \,, \nonumber
\end{eqnarray}
false alarms expected in each detection map.

\begin{figure}[t!] 
    \centering
    \includegraphics[width=\linewidth]{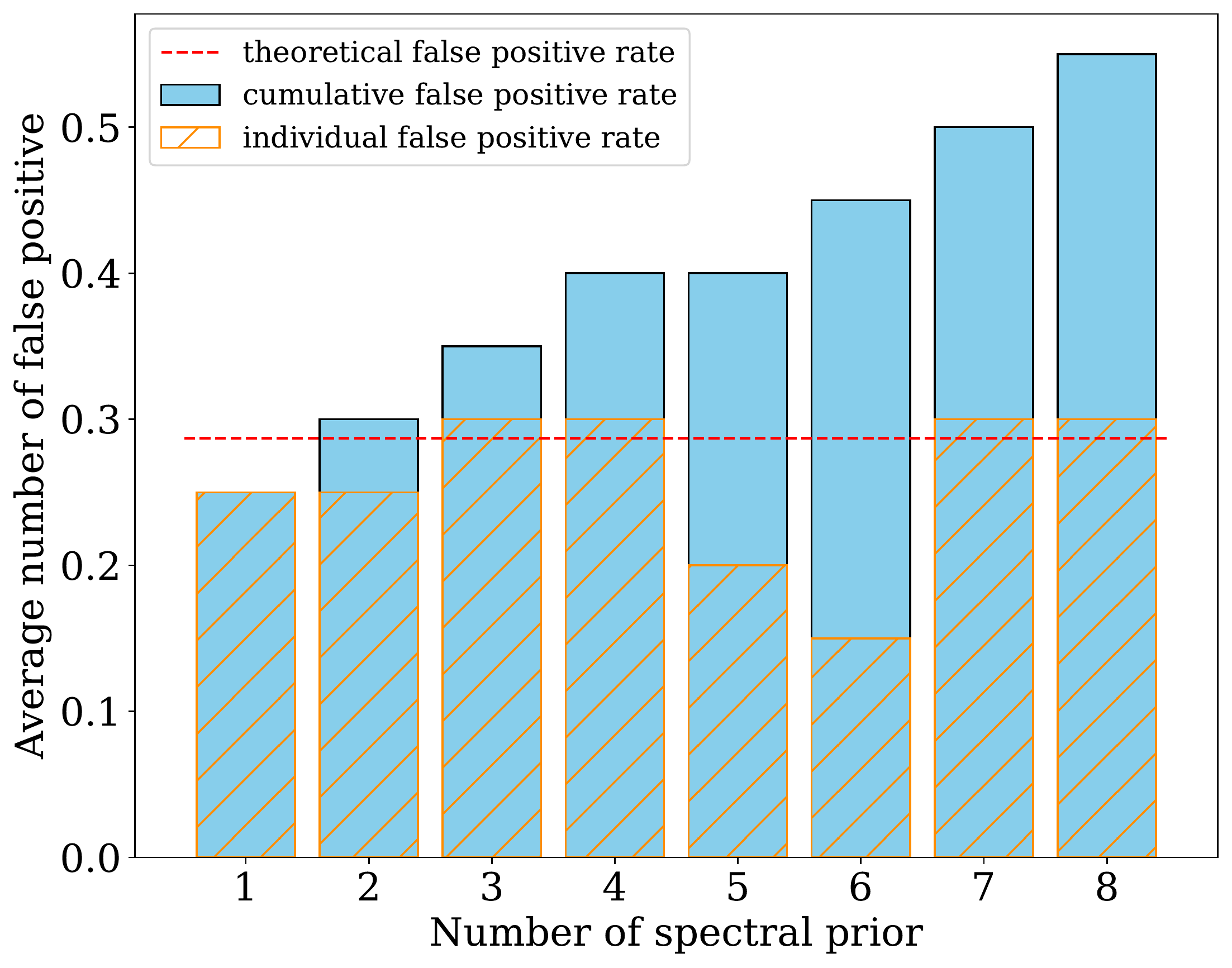}
    \caption{Average number of false positives for a detection threshold $\tau = 5$ with 1 to 8 spectral priors with IRDIS. Twenty observations were considered for these experiments. The red dashed line shows the theoretical false positive rate on a single S/N map. The mean experienced false positive rate for each S/N considered independently map is represented by the yellow dashed rectangles and the mean experienced cumulative false positive rate by the blue rectangles.}
  \label{fp_vs_priors} 
\end{figure}

Figure \ref{fp_vs_priors} shows the results for $\tau=5$ with an increasing number of spectral priors used. 
The false positive rate when considering a single S/N map corresponding to a given prior (no matter which prior is used) is in good agreement with what is expected from a Gaussian noise distribution (see Eq. (\ref{eq1})). When working with several priors, the number of detections will only increase if a new independent source (i.e. detected for the first time in the current prior) is detected. Redundant detections will only be accounted for one source. Because false positives are often redundant, the empirical cumulative false positive rate is lower than the theoretical cumulative one (i.e. if all false positives were independent).

This confirms the Gaussian nature of the S/N map produced by PACO as well as the associated statistical guarantees (i.e., control of the probability of false alarms and of detections). Based on this study, we choose to include five 
spectral priors in our library $\M \Omega_{\text{IRDIS}} \, \in \mathbb{R}^{2 \times 5}$, achieving the desired trade-off between maximizing our detection performance and lowering the number of false positives:
\begin{eqnarray}
\M \Omega_{\text{IRDIS}} = \{[1, 1]; [1, 0.7]; [1, 0.4]; [1,0.1]; [0.8, 1]\} \,. \nonumber
\end{eqnarray}

\subsubsection{Priors selection for IFS}

\begin{algorithm}[t!]
	\DontPrintSemicolon
	\KwIn{ExoREM set $\M \Omega_{\text{ER}}$ of sub-stellar spectra.}
	\KwIn{ExoREM spectrum $\V w_\text{ER}^{\text{ref}}$ of reference.}
	\KwIn{Stellar spectrum $\V s$.}
	\KwIn{Target number $N$ of spectral priors.}
	\KwIn{Gaussian kernel $\V g_L$ of width $L$.}
	\KwIn{Sampling operator $\M R_{L_\text{ER} / L}$ by $L_\text{ER} / L$.}
	\vspace{1mm}
	\KwOut{Library $\M \Omega_{\text{IFS}}$ of spectral priors.} 
	\vspace{0.5mm}
	
	\hrule
    \medskip
    
   $\blacktriangleright$ \textbf{Step 1.} Normalizing sub-stellar spectra. \vspace{0.5mm}
      
   $\M \Omega' \leftarrow \lbrace \rbrace$ \tcp*[r]{(initialization)}
   \For{$j = 1$ \KwTo $N_{\text{ER}}$}{
		$\V w_\text{ER} \leftarrow {\M \Omega_{\text{ER}}}_j$ \tcp*[r]{(get sub-stellar spectrum)}
		$w_{\ell}' \leftarrow {w_{\text{ER}}}_{\ell} / s_{\ell} \,, \forall \ell \in \llbracket 1 ; L_\text{ER} \rrbracket$ \tcp*[r]{(normalization)} \vspace{0.5mm}
		$\V w' \leftarrow \V w' \circledast \V g_L$ \tcp*[r]{(Gaussian convolution)}
		$\V w' \leftarrow \M R_{L_\text{ER} / L}(\V w')$ \tcp*[r]{(re-sampling)}
		$\M \Omega' \leftarrow \lbrace \M \Omega' \cup \V w' \rbrace$ \tcp*[r]{(storing)}
   }
   
   \medskip

   $\blacktriangleright$ \textbf{Step 2.} Building library of spectral priors. \vspace{0.5mm}

	$\V w^{\text{ref}} \leftarrow \M {\Omega'}_1$ \tcp*[r]{(get reference spectral prior)}
   $\M \Omega \leftarrow \lbrace \V w^{\text{ref}} \rbrace$ \tcp*[r]{(initialization)}  \vspace{0.5mm}
	\While{\text{card}$(\M \Omega) < N$}{
		\For{$i=1$ \KwTo \text{card}$(\M \Omega)$}{
			$\V w \leftarrow \M \Omega_i$ \tcp*[r]{(get a spectral prior)}
			\For{$j=i$ \KwTo $N_{\text{ER}}$}{
				$\V w' \leftarrow \M \Omega_j'$ \tcp*[r]{(get a candidate spectral prior)}
				$d_{i,j} \leftarrow \lVert \V w - \V w' \rVert_2^2$ \tcp*[r]{(compute distance)}
			}
			$(\_, j^{\text{max}}) \leftarrow \text{argmax}_{i,j}(d_{i,j})$ \tcp*[r]{(get max index)}
			$\M \Omega \leftarrow \lbrace \M \Omega \cup \M \Omega_{j^{\text{max}}}' \rbrace$ \tcp*[r]{(storing)}
		}
	}
        $\M \Omega_{\text{IFS}} \leftarrow \M \Omega$
	\caption{Pseudo-code of the selection procedure of IFS spectral priors for PACO.}
    \label{opti_ifs}
\end{algorithm}

In this section, we describe how we build a library $\M \Omega_{\text{IFS}} \in \mathbb{R}^{N \times L}$ of $N$ spectral priors for processing the IFS data with PACO ASDI. 

Following the notation introduced in Sect. 4.2, each element $\V w \in \mathbb{R}^{L}$ of $\M \Omega_{\text{IFS}}$ is a vector with as many components as wavelengths (i.e., $L=39$ for the IFS). In practice, we build a different set $\M \Omega_{\text{IFS}}$ for each stellar spectral type since each element $\V w$ should be expressed in contrast units, so that it depends on the spectral type of the star explicitly. For the purpose of illustration, the method is described for any given stellar spectral type without loss of generality.
The set $\M \Omega_{\text{IFS}}$ is built from a set $\M \Omega_{\text{ER}} \in \mathbb{R}^{N_\text{ER} \times L_\text{ER}}$ of sub-stellar spectra provided by the ExoREM models. The spectral resolution $L_\text{ER}$ of each element $\V w_\text{ER} \in \mathbb{R}^{L_\text{ER}}$ is  much larger than $L$, with typically $L_\text{ER} = 500$ in this study. 

For a given stellar spectrum $\V s \in \mathbb{R}^{L_\text{ER}}$ and a sub-stellar spectrum $\V w_\text{ER} \in \mathbb{R}^{L_\text{ER}}$ at the same spectral resolution $L_\text{ER}$, we obtain an (intermediate) spectral prior $\V w' \in \mathbb{R}^{L_\text{ER}}$ by dividing the two elements component-wise, i.e. $w'_{\ell} = {w_\text{ER}}_{\ell} / s_{\ell} \,, \forall \ell \in  \llbracket 1 ; L_\text{ER} \rrbracket$. Each intermediate spectral prior $\V w'$ is then normalized by the maximum value over its components. The last operations aim to reshape $\V w'$ from $L_\text{ER}=500$ to the spectral resolution $L$ of the measurements by first applying a convolution with a Gaussian kernel of standard-deviation in the order of magnitude of $L$ (typically between 30 and 50), and second re-sampling the results at the targeted spectral resolution $L$ to get a final spectral prior $\V w$.

\begin{figure}[t!]
\centering
  \includegraphics[width=\linewidth]{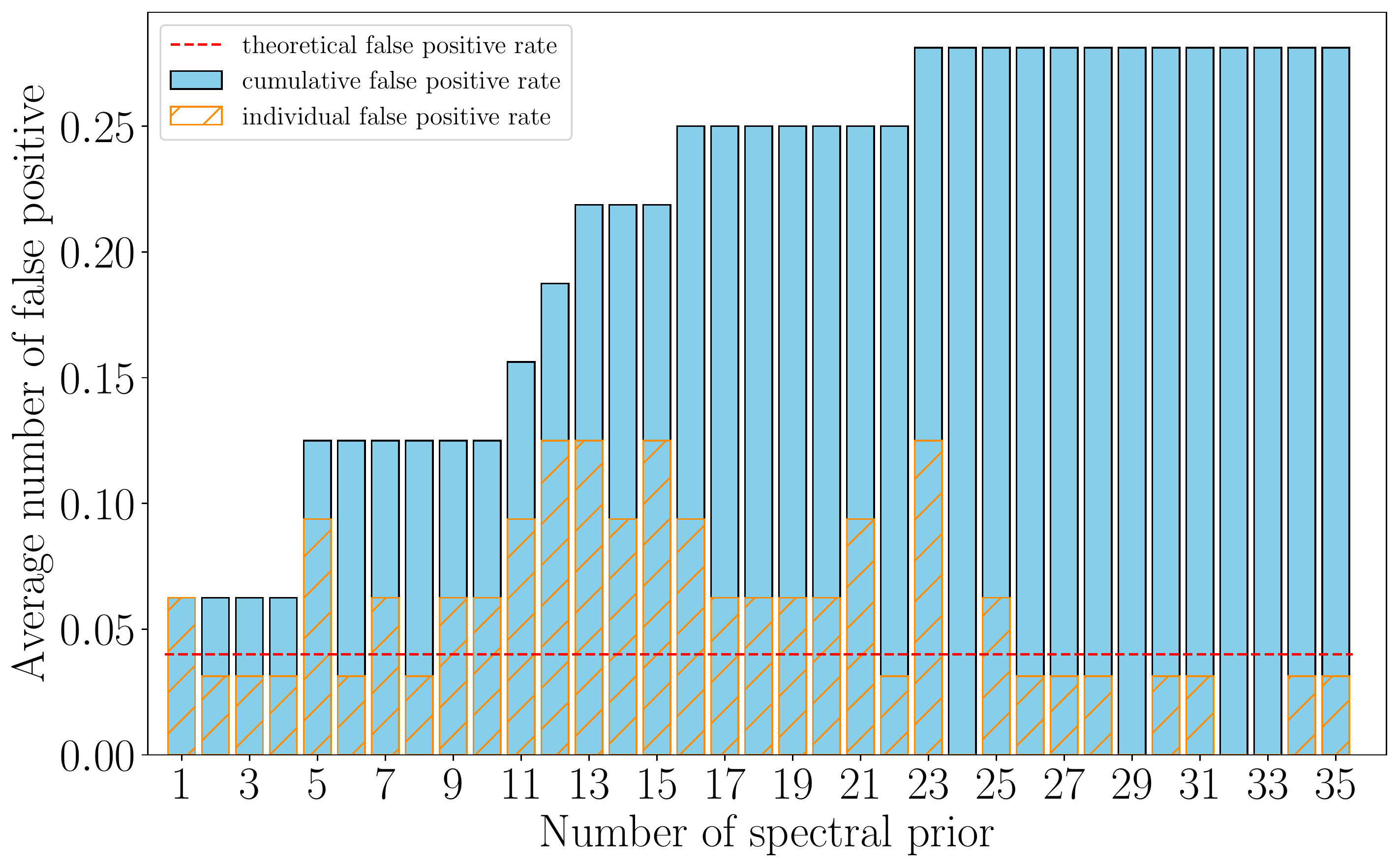}
  \caption{Number of false positives experienced at detection time when using between $N=1$ and $N=31$ spectral priors with IFS. The detection threshold is set at $\tau$=5, and the results are averaged over 32 observations. The red dashed line shows the theoretical false positive rate on a single S/N map. The mean experienced false positive rate for each S/N considered independently map is represented by the yellow dashed rectangles, and the mean experienced cumulative false positive rate by the blue rectangles.}
\label{snr5_fp_ifs}
\end{figure}

We aim to include in $\M \Omega_{\text{IFS}}$ the minimum number $N$ of spectral priors needed to represent the diversity of the observations. For that purpose, we build $\M \Omega_{\text{IFS}}$ from a large set $\M \Omega_{\text{ER}}$ (i.e., $N_\text{ER} \gg N$), and we progressively add non-redundant atoms $\V w$ in $\M \Omega_{\text{IFS}}$. We first start with a single spectral prior $\V w^{\text{ref}}$ of reference in the set $\M \Omega_{\text{IFS}}$. This spectral prior is built from a sub-stellar model $\V w_\text{ER}^{\text{ref}}$ of reference obtained from the ExoREM simulator with the following parameters: $T_{\text{eff}} = 1000$ K, Fe/H = 1.0, C/O = 0.50, and log(g) = 4.0. By looping over the (fixed) elements of $\M \Omega_{\text{ER}}$, we compute  the Euclidean distance between each resulting candidate spectral prior and the spectral priors already present in $\M \Omega_{\text{IFS}}$. We then add to the set $\M \Omega_{\text{IFS}}$ the candidate spectral prior that maximized the distance, i.e., the spectral prior that differs the most from the already selected ones. For practical reasons, we preset the number $N$ of targeted elements in $\M \Omega_{\text{IFS}}$ to perform the above selection procedure. We repeat this procedure for various values of $N$ and select the number that gives a satisfying trade-off between precision and recall, while simultaneously leading to a manageable computation time at data reduction time. As an illustration, we find that reducing the number $N$ of spectral priors in $\M \Omega_{\text{IFS}}$ from 31 to 20 decreases the detection capabilities only marginally (by less than 1\% in terms of S/N loss). In addition, we find that decreasing the number $N$ of spectral priors in the same proportion does not significantly impact  the false positive rate, see Fig. \ref{snr5_fp_ifs}. To more significantly decrease the false positive rate, it seems better to select $N \le 13$. However, doing so would result in a significant decrease in the detection capabilities (by more than 15\% in terms of S/N loss). As a conclusion of this study, we choose $N=20$ spectral priors for the IFS instrument. This number is  driven by a trade-off between precision and recall, i.e. in order to keep the S/N loss significantly small while limiting the number of false positives to a value similar to the IRDIS one.

\begin{figure}[t!]
\centering
  \includegraphics[width=\linewidth]{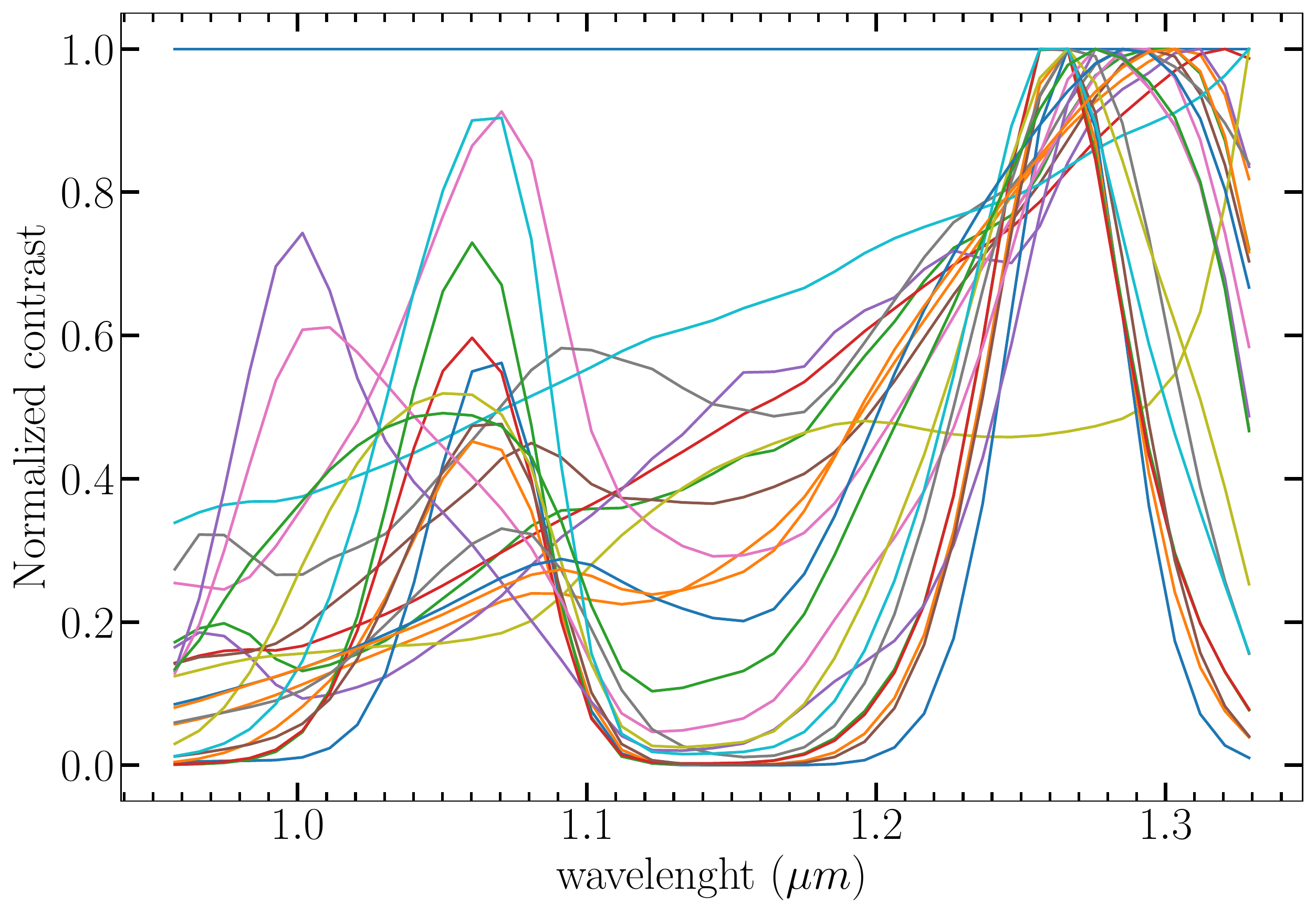}
  \caption{Example of the library $\M \Omega_{\text{IFS}}\, \in \mathbb{R}^{20 \times 39}$ of spectral priors built for a G2 star in YJ bands. Priors are expressed in contrast unit and are normalized between 0 and 1. We systematically included in the library a \textit{flat} spectral prior giving the same weights to all spectral channels.}
\label{example_library}
\end{figure}

The whole optimization process for a given stellar spectral type and for a targeted number $N$ of spectral priors is described in the form of a pseudo-code by Algorithm \ref{opti_ifs}. This selection procedure was repeated for all stellar spectral types considered in this work. Figure \ref{example_library} shows an example of such built library of spectral priors for a G2 star observed in YJ bands.

Based on the built library of spectral priors, we can now, as for IRDIS, compare the empirical false positive rate with the theoretical value. We can use again eq. \ref{eq1} with the number of processed pixels for IFS $\textit{n}_{\text{pixel}} = 140 000$. We find $\textit{N}_{\text{fp}} = 0.04$ for IFS. We can see with Fig. \ref{snr5_fp_ifs} that, as for IRDIS, the false positive rate when considering  individual S/N map is in good agreement with what is expected from a Gaussian noise distribution.

\subsection{Refined astrometric and photometric error budgets}
\label{subsec:astro_photo_err}

The PACO algorithm provides only fitting errors for both photometric and astrometric measurements. To get a complete error budget, we need to take into account other sources of errors.

For the astrometry error budget, we use both the results of the F150 analysis with the SpeCal pipeline \citep{shine2} and the calibration obtained by \cite{2021JATIS...7c5004M}. However, the F150 analysis used an average value of the typical centering error. Thanks to the  improved frame centering routine described in Sect. \ref{sec:data_reduction_frame_centering}, we are able to derive a precise estimate of this error on each data set, and we propagate it through the whole pipeline.

For the photometry error budget, we use, for the first time, the differential tip-tilt sensor (DTTS,  \cite{2010SPIE.7735E..5BB}, see Sect. \ref{subsubsec:photometric_error} for the details) measurements to derive proper photometric error bars. 

Although such precise analysis has already been done on specific targets for some particular studies, nothing was implemented routinely to perform massive analysis. 
This new analysis allows us to perform a complete, accurate, automated, and homogeneous error estimation for both astrometry and photometry.

\subsubsection{Astrometric error budget}

The PACO algorithm provides an astrometric fitting error term, hereafter denoted $\sigma_{\text{sep, PACO}}$ and $\sigma_{\text{PA, PACO}}$, associated respectively to the angular separation (sep) and to the parallactic angle (PA) of a given signal. Several additional sources of errors induced by pre-processing steps, such as the recentering of the individual frames, or  systematics related to SPHERE itself must be considered. We therefore combine several additional terms to refine the global error budget. The uncertainties associated with the separation and PA are found as follows.

\medskip

\noindent For the uncertainties on the angular separation, we combine four terms:
\begin{itemize}
    \item A distortion error of 0.4 mas at 1 as \citep{2021JATIS...7c5004M}, scaling linearly with the separation of a source:
    \begin{eqnarray}
    \sigma_{\text{dist}} \, (\text{as}) = \text{sep (as)}\times \frac{0.4}{1000} \,.
    \end{eqnarray}
    \item A plate scale error, scaling linearly with the separation of a source. This plate scale and the associated error bars are measured during each observing run (astrometric calibration, see \cite{shine2}:
    \begin{eqnarray}
    \sigma_{\text{platescale}} (\text{as}) = \frac{\text{sep (as)}}{\text{platescale}}\times \text{err}_{\text{platescale}}  \,.
    \end{eqnarray}
    \item An error on the re-centering of the individual frames, as estimated by the re-centering procedure described in Sect. \ref{sec:data_reduction_frame_centering}: 
    \begin{eqnarray}
    \sigma_{\text{recentering}} (\text{as}) = \sigma_{\text{recentering}} (\text{pxl}) \times \text{platescale} \,.
    \end{eqnarray}
    \item  PACO internal error $\sigma_{\text{sep, PACO}}$.
\end{itemize}
Those 4 terms are quadratically combined to obtain the full error budget $\sigma_{\text{sep,tot}}(\text{as})$.

\medskip

\noindent For the uncertainties on the PA, we also combine four terms:
\begin{itemize}
    \item An error on the pupil angle equal to 0.52 mas at 1 as \citep{2021JATIS...7c5004M}, scaling linearly with the separation:
    \begin{eqnarray}
    \sigma_{\text{PA~angle}}(\degree) = \arctan{\left(\frac{0.52}{1000}\times \text{sep (as)} \right)}\times \frac{180}{\pi} \,.
    \end{eqnarray}
    \item An error associated with the true North, as measured using the astrometric calibrations, $\sigma_{\text{TN}}(\degree)$.
    \item An error on the re-centering of the individual frames, as estimated by the re-centering procedure (see Sect. \ref{sec:data_reduction_frame_centering}):
    \begin{eqnarray}
    \sigma_{\text{recentering}}(\degree) = \frac{\sigma_{\text{recentering}}(\text{as})}{\text{sep (as)}} \times \frac{180}{\pi} \,.
    \end{eqnarray}
    \item PACO internal error $\sigma_{\text{PA, PACO}}(\degree)$.
\end{itemize}
Those 4 terms are also quadratically combined to obtain the full error budget $\sigma_{\text{PA,tot}}(\degree)$.

\subsubsection{Photometric error budget}
\label{subsubsec:photometric_error}

Our aim here is to estimate the relative photometric uncertainties using SPARTA data \citep{2012SPIE.8447E..2QS} as well as using information from the DTTS. 
The DTTS is a control organ of SPHERE that ensures that the star is always well centered on the coronagraph. It diverts a small fraction of the stellar light to produce an image of the star, thus allowing us to have a direct access to a PSF during the observation. While this PSF is not exactly the same as it would be on the science cameras due to non-common aberrations, it can still be used to monitor the photometric variability during the observing sequence. In the following, these series are denoted by $\text{DTTS}(t)$ as a function of the time $t$ of observation. SPARTA is the real time control computer of the adaptive optics system. During the course of an observation, SPARTA collects information on the observing conditions that are then stored. 
As for astrometry, the PACO algorithm provides a fitting photometric error $\sigma_{\text{PACO}}$, but additional terms are needed to estimate the global error budget:
\begin{itemize}
    \item  An error associated with the flux calibration of the coronagraphic frames using the PSF. Because the observing conditions vary during the observing sequence, this error is time dependent. Using datasets with bright background companions, detectable with a high S/N on each individual frame, we found that the flux variations are well correlated with the Strehl ratio (SR) variations as provided by SPARTA\footnote{\samepage \href{https://www.eso.org/sci/facilities/develop/ao/tecno/sparta.html}{https://www.eso.org/sci/facilities/develop/ao/tecno/sparta.html}}.
    Besides, the photometry of our faint sources cannot be estimated on each frame. Hence, we use the time series SR($t$) data taken during the observation to estimate an average photometric error over the whole coronagraphic sequence.   
    
    Because SPARTA provides the SR at 1.6 \text{$\upmu$m}, we compute the Strehl ratio $\text{SR}_{\lambda_\text{i}}(t)$ at the working wavelength $\lambda_\text{i}$ by using the following approximation (based solely on the adaptive optics fitting error and the Maréchal approximation \citep{marechal_approx}) to capture the wavelength dependency for good conditions: 
    \begin{eqnarray}
    \text{SR}_{\lambda_\text{i}}(t) = \text{SR}(t)^{\left(\frac{1.6}{\lambda_i}\right)^2} \,.
    \end{eqnarray}
    Then we compute the standard deviation of the SR, after removing the values for rejected frames, as follows:
    \begin{eqnarray}
    \text{var}_{\text{flux}, \lambda_i} = \frac{\sigma(\text{SR}_{\lambda_\text{i}}(t))}{\text{SR}_0} \,,
    \end{eqnarray}
    where $\text{SR}_0$ is the SR of the PSF used. In  PACO, it is the average between the off-axis PSF taken before the observing coronagraphic sequence and the one after. The error associated to the extracted spectra at each wavelength $\lambda_i$ is therefore:
    \begin{eqnarray}
    \sigma_{\text{norm},\lambda_i} = \text{spectra}_{\lambda_i} \times \text{var}_{\text{flux}, \lambda_i} \,.
    \end{eqnarray}
   
    It sometimes happens that no SPARTA data are available. In such a case, the error is computed using the difference between the two available PSFs.
    
    \item The sky transparency: this term is measured thanks to the DTTS. 
    While the peak of the DTTS PSF is also linked to the SR variation recorded by SPARTA, the total integrated flux from the DTTS PSF directly relates to the evolution of the sky transparency, but does not provide an absolute photometric measurement. By estimating the DTTS flux, we estimate the median and standard-deviation of the sky transparency variations over the coronagraphic sequence:
    \begin{eqnarray}
    \text{var}_{\text{transp}} &=& \frac{\sigma(\text{DTTS}(t))}{\text{median}(\text{DTTS}(t))} \,, \\
    \sigma_{\text{transp},\lambda_i} &=& \text{spectrum}_{\lambda_i} \times \text{var}_{\text{transp}} \,.
    \end{eqnarray}
    \item  PACO internal error $\sigma_{\text{PACO}}$.
\end{itemize}
Those three terms are then quadratically summed to obtain the full photometric error budget.

\subsection{Multi-epoch discrimination tool}
\label{subsec:MEDT}

Second epoch observations are  crucial to test whether a signal is due to a source gravitationally bound to the target star. We developed a tool that automatically performs  this analysis. In the case where the candidate is not recovered, it computes the false positive probability, given the detection limits achieved for the second epoch. 
\medbreak
In practice, we first check whether a source is detected in the second epoch data at the  position expected from a background object, knowing the proper motion of the star\footnote{Provided by Simbad: \href{http://simbad.u-strasbg.fr/simbad/}{http://simbad.u-strasbg.fr/simbad/}.}. This step allows us to identify the background sources (providing that the detection limits of the second epoch data set are good enough). If the signal cannot be associated with a background source, we search for a fainter signal (detected at $\tau \ge 4$) within a disk $D$ centered on the position of the source at the first epoch, and with a radius corresponding to the motion of a gravitionally bound object orbiting on a circular pole-on orbit (corresponding to the maximum possible motion in projected separation). If a signal is found, we attribute it to a possible companion. In case of an ambiguous choice (i.e., the motion of the object can be associated either with a background source or a gravitionally bound source) a flag is raised to report the ambiguity.
If no signal is found, we check whether the non-detection in the second epoch is due to poorer conditions or to the fact that the first detection was a false positive. To do so, we process as follows:
\begin{itemize}
    \item We compute the S/N (called $\text{S/N}_2$) that the signal should have in the second epoch using the second epoch contrast\footnote{The contrast is estimated by averaging the contrast in the disk $D$.} map.
    \item We measure the maximum S/N in the disk area in the second epoch data  ($\text{S/N}_{\text{max}}$).
    \item As the S/N maps follow a centered Gaussian distribution with a unit variance, we can estimate the probability $p$ of the source to be a real signal:
    \begin{eqnarray}
    p = \int_{-\infty}^{\text{S/N}_{\text{max}} - \text{S/N}_{\text{2}}} \frac{1}{\sqrt{2\pi}} e^{-0.5x^2} dx \,.
    \end{eqnarray}
\end{itemize}

\section{Contrast comparison with other algorithms}
\label{sec:contrast_comparison}

\begin{figure}[t!] 
\centering
  \includegraphics[trim={0.1cm 0.1cm 0.1cm 0.1cm},clip,width=\linewidth]{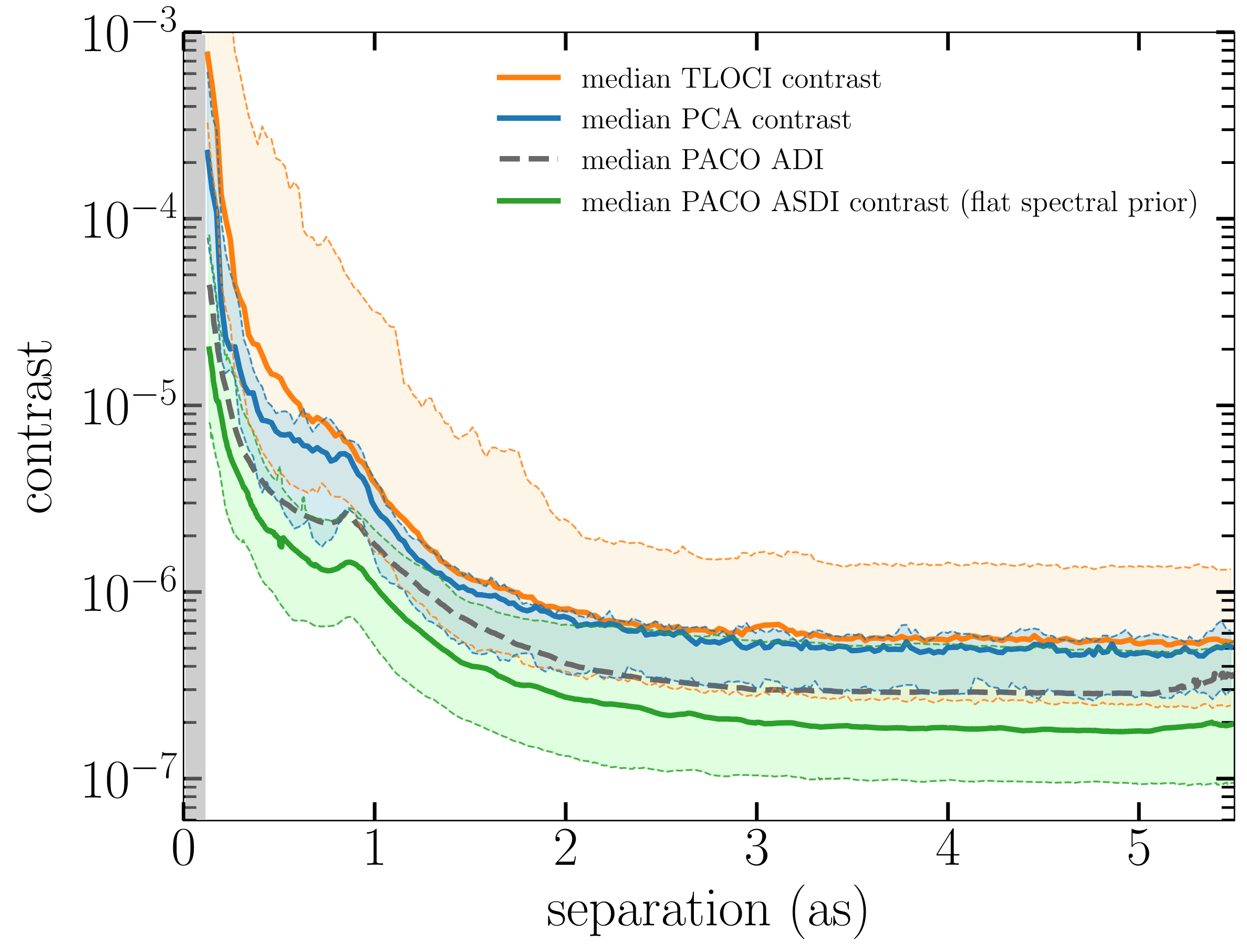}
  \caption{Contrast comparison at $5\sigma$ between PACO, TLOCI, and PCA for IRDIS for the 24 science stars considered in this paper. A $flat$ prior was used for PACO. Dashed lines show the 95\% completeness interval. The grey area represent the coronographic mask. Contrast curves provided by PCA and TLOCI do not strictly correspond to a $5\sigma$ false alarm rate contrarily to the contrast curves of PACO. The achievable contrasts are thus significantly over-optimistic for PCA and TLOCI, see discussion in the text.}
  \label{contrast_vs_TLOCI} 
\end{figure}

Here, we present the contrast performance achieved by PACO on both IRDIS and IFS and compare them with the performance achieved by TLOCI \citep{MAROIS_TLOCI} and PCA ADI \citep{SOUMMER_PCA}; \cite{AMARA_PYNPOINT}) for IRDIS, and TLOCI and PCA ASDI for IFS \citep{pca_padova}. Those algorithms were used for the analysis of the SHINE F150 survey  \citep{shine2}. They are implemented in the SpeCal package \citep{galicher2018astrometric} dedicated to the analysis of the SHINE data with the following characteristics:

For TLOCI, the stellar profile is estimated frame by frame for each pixel in the field of view. The estimation uses a linear combination of all data to minimize the residuals after subtraction. The area on which the optimization is computed is much bigger than the subtraction area, thus mitigating as best as possible the self subtraction of point-like sources. The Specal implementation of TLOCI also assumes a flat planet spectrum in contrast. The parameters for TLOCI were set as they were in the F150 reduction which are:
\begin{itemize}
    \item The optimization zone is separated by 0.5 full width at half maximum (FWHM) from the region of interest to avoid bias in the linear combination in case of the presence of a source in the region of interest;
    \item The radial width (in radius) of the subtraction zone is set to one FWHM;
    \item The radial to azimuthal ratio of the subtraction zone is set to 1.5;
    \item The optimization zone area is set to 20 PSF FWHM.
\end{itemize}

Both PCA algorithms are based on the equations described in \cite{SOUMMER_PCA}. For IRDIS (ADI processing), the principal components (PCs) are computed independently for each spectral channel. For IFS (ASDI processing) the PCs are computed using the spatial and spectral dimensions simultaneously. For IRDIS, 5 PCs were used, and for IFS, 50, 100 and, 150 PCs were used.

The throughput of both TLOCI and PCA is estimated internally by SpeCal at each location in the field of view by generating a datacube with fake planets. The 1D throughput curve is then applied to the residual maps of both algorithms.

The contrast curves/maps are estimated for each spectral channel by computing the pixel by pixel azimuthal  standard deviation in an annulus of 0.5 FWHM on the residual maps, once corrected from the throughput. The 5$\sigma$ detection limits are derived from this estimation by taking into account several corrections: the flux loss from ADI subtraction, the coronograph transmission, and the neutral density of the off-axis PSF. Finally, these detection limits are normalized by the off-axis PSF flux. S/N maps are directly derived from the estimated flux and its associated standard-deviation (i.e., contrast at 1$\sigma$).

We present in Sect. \ref{subsec:contrast_comp} the contrast comparison between PACO and the algorithms described above. Since this direct comparison of contrast is biased by the diverse hypotheses made by each algorithm, we then present in Sect. \ref{subsec:reliability} a set of numerical experiments resorting to massive injections of FPSs. This demonstrates the reliability of the  contrast curves obtained with PACO, as well as the gain in sensitivity and control of the probability of false alarms compared with the two other algorithms.

\subsection{Contrast performance}
\label{subsec:contrast_comp}

\begin{figure}[t!] 
\centering
  \includegraphics[trim={0.1cm 0.1cm 0.1cm 0.1cm},clip,width=\linewidth]{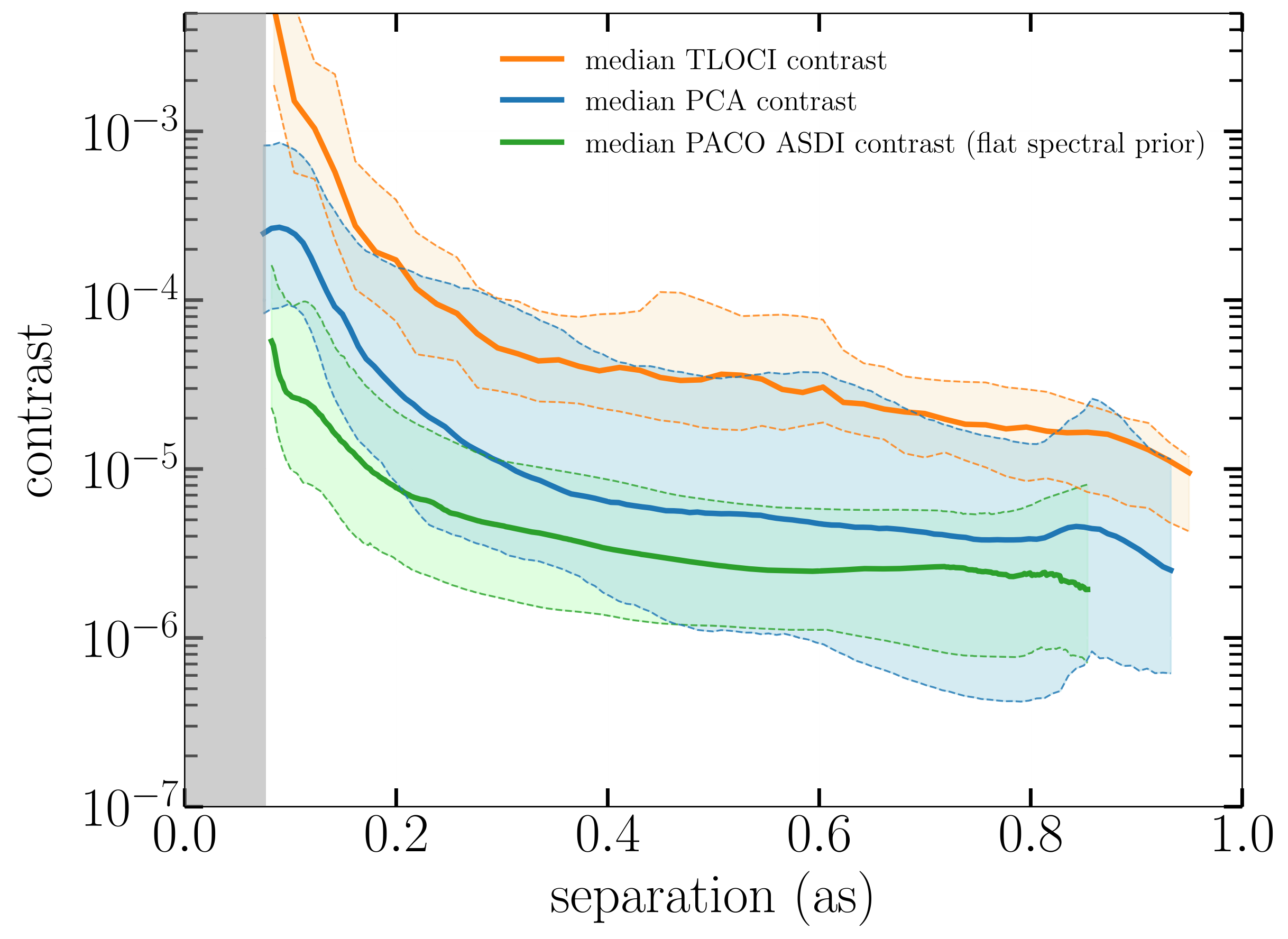}
  \caption{Contrast comparison at $5\sigma$ between PACO, TLOCI, and PCA for IFS for the 24 science stars considered in this paper. Dashed lines show the 95\% completeness interval. The grey area represents the coronographic mask. Contrast curves provided by PCA and TLOCI do not correspond to a $5\sigma$ false alarm rate contrarily to the contrast curves of PACO. The achievable contrasts are thus significantly over-optimistic for PCA and TLOCI, see discussion in the text.}
  \label{contrast_vs_TLOCI_ifs} 
\end{figure}

Figure \ref{contrast_vs_TLOCI} (respectively, Fig.  \ref{contrast_vs_TLOCI_ifs}) shows the predicted $5\sigma$ contrast as a function of the angular separation on IRDIS (respectively, IFS) obtained with the three considered algorithms for all epochs of the 24  stars considered in this study. 
For this comparison, we use the combined contrast of TLOCI  and PCA, as well as the contrast obtained with the $flat$ prior with PACO (hereafter noted \textit{PACO-flat}) to possibly make the most direct comparison. Although not used in this study, we also included the median PACO ADI contrast curve (i.e., obtained without joint processing of the spectral channels) in Fig. \ref{contrast_vs_TLOCI} for reference. The gain offered by the ASDI mode is very close $\sqrt{2}$, which corresponds to the expected theoretical value when combining the information of two (independent) channels. In order to compare results from the three considered algorithms, it has to be noted that, given the non-statistical nature of TLOCI and PCA, the $5\sigma$ detection limits are not statistically grounded, i.e. we experienced in practice many more false alarms as theoretically expected for the targeted confidence level, especially at short angular separations from the target star. Moreover, the $flat$ prior represents the most difficult case for PACO (especially for IFS) because we try to detect a signal with the same spectra as the host star, which means that the spectral prior does not explicitly help in disentangling the two components.

For IRDIS, at close angular separations, PACO performs better than TLOCI (resp. PCA) by a factor of about 7 (resp. 5) at 0.5$\arcsec$, and by 5 at 1$\arcsec$ and beyond compare to both TLOCI and PCA. Moreover, due to the statistical nature of PACO, the number of false positives follows what is expected at a $5\sigma$ confidence under a multivariate Gaussian hypothesis (see Fig. \ref{fp_vs_priors}), unlike TLOCI and PCA.

For IFS, the gain ranges between a factor 3 and a factor 5, depending on the separation, compared to PCA. It is much larger compared to TLOCI by about a factor 10 for all separations, but this result is expected as TLOCI performs worse on IFS as compared to PCA. However, the achieved performance in terms of contrast is much more consistent with PACO than it is with PCA. We remind that using the flat prior as a benchmark allows us to compare PACO with PCA/TLOCI as fairly as possible. It does not however represent the full capability of PACO to detect faint sources because this is the \textit{worst} possible case, as we are trying to detect a highly correlated planetary spectrum with respect to the star spectrum.

\subsection{Validation of the reliability of the contrast curves}
\label{subsec:reliability}

Each algorithm uses different hypothesis to compute the contrast limits.
To further assess the gain in contrast obtained with PACO, and the comparison with TLOCI or PCA, we present a set of numerical experiments. With PACO, the contrast is estimated assuming that the statistical parameters (mean and covariance matrices) characterizing the stellar leakages are computed from pure noise realizations. In practice, the underlying presence of the (unknown) sought objects corrupts these estimates, which leads to a (slight) bias in the estimated performance. In this section, we aim at quantifying this bias via numerical experiments. For that purpose, we resort to massive injections of FPSs at contrast levels predicted by PACO and we re-run the algorithm to quantify the real detection confidence experienced for such levels of contrast. Ensuring such confidence is key to allow an unsupervised selection of candidate companions through simple thresholding of the derived S/N map. Given computational constraints, injection tests (as presented in Sect. \ref{subsubsec:ird_inj_data} \& \ref{subsubsec:ifs_inj_data}) were performed on a dataset of HD 377 (star included in this study) for both IRDIS and IFS.

\subsubsection{IRDIS data}
\label{subsubsec:ird_inj_data}

\begin{figure}[t!]
\centering
\includegraphics[trim={0.1cm 0.1cm 0.1cm 0.1cm},clip,width=\linewidth]{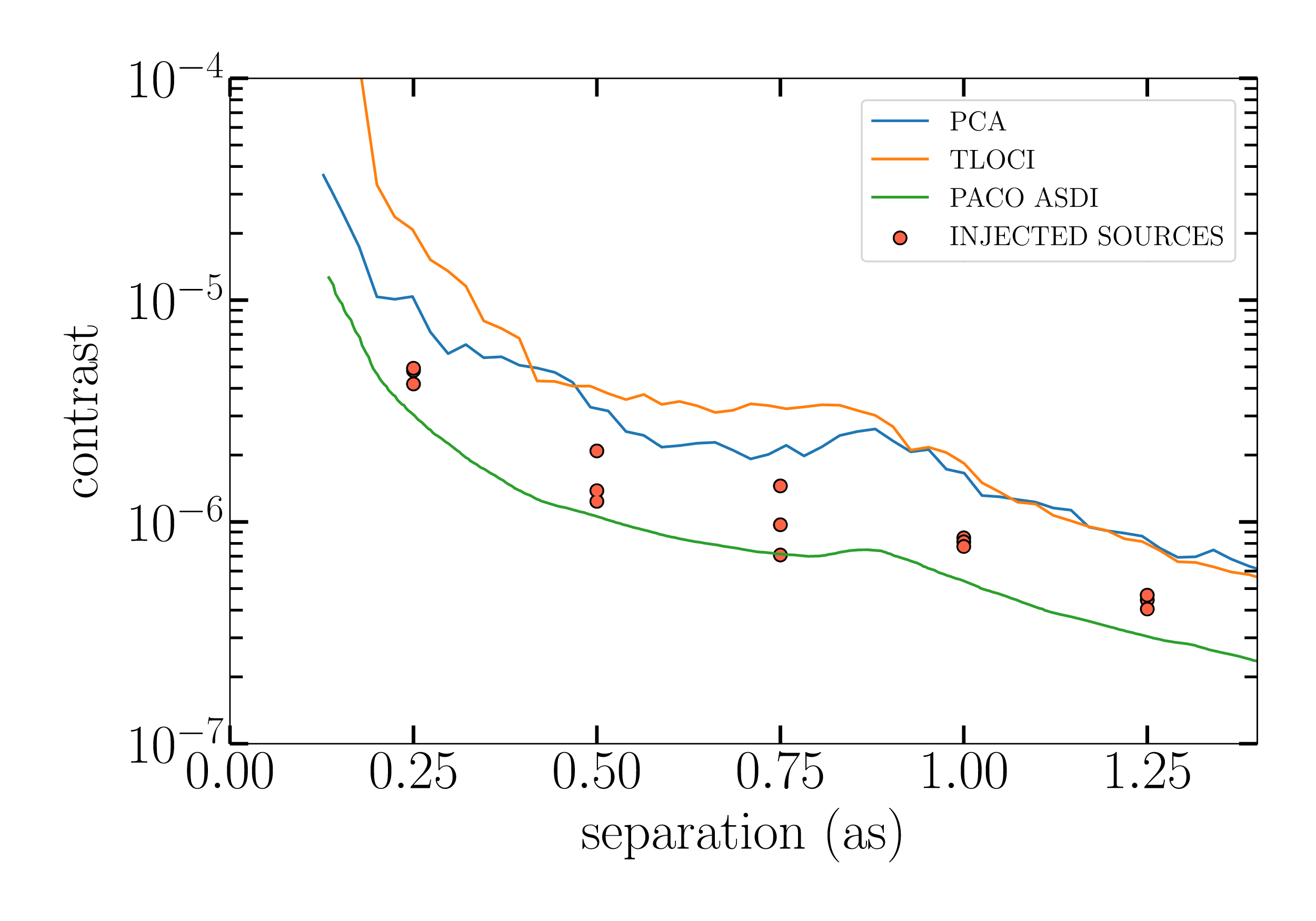}
\caption{Comparison of the contrast curves at $5\sigma$ obtained with PCA (5 modes), TLOCI, and \textit{PACO-flat} for IRDIS on HD 377. The contrasts of the injected fake planets were computed using 2-D contrast maps, hence the differences with the $5\sigma$ curve: local variations of the achieved contrast are averaged azimuthally.}
\label{contrast_curve_comp_ird_close}
\end{figure}

\begin{figure*}[t!]
    \centering
        \begin{minipage}{.3\textwidth}
            \begin{subfigure}{\textwidth}
            \centering
            \includegraphics[width=\textwidth]{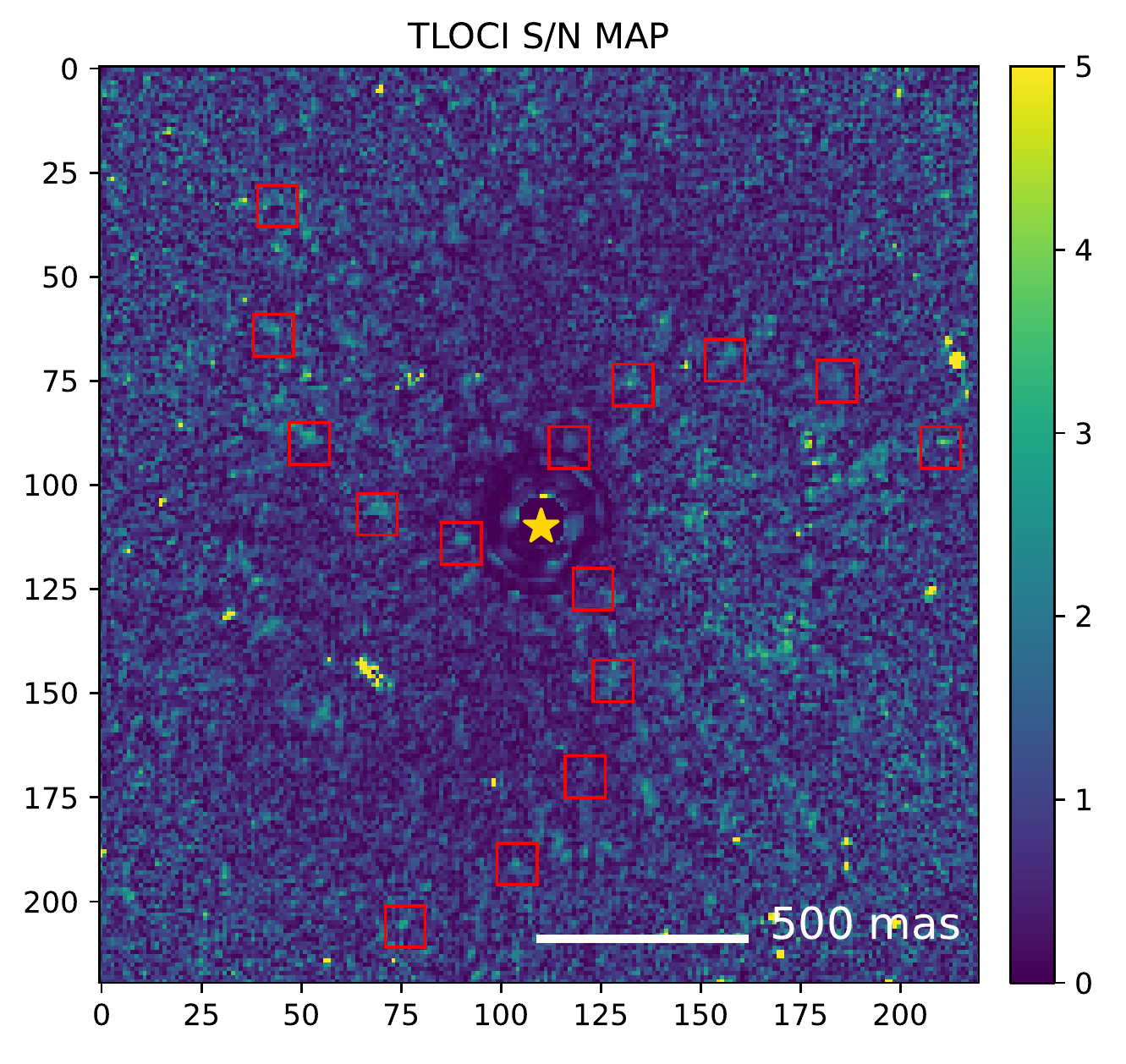}
            
            \end{subfigure}\\
            \begin{subfigure}{\textwidth}
            \centering
            \includegraphics[width=1.05\textwidth]{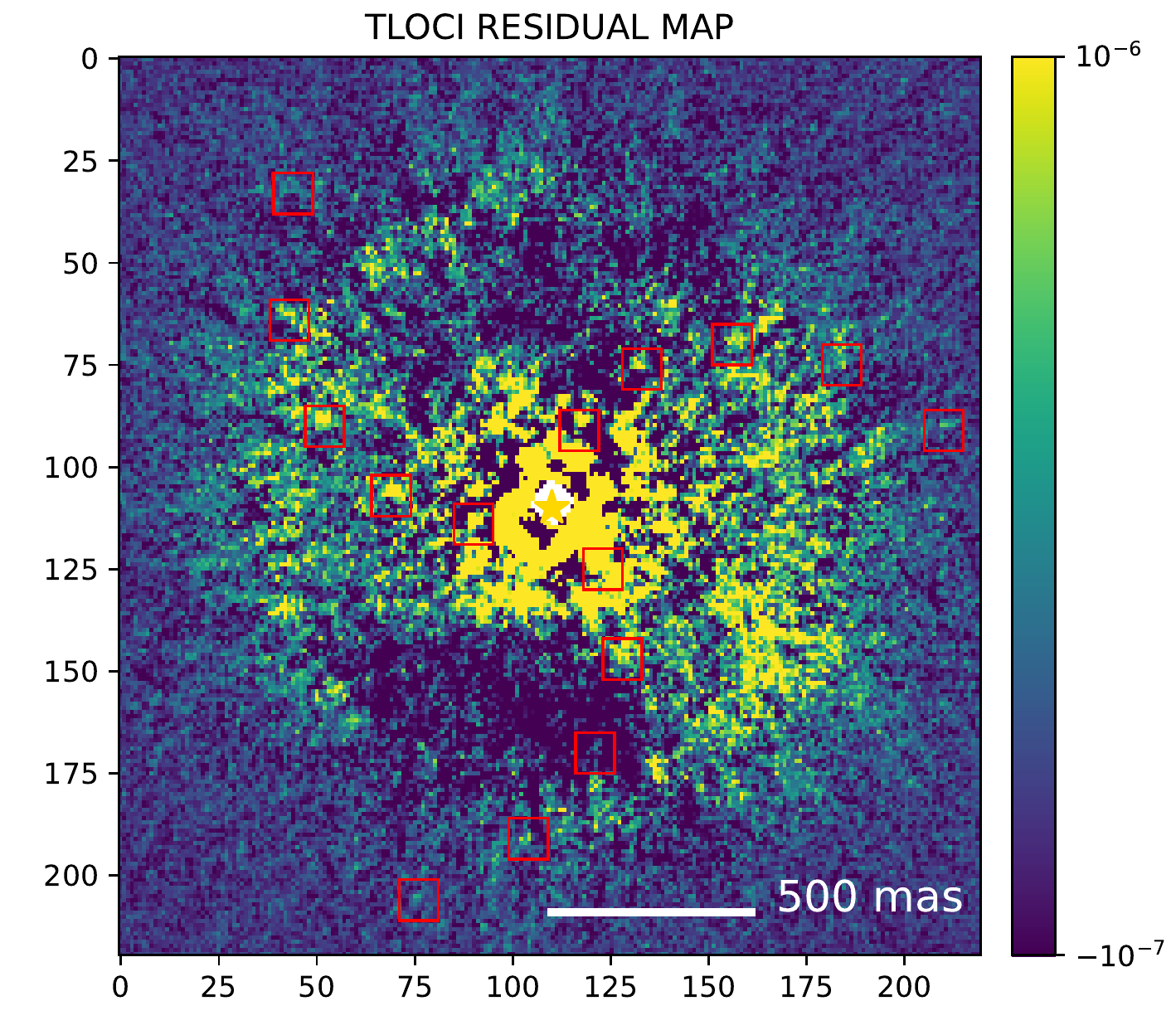}
            
            \end{subfigure}%
            
        \end{minipage}
        \hfill
        \begin{minipage}{.3\textwidth}
            \begin{subfigure}{\textwidth}
            \centering
            \includegraphics[width=0.98\textwidth]{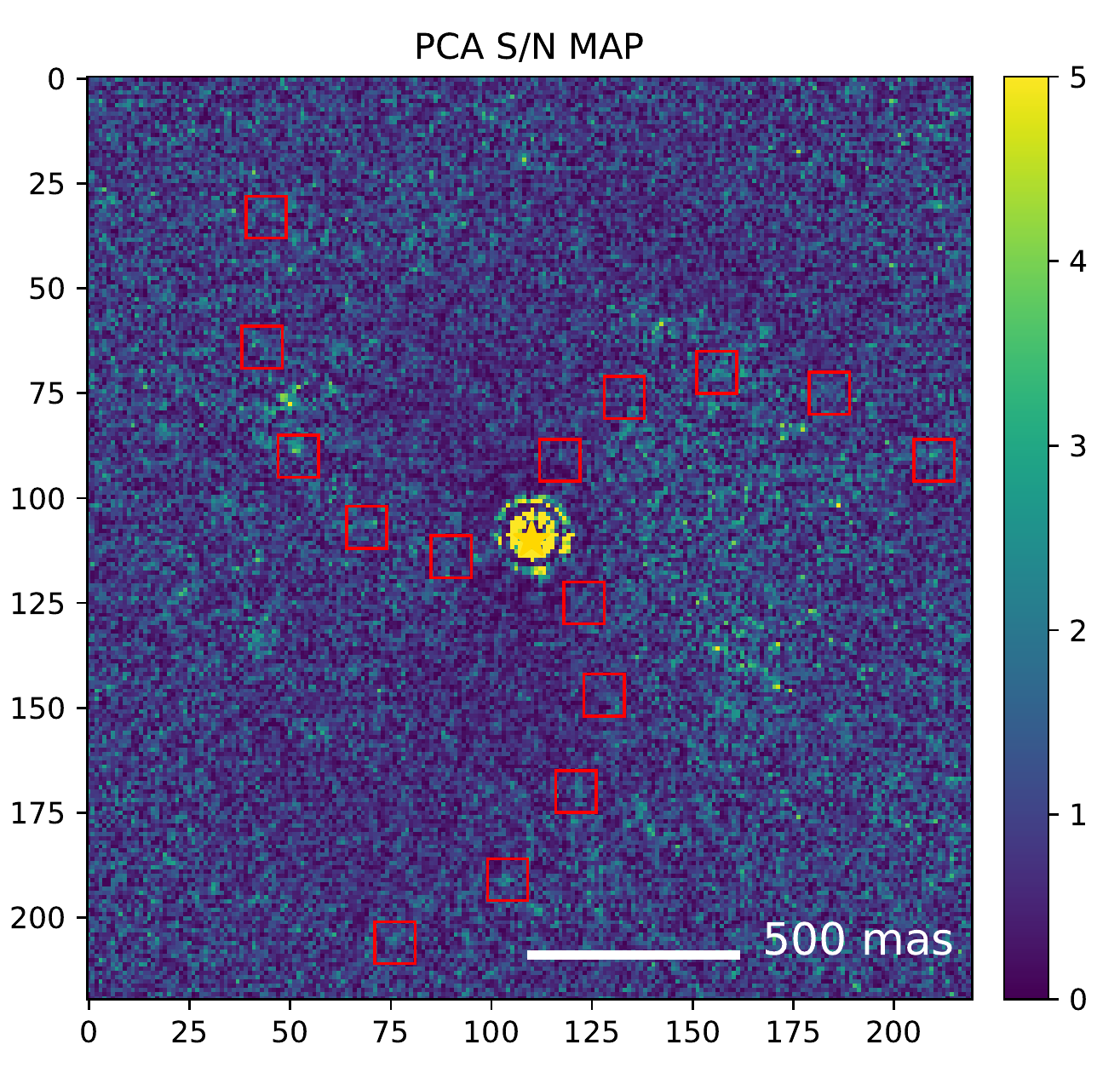}
            
            \end{subfigure}\\
            \begin{subfigure}{\textwidth}
            \centering
            \includegraphics[width=1.05\textwidth]{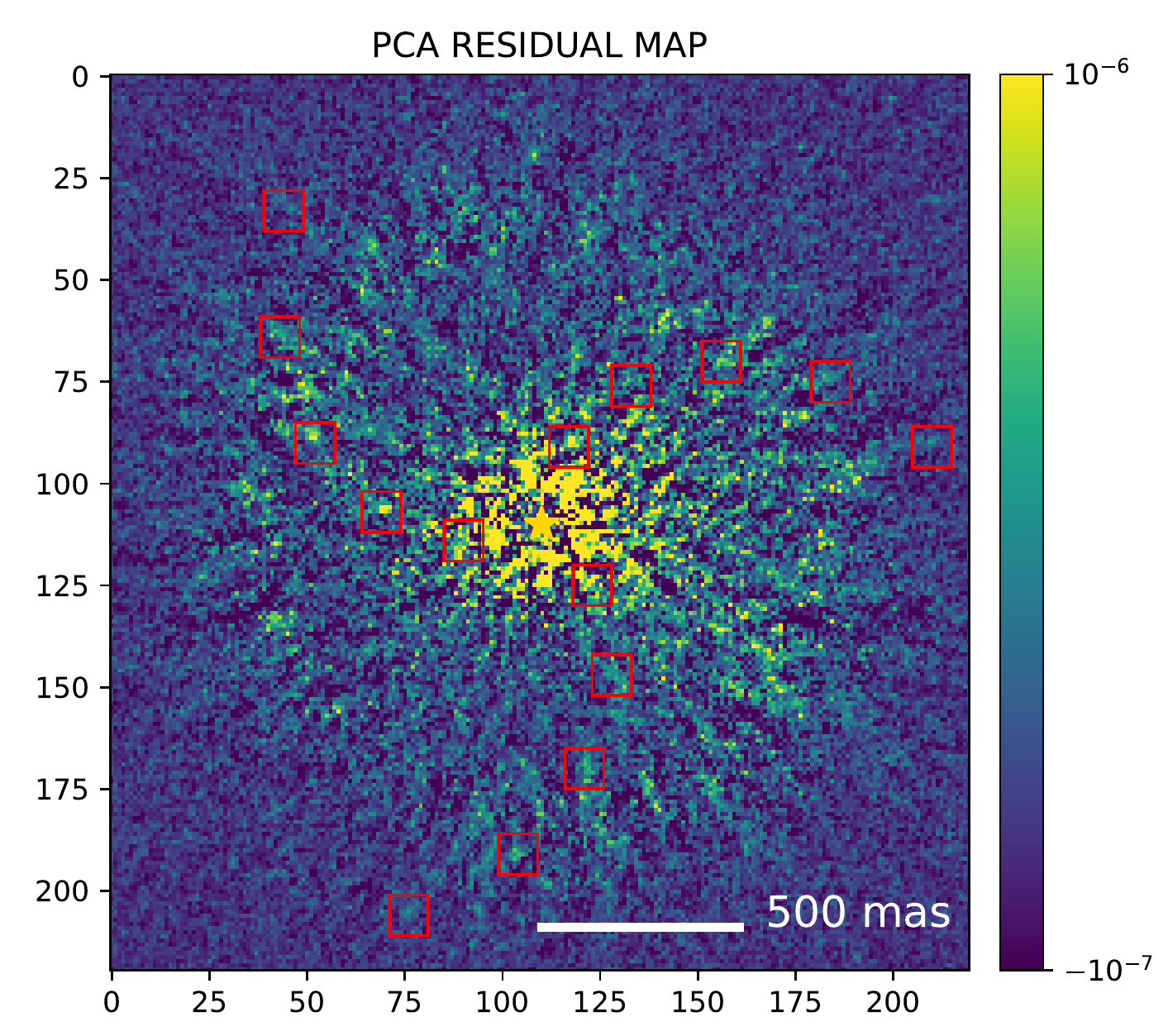}
            
            \end{subfigure}%
            
        \end{minipage}
        \hfill
        \begin{minipage}{.3\textwidth}
            \begin{subfigure}{\textwidth}
            \centering
            \includegraphics[width=\textwidth]{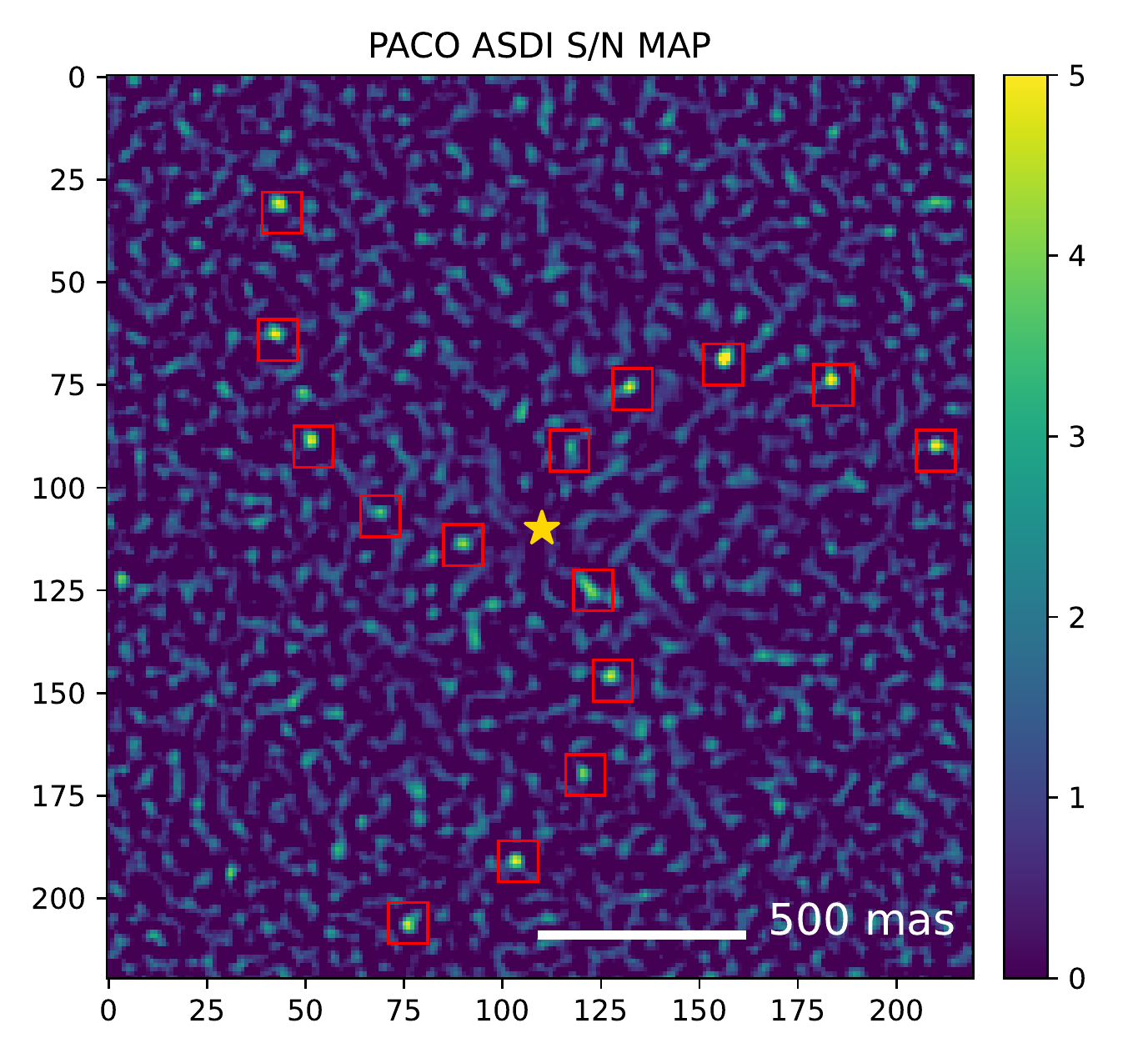}
            
            \end{subfigure} 
            \vspace{0.95\textwidth}
            
        \end{minipage}%
    \caption{S/N map provided by PACO (top right), residual map provided by SpeCal PCA using 10 modes (bottom middle), associated S/N map (top middle), residual map from TLOCI (bottom left) and associated S/N map (top left). The injected fake planets are clearly visible on the S/N maps from PACO. None of the high S/N detections on the PCA/TLOCI map corresponds to  injected sources. The locations of injected sources are highlighted by the red boxes.}
\label{inj_comp_ird_close}
\end{figure*}

For IRDIS, two sets were created; one with injections at close separations (< 1.25 as), and one at large separations (> 1.5 as), see Appendix \ref{appendix_inj}. For both, from the 2-D contrast maps provided by PACO, we set the contrast of the FPSs in order to achieve (theoretically) a detection slightly above the $5\sigma$ threshold on the S/N map. More detailed information on the close-in injected sources parameters can be found in Table \ref{table_inj_ird_close}. Figure \ref{contrast_curve_comp_ird_close} shows the injected sources contrast (red dots) compared to the contrast curve provided by PACO, as well as the contrast curves provided by TLOCI and PCA for this particular target in the first case (close separation).  Figure \ref{inj_comp_ird_close} shows the S/N maps obtained with the three algorithms as well as residual maps\footnote{PACO does not produce residual maps since it describes the stellar component through a statistical model rather than resorting to explicit combinations and/or subtractions of images.} obtained after subtraction of the estimated stellar component for PCA and TLOCI for the injections at close separations. 

We also created a set of injections of companions at larger separations. Detailed information on the injected sources parameters can be found in Table \ref{table_inj_ird_far}. Figure \ref{contrast_curve_comp_ird_far} shows the contrast of the injected FPSs compared to the contrast curve of the three algorithms. The corresponding S/N and residual maps obtained after subtraction of the estimated stellar component for PCA and TLOCI are shown in Figs. \ref{paco_snr_map_inj_far} to \ref{tloci_snr_map_inj_far}. 

\begin{figure}[t!]
\centering
\includegraphics[width=\linewidth]{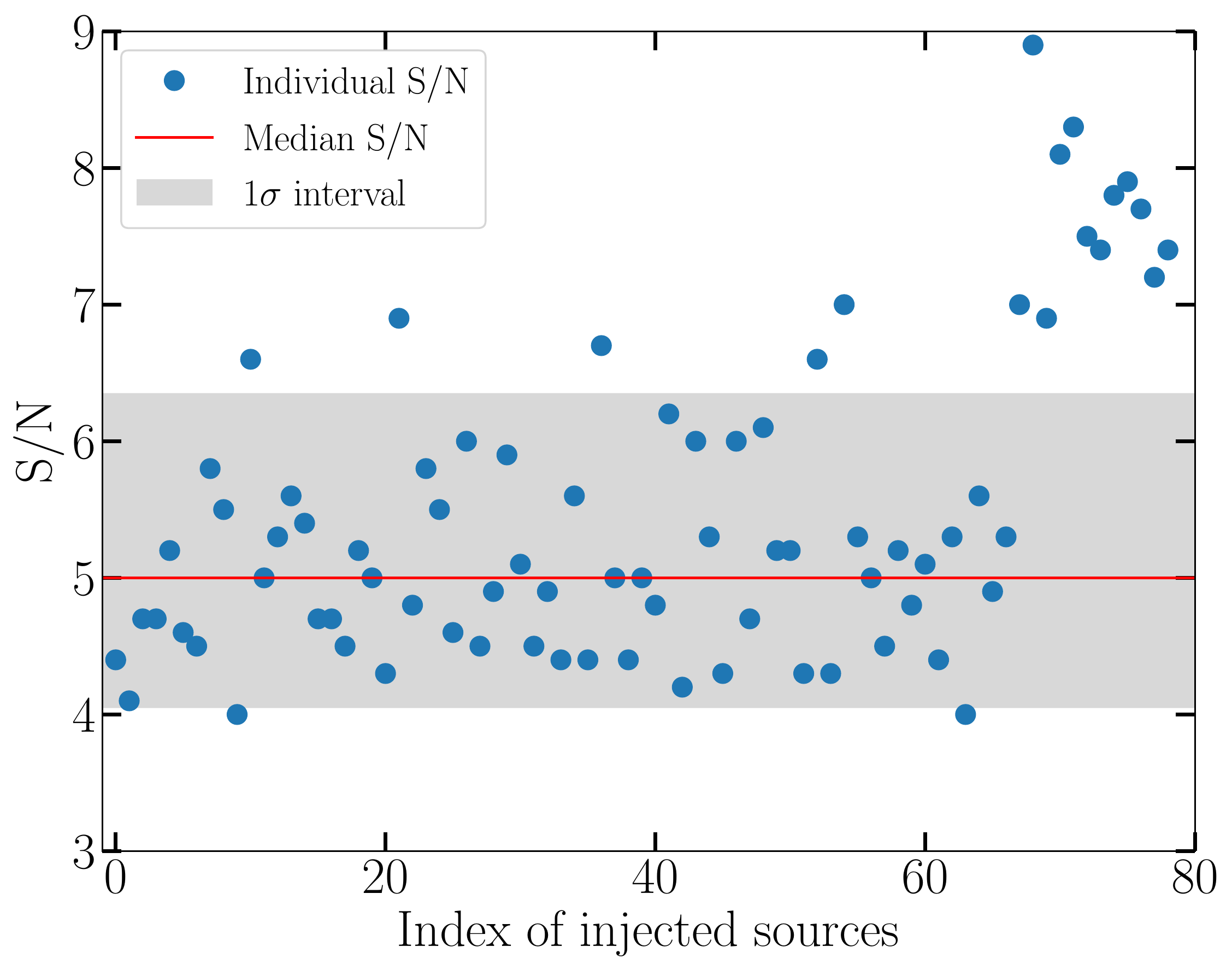}
\caption{Retrieved S/N for the injected sources for IRDIS. The detection threshold $\tau$ was set to 4. The red line shows the median S/N of the injected sources and the grey area the $1\sigma$ interval containing 68\% of the detected injected sources.}
\label{retrieved_inj_snr_ird}
\end{figure}

Figure \ref{retrieved_inj_snr_ird} shows the retrieved S/N of FPSs in both cases (for a total of 90 injected sources)\footnote{Because we use a 4$\sigma$ threshold for the analysis, we did not recover injections with a S/N below 4$\sigma$. We can however still find the median S/N by computing the S/N for which half of the injected sources (the 45 higher S/N sources in that case) have a higher S/N}. The median S/N is 5.0 which is in very good agreement with the contrast of injection. We also calculated the empirical standard deviation value $\sigma = 1.15$. This results is also close to the theoretical value predicted by a Gaussian distribution, which is 1.

\comment{
\begin{figure}[t!]
\centering
\includegraphics[width=\linewidth]{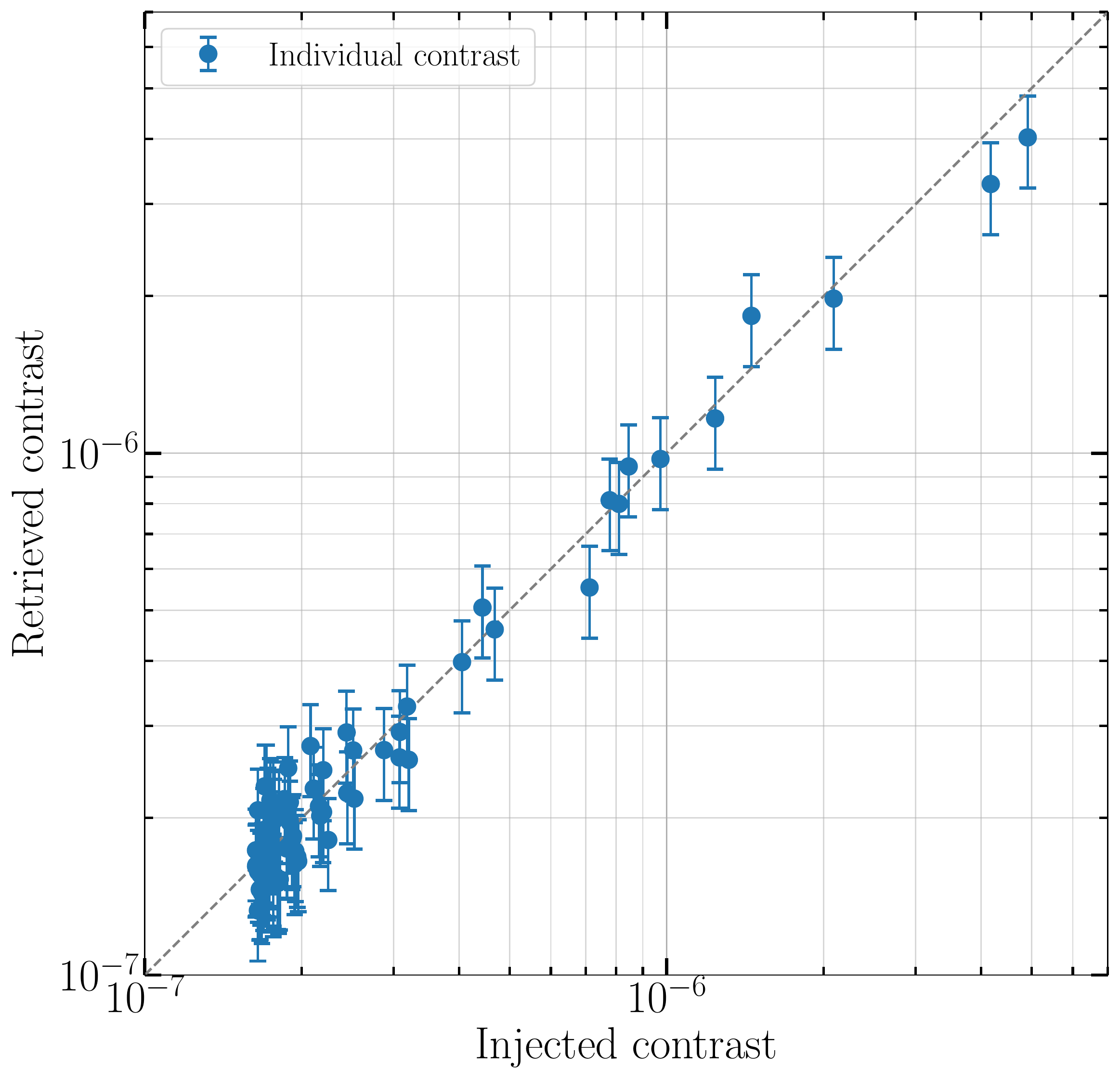}
\caption{Retrieved contrast for the injected sources for IRDIS.}
\label{retrieved_inj_contrast}
\end{figure}
}
These results confirm that the contrast estimates produced by PACO (e.g., shown in Fig. \ref{contrast_vs_TLOCI}) are reliable, statistically grounded, and that the detection sensitivity is improved compared to TLOCI and PCA, for the whole range of angular separations.

\subsubsection{IFS data}
\label{subsubsec:ifs_inj_data}

\begin{figure*}[t!]
\centering

\includegraphics[width=.3\textwidth]{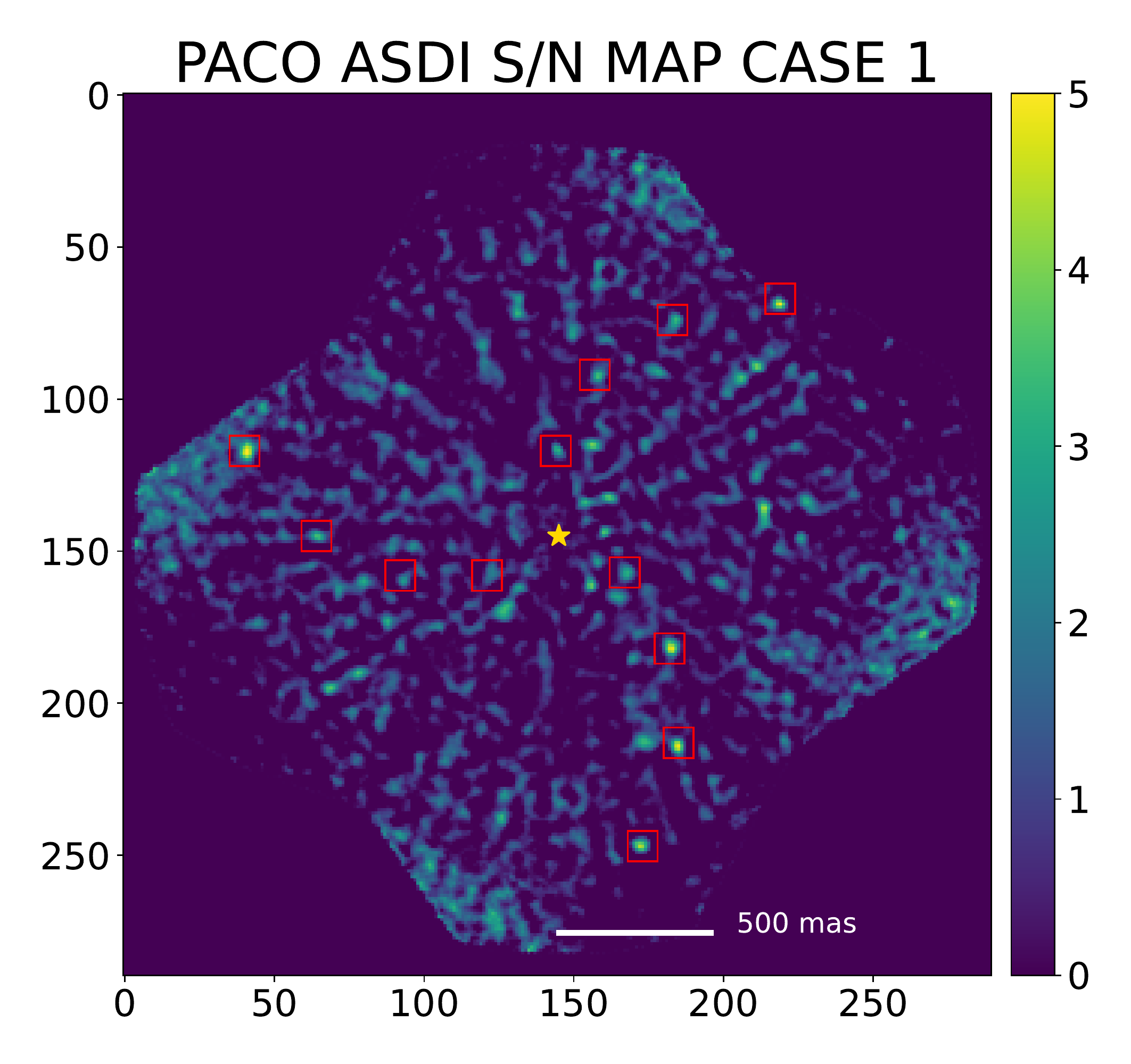}\quad
\includegraphics[width=.3\textwidth]{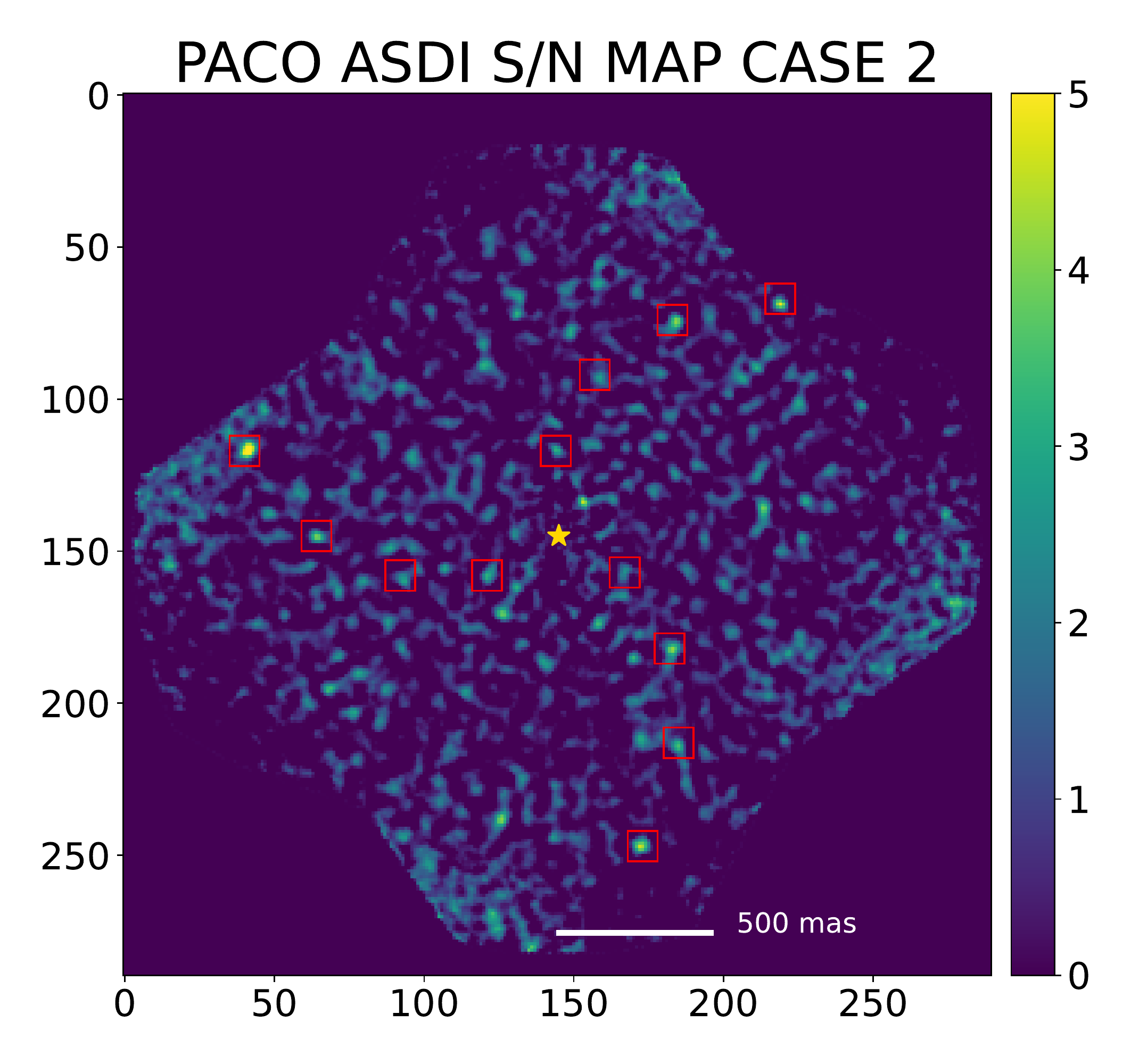}\quad
\includegraphics[width=.3\textwidth]{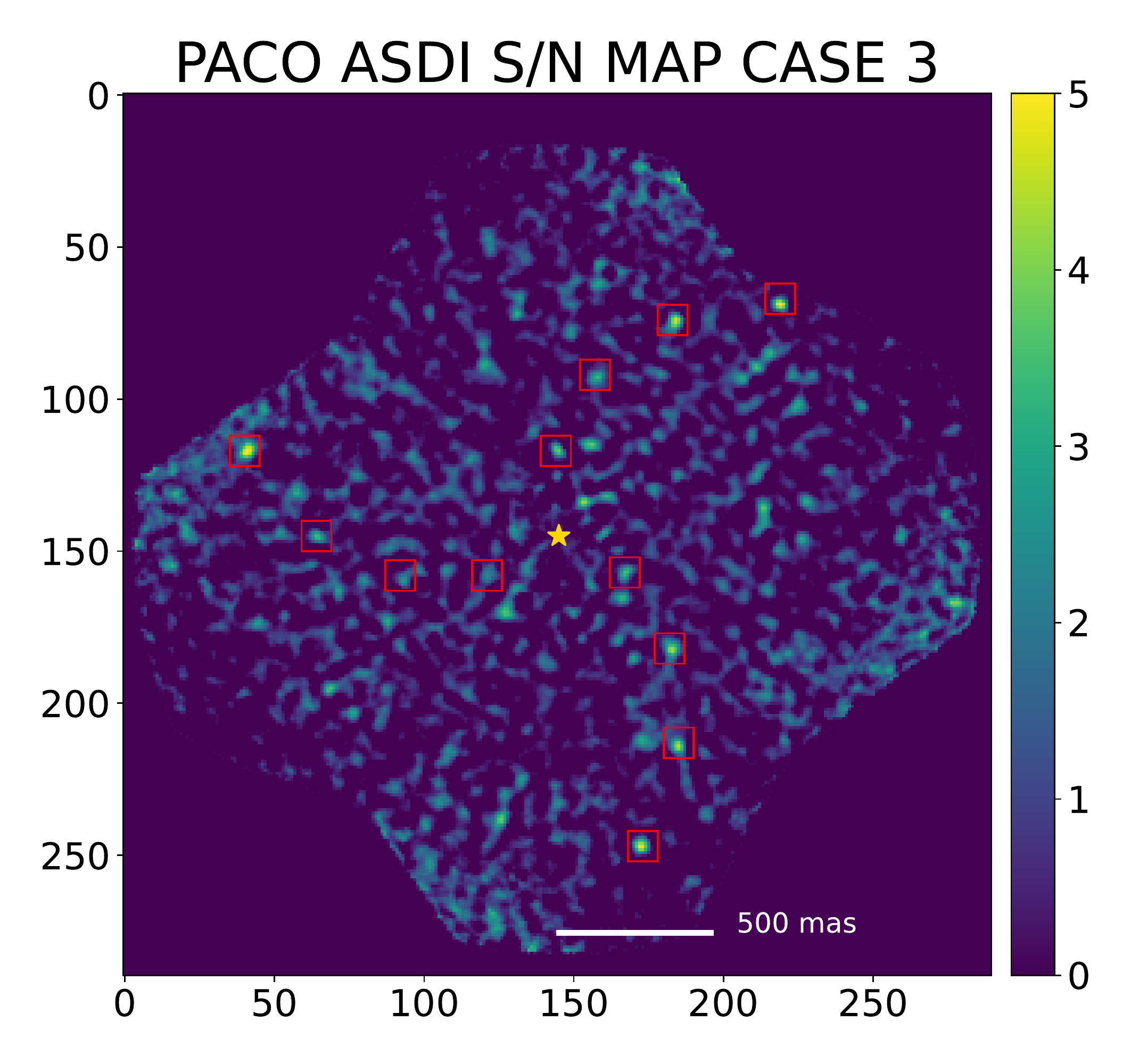}

\medskip

\includegraphics[width=.3\textwidth]{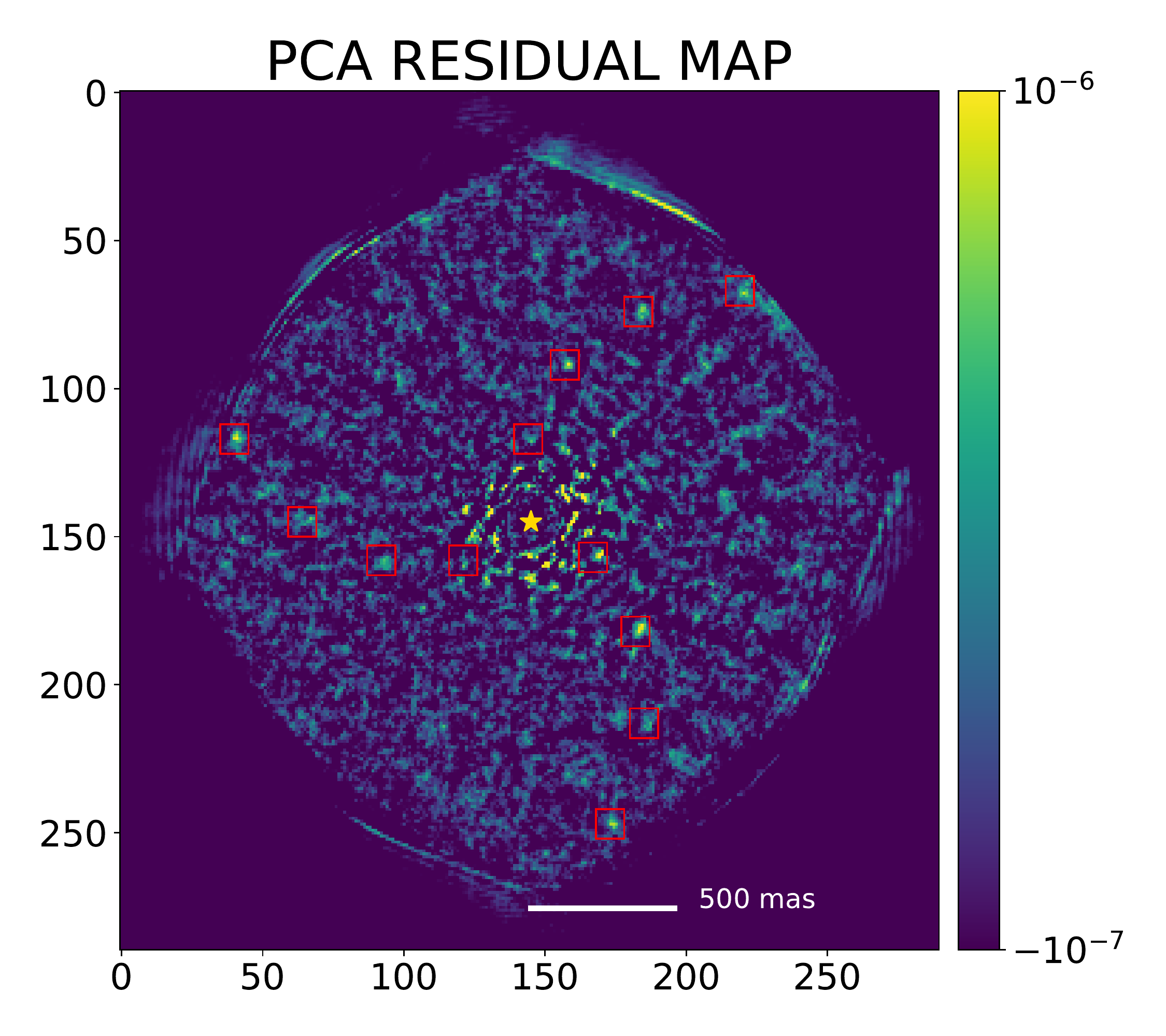}\quad
\includegraphics[width=.3\textwidth]{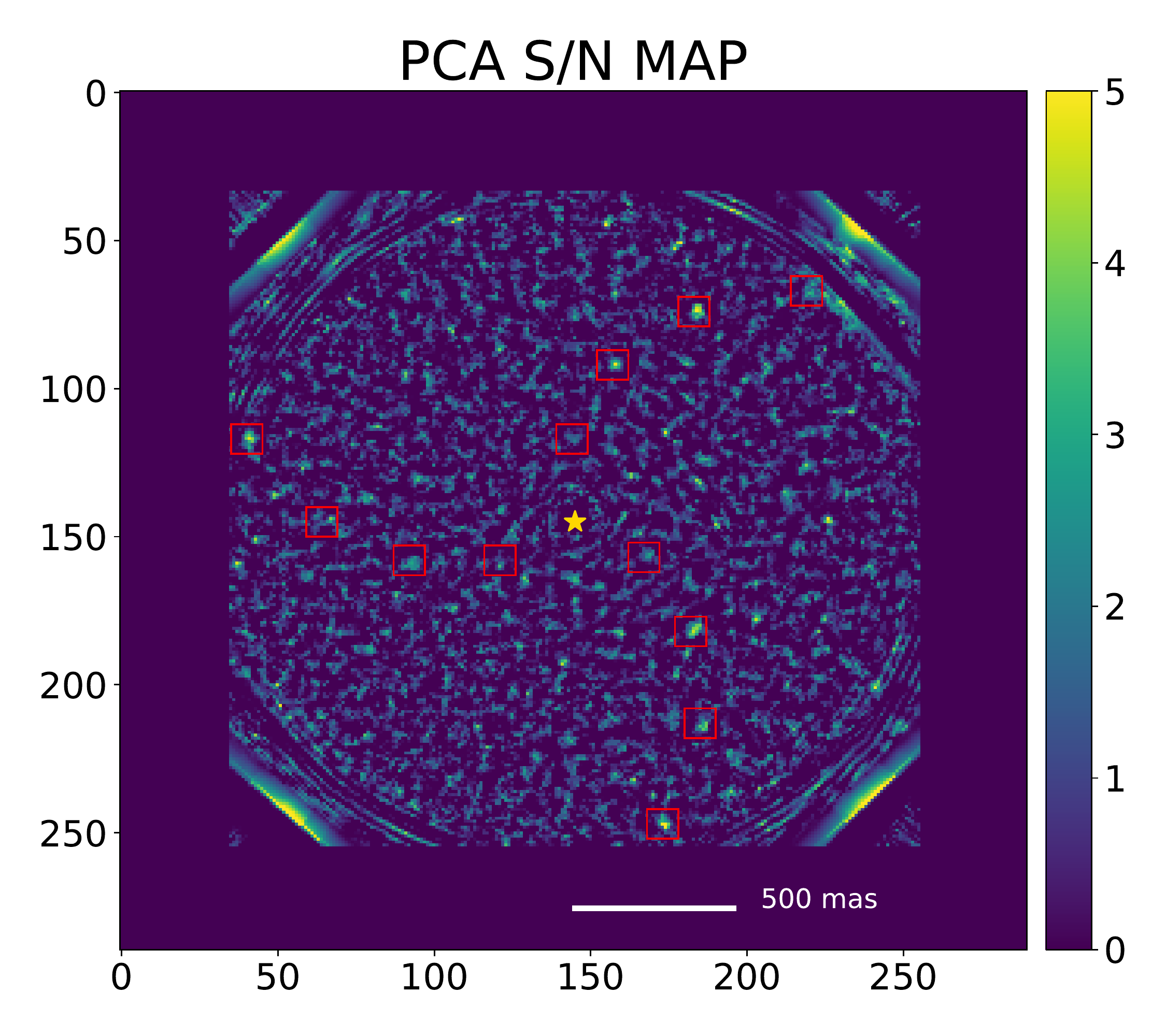}

\medskip

\includegraphics[width=.3\textwidth]{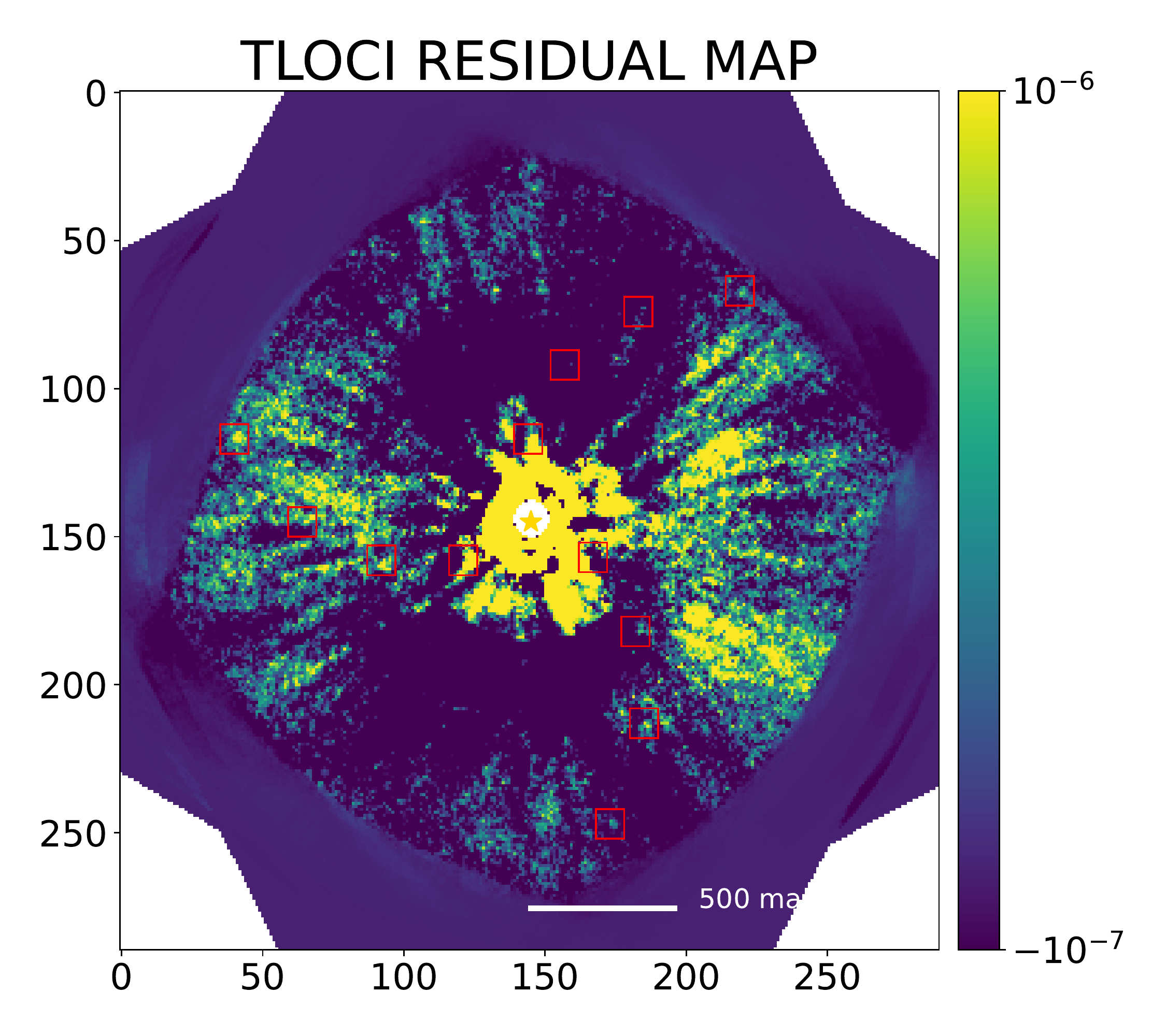}\quad
\includegraphics[width=.3\textwidth]{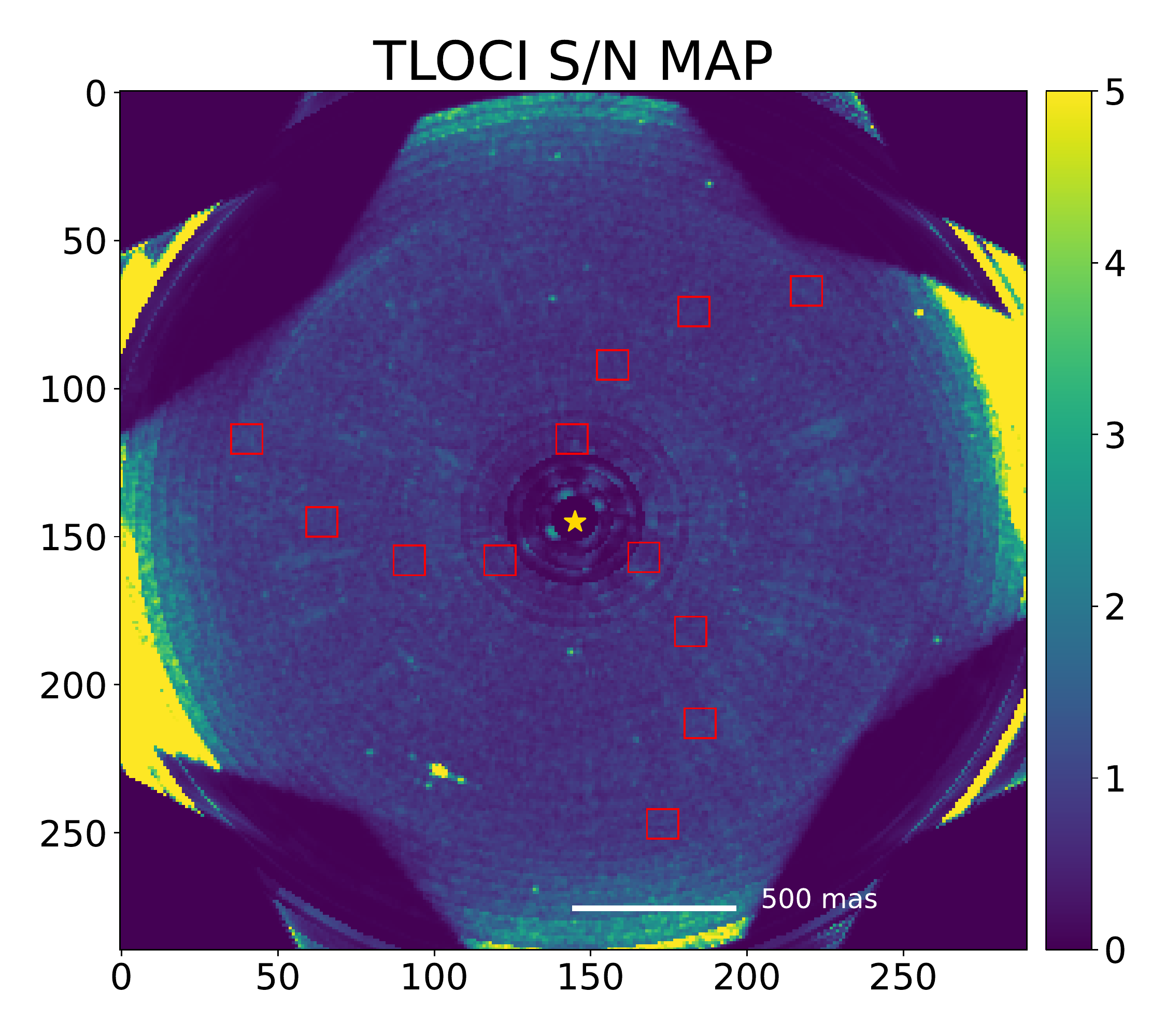}
\caption{S/N and residual maps for the injected fake sources on IFS. Top row: PACO ASDI S/N maps corresponding to the various injected spectra (see Fig. \ref{normalized_spectra}). Middle row: residual and S/N maps using PCA. Bottom row: residual and S/N maps using TLOCI. The locations of injected sources are highlighted by  red boxes.}
\label{results_inj_ifs}
\end{figure*}

For IFS, we injected 72 sources with three different shapes of spectra (24 sources per shape) corresponding to three of the spectral priors (flat, L-type, T-type) used during the reduction (hereafter denoted \textit{cases}), for which the contrast injected was also computed using the 2-D contrast maps. 
We also consider a \textit{flat} injection which corresponds (as the priors are expressed in contrast unit) to a SED with a same shape than that of the star. The normalized spectra injected for the three cases can be found in Fig. \ref{normalized_spectra} as well as complete information on the injected sources in Table \ref{table_inj_ifs}. Corresponding S/N and residual maps are given in Fig. \ref{results_inj_ifs}. 

\begin{figure}[t!]
\centering
\includegraphics[width=\linewidth]{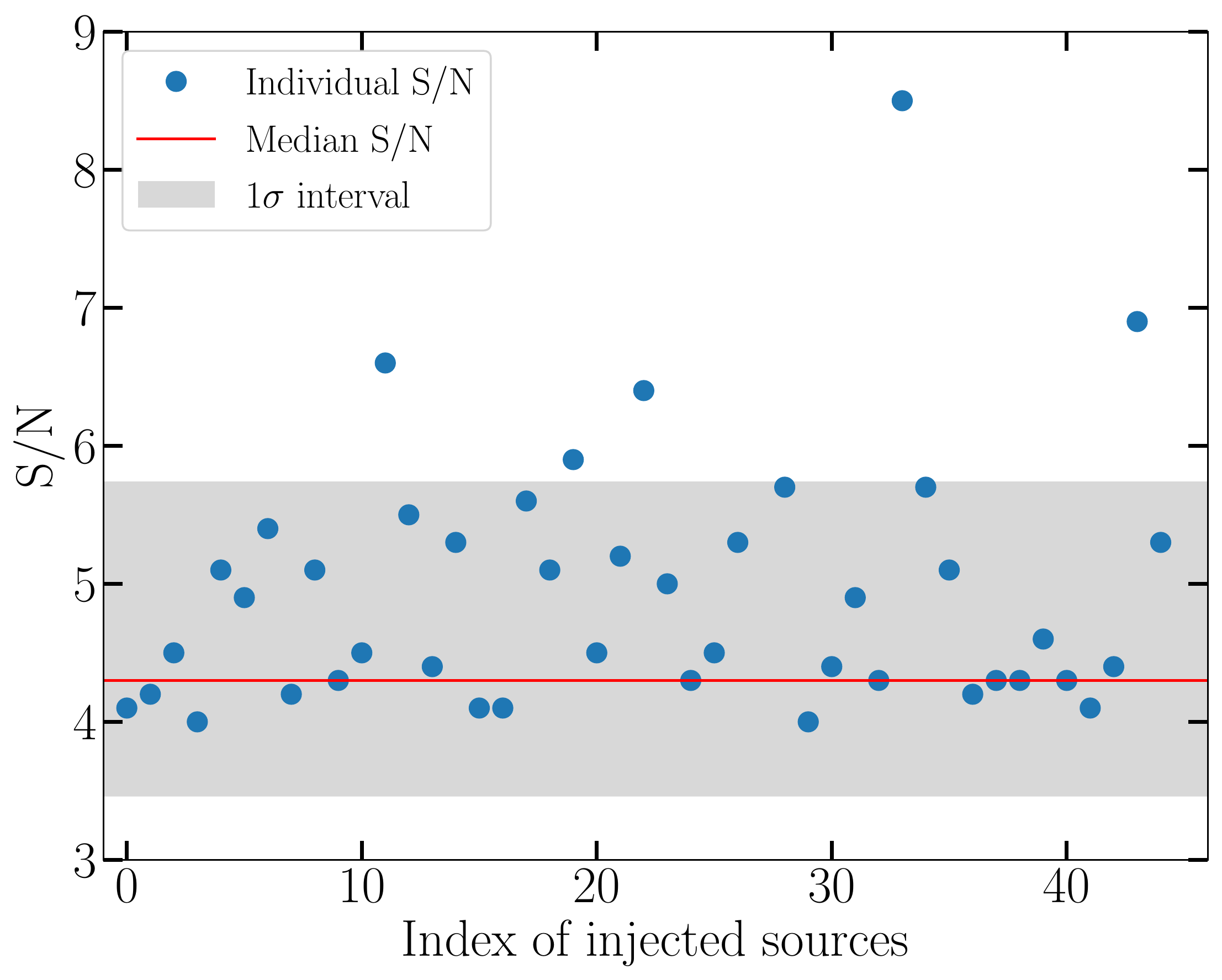}
\caption{Retrieved S/N for the injected sources for IFS. The detection threshold $\tau$ was set to 4. The red line shows the median S/N of the injected sources and the grey area the $1\sigma$ interval containing 68\% of the detected injected sources.}
\label{retrieved_inj_snr_ifs}
\end{figure}

The S/N of the detected injections are shown in Fig. \ref{retrieved_inj_snr_ifs}. As done for IRDIS, we can compute the median S/N of the injected sources, which is 4.3, with a standard deviation of 1.16. This suggests that the estimated contrast is slightly optimistic by about 15\%. 

As expected, we conclude that the detection limits in contrast derived by PACO for IFS are slightly optimistic without questioning the previously mentioned results, because the equivalent number of independent spectral channels that are recombined is about L/2. The detection sensitivity is still improved at all angular separation, coupled with the false positive rate consistent with the chosen S/N threshold.

\section{Results on the mini-survey}
\label{sec:results}

\subsection{Identified point sources and status}

Running the PACO algorithm along  with the analysis tools described in Sects. \ref{subsec:prior_asdi} and \ref{subsec:MEDT} over the 40 datasets of this survey allows us to identify 60 (58 IRDIS, 3 IFS) point-like sources with an S/N above the $5\sigma$ detection confidence. Note that one source, PZ TEL B, is detected with both instruments. For comparison, only 40 sources were detected by the F150 analysis. This again illustrates the enhanced detection capabilities of PACO with respect to more classical algorithms. We classify the sources as follows:
\begin{itemize}
    \item BCKG (background source): source classified as background in the F150 or in this analysis using a proper motion analysis.
    \item KC (known companion): either planet, brown dwarf or stellar companion.
    \item SUSP BCKG CMD (suspected background using a color magnitude diagram): source detected in the present analysis, that was not detected in the F150, with only one epoch available and a color consistent with a background nature.
    \item CC (candidate companion): source detected in the present analysis, that was not detected in the F150, with only one epoch and with a color compatible with a planetary/brown dwarf companion.
    \item FP: false positive identified using the multi-epoch tool described in Sect. \ref{subsec:MEDT}. This classification is only possible for stars with multiple epochs with similar quality data.
\end{itemize}

\begin{figure}[t!]
\centering
  \includegraphics[width=\linewidth]{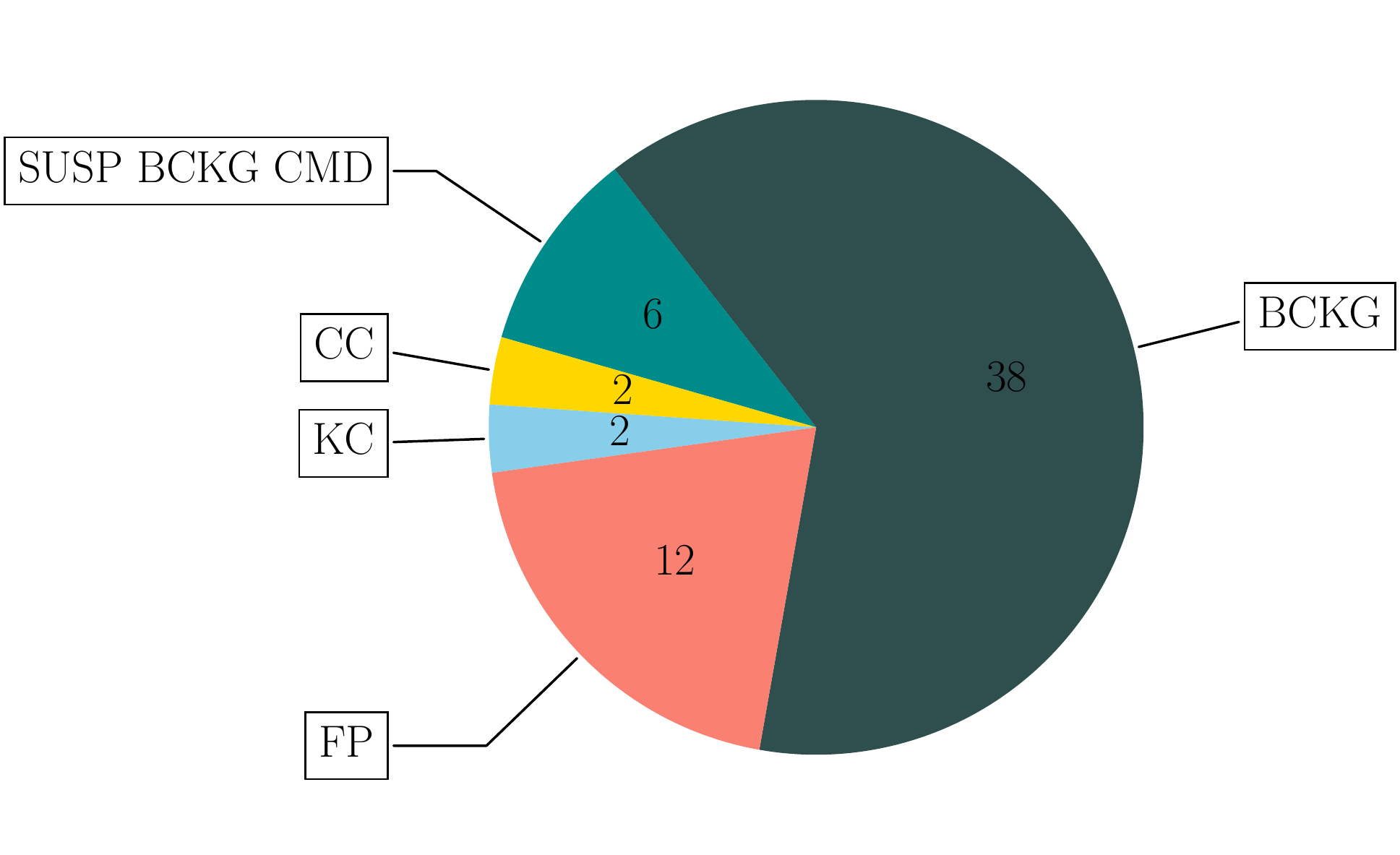}
  \caption{Classification of the 60 sources detected during the survey. Note that PZ Tel B is visible in both IRDIS and IFS, thus accounted for only one source in this plot.}
\label{pie_chart}
\end{figure}

\begin{figure}[t!]
\centering
  \includegraphics[width=\linewidth]{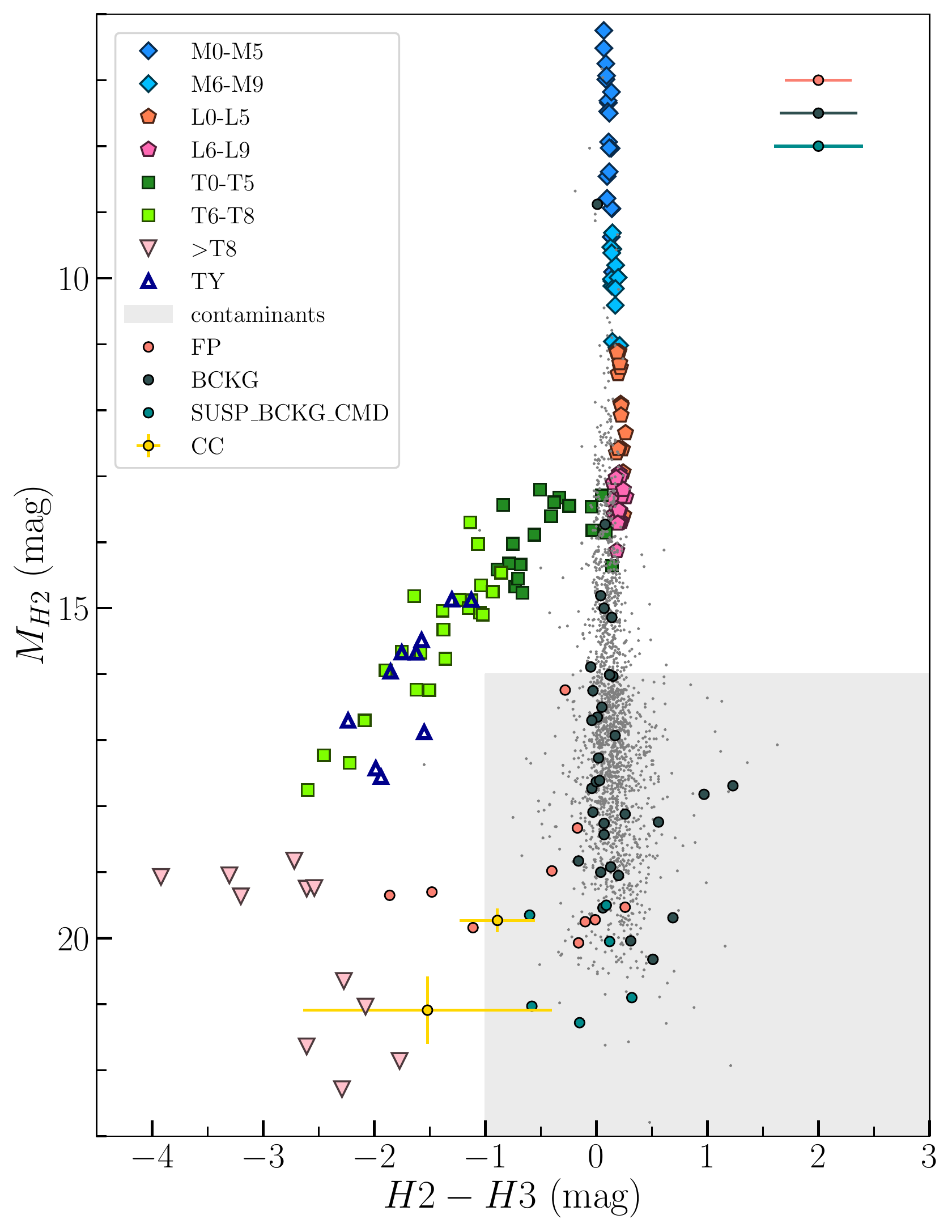}
  \caption{CMD plot with the classification of the 60 sources detected during the survey. Note that individual error bars are not shown for clarity purposes except for the two interesting candidates (yellow circles). Typical error bars are indicated in the top right. The grey dots represent the background sources identified during the SHINE F150 survey.}
\label{CMD_plot}
\end{figure}

Figure \ref{pie_chart} shows a pie-chart diagram of the sources classification. Individual information will be provided in a VizieR table, see Table \ref{det_table} for an example of the parameters provided. Finally, the targets are plotted on a CMD in Fig. \ref{CMD_plot}.  

Among the 60 identified signal of interest, eight are not characterized because of a lack of a second epoch of observation. Among those eight sources, six fall into the domain of probable background sources (grey area in  Fig. \ref{CMD_plot}).  We classify them as SUSP BCKG CMD. Two are more promising giving their colors. They require additional observations to definitely distinguish between a background source, a false positive or a bound companion. We  therefore classify them as candidate companions (CC) at this stage. 

Finally, we find twelve false positives during this survey. Given the number of datasets for IRDIS (37) and the number of priors used (5), we should expect around 16 false positives (see Fig. \ref{fp_vs_priors}). Taking also into account that FPs could be present amongst the eight sources classified as candidates companions (either "CC", or "SUSP BCKG CMD"), we  conclude that this number is in good agreement with the theory predictions, confirming the Gaussian nature of noise in the detection maps of PACO. One IFS false positive was also found.

\begin{table*}[t!]
\vspace{0.1cm}
\centering
\makebox[\linewidth]{ 
\begin{tabular}{ccccccc}
EPOCH & SEP (mas) & PA ($\degree$) & $\text{M}_{\text{H2}}$ & $\text{M}_{\text{H3}}$ & S/N & STATUS\\ 
\hline
\multicolumn{7}{c}{\textbf{HD 987}} \\
2015-09-25 & 1133.0$\pm$2.1 & 42.3$\pm$0.1 & 17.73$\pm$0.23 & 17.78$\pm$0.23 & 11.9 & BCKG \\
\hline
\multicolumn{7}{c}{\textbf{HD 61005}} \\
2015-02-02 & 5870.0$\pm$13.3 & 314.4$\pm$0.1 & 17.82$\pm$0.12 & 17.69$\pm$0.13 & 11.2 & BCKG \\
2015-02-02 & 3046.0$\pm$7.1 & 327.0$\pm$0.1 & 17.88$\pm$0.13 & 17.86$\pm$0.13 & 11.1 & BCKG \\
\hline
\hline 
\end{tabular}} \par
\caption{Properties of the signals detected in the IRDIS images. $\text{M}_{\text{H2/3}}$ is the absolute magnitude of the candidate in both filters.}
\label{det_table}
\vspace{0.2cm}
\end{table*}

\subsection{Searching for additional companions } 

Here, we want to put constraints on the properties of possible, yet unseen companions. We used the MESS2 tool \citep{2015IAUGA..2255268L} that uses for each target, all the detection limit maps derived from the PACO analysis, once expressed in masses, and, whenever available in the ESO HARPS \citep{MAYOR_HARPS} archive, radial velocity data.  
Indeed, the combination of direct imaging data with RV data provides a large  exploration of the star's environments, from a fraction of au out to 100 au. To convert the PACO contrasts into detection limits expressed in masses, we use luminosity-mass relationships given by the COND atmospheric model \citep{CondModel}, and the stars' ages and masses provided by  \cite{2021A&A...651A..70D}. 
Finally, to run MESS2, we assume a uniform distribution of eccentricities between 0 and 0.5, and a uniform orbital plane inclination between 0 and 90 degrees.

\begin{figure*}[t!]
\centering

\includegraphics[width=.45\textwidth]{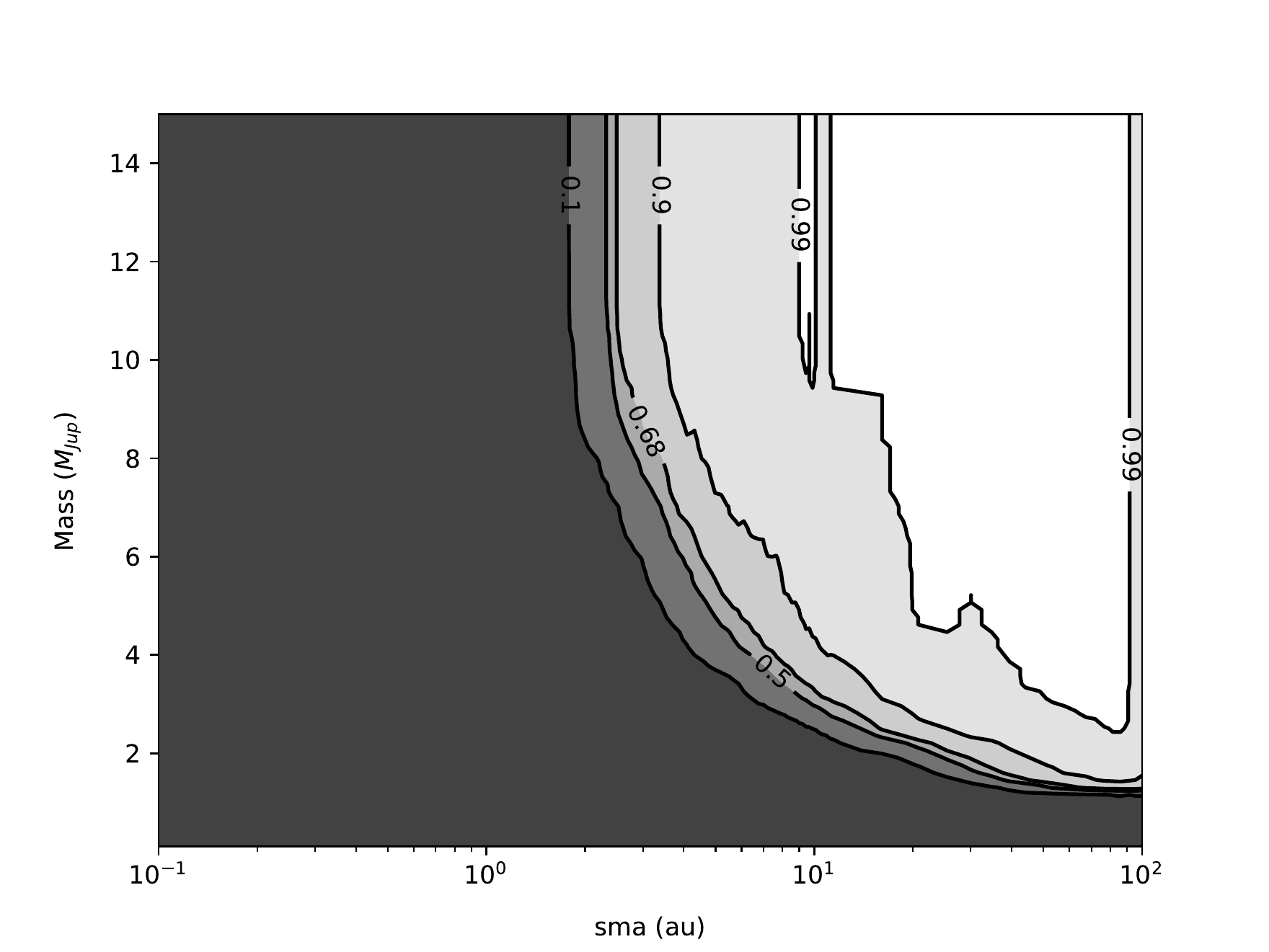}\quad
\includegraphics[width=.45\textwidth]{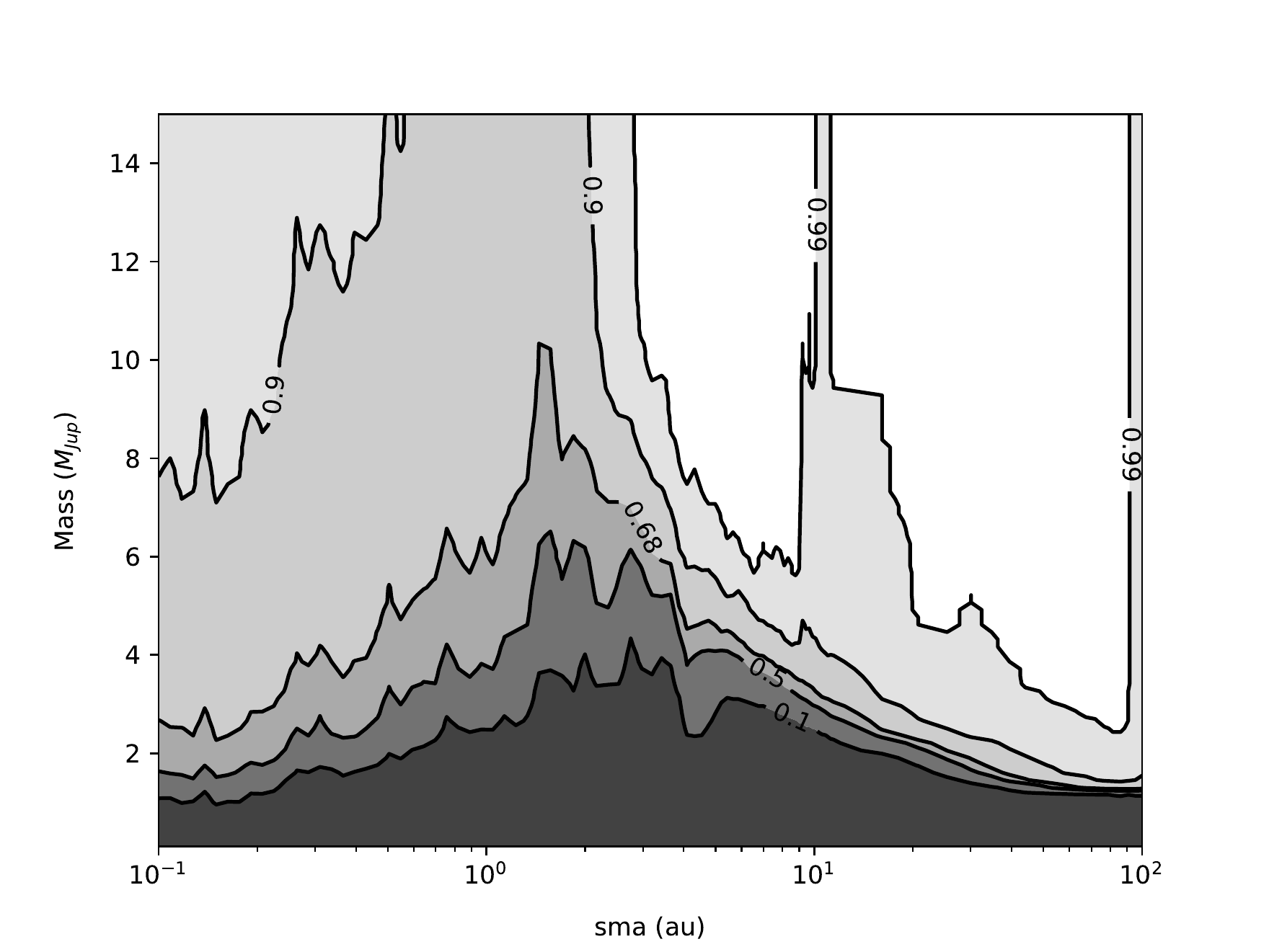}
\caption{Detection maps using the MESS2 \citep{2015IAUGA..2255268L} tool for HD 202917, DI only (left), DI+RV (right). The eccentricity is assumed between 0 and 0.5 and the inclination between 0 and 90 degrees. Contrast maps are converted to mass maps using the COND atmospheric model \citep{CondModel} and the stellar parameters of the system are extracted from \cite{2021A&A...651A..70D}. Two epochs were used.}
\label{EX_MESS2}
\end{figure*}

An example of detection limits obtained using the imaging data alone on the one hand and using both the imaging and RV data on the other hand is provided in Fig. \ref{EX_MESS2}.
The results obtained for all stars of the sample are presented in Appendix \ref{MESS2}. 

\begin{figure*}[t!]
\centering
\includegraphics[width=.45\textwidth]{MESS2_fig/hip_105388_limdet_DI.pdf}\quad
\includegraphics[width=.45\textwidth]{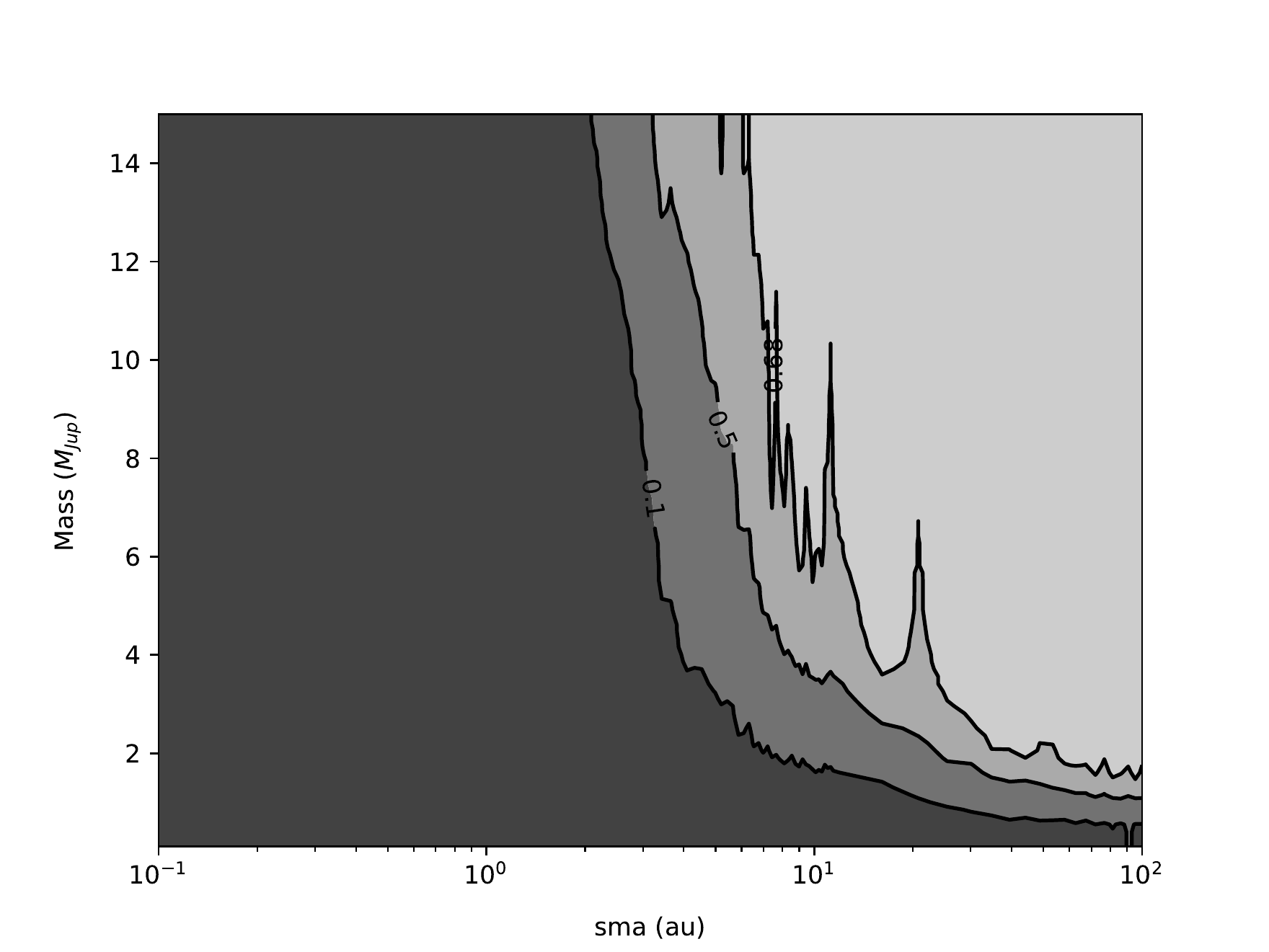}
\caption{Comparison between the detection limits obtained with PACO ASDI (left) and TLOCI (right) for HD 202917 using the MESS2 tool \citep{2015IAUGA..2255268L}. The eccentricity is assumed between 0 and 0.5 and the inclination between 0 and 90 degrees. Contrast maps are converted to mass maps using the COND atmospheric model \citep{CondModel} and the stellar parameters of the system are extracted from \cite{2021A&A...651A..70D}. Two epochs were used.}
\label{comp_limdet}
\end{figure*}

The  detection limits in terms of masses are significantly improved with respect to previous analysis, thanks to improved detection limits. For instance, we get a detection limit of about 5 $\text{M}_{\text{Jup}}$ (68 $\% $ probability) at 5 au for HD 202917, while the detection limit with TLOCI is 10 $\text{M}_{\text{Jup}}$ at the same separation (see Fig. \ref{comp_limdet}). At 10 au, 3 $\text{M}_{\text{Jup}}$ planets could be detected, compared to 6 $\text{M}_{\text{Jup}}$ with TLOCI. 
This represents a substantial  improvement in the detection limits.  
Nonetheless, Jupiter siblings are still out of reach in the present data. Improved adaptive optics systems as that of the SPHERE+ project \citep{sphereplus} on the VLT are needed to reach such objectives. 
Finally, we see that, in most cases, combining the radial velocity (RV) and direct imaging (DI) allows us to bridge the gap between the two techniques. The detection limits will be further improved using Hipparcos-Gaia data in the typical 3-10 au region.

\section{Conclusion}
\label{sec:conclusion}
We have presented here a number of upgrades made to our reduction and analysis pipeline using PACO so as to improve the sensitivity and the characterization of the detected sources. In particular, we have improved the precision and robustness of the astrometric and photometric error bars for the detected sources. We have developed custom built spectral prior libraries to optimize the detection capability of the ASDI mode for both IRDIS and IFS.
The contrast performance are significantly improved compared to those obtained with classical algorithms such as TLOCI and PCA. Also, we have showed that PACO provides statistically meaningful S/N maps.  This work paves the way to an end-to-end, homogeneous, and unsupervised massive re-reduction of archival SPHERE direct imaging data in the quest of  exoJupiters.

We used PACO ASDI to search for exoJupiters in a sample of 24 selected young, solar-type targets observed with SPHERE IRDIS and IFS that are part of the SHINE survey. 
This new analysis allowed us to identify two candidate companions in this small sample. Second epochs are necessary to unveil their nature.

\begin{acknowledgements}
This project has also received funding from the European Research Council (ERC) under the European Union's Horizon 2020 research and innovation programme (COBREX; grant agreement n° 885593).
SPHERE is an instrument designed and built by a consortium
consisting of IPAG (Grenoble, France), MPIA (Heidelberg, Germany),
LAM (Marseille, France), LESIA (Paris, France), Laboratoire Lagrange
(Nice, France), INAF - Osservatorio di Padova (Italy), Observatoire de
Genève (Switzerland), ETH Zürich (Switzerland), NOVA (Netherlands), ONERA
(France) and ASTRON (Netherlands) in collaboration with ESO. SPHERE
was funded by ESO, with additional contributions from CNRS (France),
MPIA (Germany), INAF (Italy), FINES (Switzerland) and NOVA (Netherlands).
SPHERE also received funding from the European Commission Sixth and Seventh
Framework Programmes as part of the Optical Infrared Coordination Network
for Astronomy (OPTICON) under grant number RII3-Ct-2004-001566 for
FP6 (2004-2008), grant number 226604 for FP7 (2009-2012) and grant number
312430 for FP7 (2013-2016). 
This work has made use of the SPHERE Data
Centre, jointly operated by OSUG/IPAG (Grenoble), PYTHEAS/LAM/CeSAM
(Marseille), OCA/Lagrange (Nice), Observatoire de Paris/LESIA (Paris), and
Observatoire de Lyon (OSUL/CRAL).
This research has made use of the SIMBAD database and VizieR catalogue access tool, operated at CDS, Strasbourg, France. This work is supported by the French National Research Agency in the framework of the Investissements d’Avenir program (ANR-15-IDEX-02), through the funding of the "Origin of Life" project of the Univ. Grenoble-Alpes.
This work was supported by the Action Spécifique Haute Résolution Angulaire (ASHRA) of CNRS/INSU co-funded by CNES.
This work is based on observations collected at the European Southern Observatory under ESO programme(s) 096.C-0241(A/B/C/F/G), 095.C-0298(A/B/D/H), 1100.C-0481(F/G), 097.C-0865(A/B/C/D), 198.C-0209(A/E/G).

\end{acknowledgements}

\bibliographystyle{aa}
\bibliography{biblio.bib}

\onecolumn

\appendix

\section{Target parameters and observing logs}

\begin{table*}[h!]
\vspace{0.4cm}
\centering
\small 
\makebox[\textwidth]{ 
\begin{tabular}{ccccccccc}
NAME & OTHER ID & RA$^a$ & DEC$^a$ & H (mag) & dist (pc)$^b$  & age (Myr)$^c$ & ST$^d$ & MG$^e$\\ 
\hline 
\hline 
BD-12 243 & - & 01 20 32.2680 & -11 28 03.727 & 6.65 & 35.3 & $149^{+31}_{-49}$ & G9V & ABDO\\ 
CD-31 16041 & - & 18 50 44.4830 & -31 47 47.382 & 7.67 & 49.5 & $24^{+5}_{-5}$ & K8Ve & BPIC\\ 
CD-61 1439 & - & 06 39 50.0234 & -61 28 41.530 & 6.64 & 22.2 & $149^{+31}_{-49}$ & K7V(e)& ABDO\\
HD105 & - & 00 05 52.5447 & -41 45 11.044 & 6.19 & 38.8 & $45^{+5}_{-10}$ & G0V & TUC\\ 
HD118100 & - & 13 34 43.2063 & -08 20 31.338 & 6.31 & 20.5 & $150^{+50}_{-50}$ & K5Ve&- \\ 
HD1466 & - & 00 18 26.1235 & -63 28 38.980 & 6.25 & 42.8 & $45^{+5}_{-10}$ & F8V & TUC\\ 
HD164249 & - & 18 03 03.4097 & -51 38 56.432 & 6.02 & 49.3 & $24^{+5}_{-5}$ & F6V & BPIC\\ 
HD174429 & V*PZTel & 18 53 05.8735 & -50 10 49.897 & 6.49 & 47.3 & $24^{+5}_{-5}$ & G9IV & BPIC\\ 
HD17925 & - & 02 52 32.1281 & -12 46 10.968 & 4.13 & 10.4 & $150^{+150}_{-80}$ & K1V & - \\ 
HD181327 & - & 19 22 58.9437 & -54 32 16.975 & 5.98 & 47.8 & $24^{+5}_{-5}$ & F6V & BPIC\\ 
HD189245 & - & 20 00 20.2493 & -33 42 12.424 & 4.64 & 22.1 & $150^{+150}_{-70}$ & F7V & - \\ 
HD197890 & V*BOMic & 20 47 45.0056 & -36 35 40.769 & 6.93 & 51.0 & $45^{+55}_{-35}$ & K3V(e) & - \\ 
HD202917 & - & 21 20 49.9576 & -53 02 03.155 & 7.03 & 46.7 & $45^{+5}_{-10}$ & G7V & TUC\\ 
HD218860 & - & 23 11 52.0534 & -45 08 10.631 & 7.11 & 47.8 & $149^{+31}_{-49}$ & G8V & ABDO\\ 
HD224228 & - & 23 56 10.6732 & -39 03 08.409 & 6.01 & 22.0 & $149^{+31}_{-49}$ & K2V & ABDO\\ 
HD377 & - & 00 08 25.7454 & +06 37 00.489 & 6.15 & 38.4 & $150^{+50}_{-50}$ & G2V & - \\ 
HD43989 & V*V1358Ori & 06 19 08.0574 & -03 26 20.361 & 6.59 & 51.7 & $42^{+8}_{-7}$ & G0V & TUC\\
HD44627 & - & 06 19 12.9130 & -58 03 15.527 & 7.09 & 50.1 & $45^{+5}_{-10}$ & K1V(e) & CAR\\ 
HD45270 & - & 06 22 30.9408 & -60 13 07.147 & 5.16 & 23.9 & $149^{+31}_{-49}$ & G1V & ABDO\\ 
HD49855 & - & 06 43 46.2456 & -71 58 35.390 & 7.38 & 55.4 & $45^{+5}_{-10}$ & G6V & CAR\\ 
HD61005 & - & 07 35 47.4623 & -32 12 14.045 & 6.58 & 36.5 & $50^{+20}_{-10}$ & G8Vk & - \\  
HD8558 & - & 01 23 21.2547 & -57 28 50.688 & 6.95 & 45.3 & $45^{+5}_{-10}$ & G7V & TUC\\ 
HD90712 & - & 10 27 47.7769 & -34 23 58.130 & 6.15 & 37.7 & $150^{+50}_{-80}$ & G0V & - \\
HD987 & - & 00 13 53.0108 & -74 41 17.850 & 7.09 & 45.9 & $45^{+5}_{-10}$ & G8V & TUC\\ 
\hline 
\hline 
\end{tabular}} \par 
\caption{Summary of the main parameters of the targeted stars.}
\vspace{0.2cm}
\textbf{Notes:} $^a$: Coordinates   in J2000 IRCS. $^b$: Distances as derived from the parallaxes provided by Simbad. $^c$: Ages as extracted from \cite{2021A&A...651A..70D}. $^d$: Spectral type. $^e$: Moving group.
\label{star_table}
\end{table*}

\begin{small} 
\onecolumn 
{\setlength\tabcolsep{2pt} 
\begin{longtable}{cccccccccc}
STAR & DATE OBS & FILTER & DIT(s)$\times$Nframe & $\Delta$PA ($\degree$)$^a$ & Seeing (")$^b$ & Airmass$^b$ & $\tau_0$ (ms)$^{a,b}$ & Program ID \\ 
\hline 
\hline 
BD-12 243 & 2015-12-01 & DB\_H23 & 64x64 & 51.0 & 1.1 & 1.04 & 7 & 096.C-0241(B) \\ 
BD-12 243 & 2015-12-01 & OBS\_YJ & 64x64 & 51.1 & 1.1 & 1.04 & 7 & 096.C-0241(B) \\ 
\hline
CD-31 16041 & 2015-05-15 & DB\_H23 & 64x64 & 81.2 & 1.1 & 1.02 & 3.4 & 095.C-0298(A) \\ 
CD-31 16041 & 2015-05-15 & OBS\_YJ & 64x64 & 81.3 & 1.08 & 1.02 & 3.4 & 095.C-0298(A) \\ 
CD-31 16041 & 2018-04-17 & DB\_H23 & 96x42 & 84.5 & 0.88 & 1.01 & 10.3 & 1100.C-0481(F) \\ 
CD-31 16041 & 2018-04-17 & OBS\_YJ & 96x42 & 83.8 & 0.88 & 1.01 & 10.3 & 1100.C-0481(F) \\ 
\hline
CD-61 1439 & 2016-01-02 & DB\_H23 & 64x64 & 26.1 & 0.9 & 1.25 & 2.4 & 096.C-0241(C) \\ 
CD-61 1439 & 2016-01-02 & OBS\_YJ & 64x64 & 26.1 & 0.91 & 1.25 & 2.4 & 096.C-0241(C) \\ 
\hline 
HD105 & 2015-09-26 & DB\_H23 & 64x72 & 45.0 & 1.06 & 1.07 & 17.4 & 095.C-0298(D) \\ 
HD105 & 2015-09-26 & OBS\_YJ & 64x68 & 109.9 & 0.98 & 1.07 & 17.4 & 095.C-0298(D) \\ 
\hline 
HD118100 & 2016-06-27 & DB\_H23 & 64x64 & 52.0 & 0.8 & 1.05 & 3.8 & 097.C-0865(C) \\ 
\hline 
HD1466 & 2015-10-26 & DB\_H23 & 64x64 & 25.0 & 1.07 & 1.29 & 1.4 & 096.C-0241(A) \\ 
HD1466 & 2015-10-26 & OBS\_YJ & 64x64 & 25.1 & 1.08 & 1.29 & 1.4 & 096.C-0241(A) \\ 
HD1466 & 2016-09-18 & DB\_H23 & 64x80 & 31.1 & 0.8 & 1.29 & 4.9 & 097.C-0865(D) \\ 
HD1466 & 2016-09-18 & OBS\_YJ & 64x80 & 31.3 & 0.8 & 1.29 & 4.9 & 097.C-0865(D) \\ 
\hline 
HD164249 & 2015-05-10 & DB\_H23 & 64x56 & 34.3 & 1.87 & 1.13 & 1.2 & 095.C-0298(A) \\ 
HD164249 & 2015-05-10 & OBS\_YJ & 64x56 & 34.5 & 1.86 & 1.13 & 1.2 & 095.C-0298(A) \\ 
HD164249 & 2015-06-01 & DB\_H23 & 64x64 & 33.9 & 1.24 & 1.13 & 1.1 & 095.C-0298(B) \\ 
HD164249 & 2015-06-01 & OBS\_YJ & 64x64 & 34.1 & 1.25 & 1.13 & 1.1 & 095.C-0298(B) \\ 
HD164249 & 2016-04-17 & DB\_H23 & 64x61 & 37.3 & 1.72 & 1.13 & 1.5 & 097.C-0865(A) \\ 
HD164249 & 2016-04-17 & OBS\_YJ & 64x61 & 37.3 & 1.72 & 1.13 & 1.5 & 097.C-0865(A) \\ 
HD164249 & 2018-04-11 & DB\_H23 & 96x40 & 31.9 & 0.52 & 1.13 & 5.6 & 1100.C-0481(F) \\ 
HD164249 & 2018-04-11 & OBS\_YJ & 96x40 & 32.0 & 0.52 & 1.13 & 5.6 & 1100.C-0481(F) \\ 
\hline 
HD174429 & 2015-05-06 & DB\_H23 & 16x82 & 11.4 & 1.43 & 1.14 & 1.3 & 095.C-0298(A) \\ 
HD174429 & 2015-05-06 & OBS\_YJ & 32x2 & 0.3 & 1.43 & 1.14 & 1.3 & 095.C-0298(A) \\
HD174429 & 2015-05-31 & DB\_H23 & 32x32 & 9.3 & 1.28 & 1.11 & 1 & 095.C-0298(B) \\ 
HD174429 & 2015-05-31 & OBS\_YJ & 32x32 & 9.3 & 1.28 & 1.11 & 1 & 095.C-0298(B) \\ 
HD174429 & 2016-09-17 & DB\_H23 & 32x80 & 29.0 & 0.56 & 1.11 & 14.3 & 097.C-0865(D) \\
HD174429 & 2016-09-17 & OBS\_YJ & 64x40 & 28.8 & 0.56 & 1.11 & 14.3 & 097.C-0865(D) \\ 
HD174429 & 2017-05-18 & DB\_H23 & 32x60 & 33.1 & 0.81 & 1.11 & 2.8 & 198.C-0209(G) \\ 
HD174429 & 2017-05-18 & OBS\_YJ & 64x57 & 36.8 & 0.81 & 1.11 & 2.8 & 198.C-0209(G) \\
HD174429 & 2018-05-13 & DB\_H23 & 16x30 & 8.9 & 0.81 & 1.12 & 3.4 & 1100.C-0481(G) \\ 
HD174429 & 2018-05-13 & OBS\_YJ & 32x30 & 9.5 & 0.85 & 1.12 & 3.4 & 1100.C-0481(G) \\ 
\hline 
HD17925 & 2016-10-15 & DB\_H23 & 32x160 & 70.8 & 0.66 & 1.03 & 2.5 & 198.C-0209(A) \\ 
HD17925 & 2016-10-15 & OBS\_YJ & 64x80 & 70.0 & 0.66 & 1.03 & 2.5 & 198.C-0209(A) \\ 
\hline 
HD181327 & 2015-05-10 & DB\_H23 & 64x56 & 31.1 & 1.2 & 1.16 & 1.7 & 095.C-0298(A) \\ 
HD181327 & 2015-05-10 & OBS\_YJ & 64x52 & 31.2 & 1.2 & 1.16 & 1.7 & 095.C-0298(A) \\ 
\hline 
HD189245 & 2015-05-14 & DB\_H23 & 32x128 & 16.2 & 0.75 & 1.13 & 6.9 & 095.C-0298(A) \\ 
HD189245 & 2015-05-14 & OBS\_YJ & 32x128 & 18.2 & 0.73 & 1.13 & 6.9 & 095.C-0298(A) \\ 
\hline 
HD197890 & 2015-06-04 & DB\_H23 & 64x64 & 61.5 & 1.25 & 1.03 & 2.9 & 095.C-0298(B) \\ 
HD197890 & 2015-06-04 & OBS\_YJ & 64x64 & 62.4 & 1.23 & 1.03 & 2.9 & 095.C-0298(B) \\ 
\hline 
HD202917 & 2015-05-31 & DB\_H23 & 64x86 & 49.6 & 1.25 & 1.15 & 1.1 & 095.C-0298(B) \\ 
HD202917 & 2015-05-31 & OBS\_YJ & 64x64 & 32.8 & 1.44 & 1.14 & 0.9 & 095.C-0298(B) \\ 
HD202917 & 2016-05-31 & DB\_H23 & 64x64 & 32.7 & 0.83 & 1.14 & 2.6 & 097.C-0865(B) \\ 
HD202917 & 2016-05-31 & OBS\_YJ & 64x64 & 32.8 & 0.84 & 1.14 & 2.6 & 097.C-0865(B) \\ 
\hline 
HD218860 & 2015-09-30 & DB\_H23 & 64x64 & 43.8 & 0.68 & 1.07 & 3.9 & 095.C-0298(D) \\ 
HD218860 & 2015-09-30 & OBS\_YJ & 64x64 & 44.1 & 0.7 & 1.07 & 3.9 & 095.C-0298(D) \\ 
\hline 
HD224228 & 2015-10-25 & DB\_H23 & 64x64 & 59.3 & 1.48 & 1.04 & 1.1 & 096.C-0241(A) \\ 
HD224228 & 2015-10-25 & OBS\_YJ & 64x64 & 59.4 & 1.51 & 1.04 & 1.1 & 096.C-0241(A) \\ 
\hline 
HD377 & 2016-10-14 & DB\_H23 & 64x80 & 35.8 & 0.63 & 1.18 & 2.8 & 198.C-0209(A) \\ 
HD377 & 2016-10-14 & OBS\_YJ & 64x80 & 35.9 & 0.62 & 1.18 & 2.8 & 198.C-0209(A) \\ 
\hline 
HD43989 & 2015-10-28 & DB\_H23 & 64x80 & 48.5 & 1.01 & 1.08 & 1.9 & 096.C-0241(F) \\ 
HD43989 & 2015-10-28 & OBS\_YJ & 64x80 & 48.3 & 1.0 & 1.08 & 1.9 & 096.C-0241(F) \\ 
HD43989 & 2017-02-09 & DB\_H23 & 64x56 & 36.5 & 0.63 & 1.09 & 8.7 & 198.C-0209(E) \\ 
HD43989 & 2017-02-09 & OBS\_YJ & 64x56 & 36.4 & 0.62 & 1.09 & 8.7 & 198.C-0209(E) \\ 
\hline 
HD44627 & 2015-02-06 & DB\_H23 & 64x64 & 28.3 & 1.08 & 1.2 & 5.3 & 095.C-0298(H) \\ 
HD44627 & 2015-02-06 & OBS\_YJ & 64x64 & 28.4 & 1.07 & 1.2 & 5.3 & 095.C-0298(H) \\ 
\hline 
HD45270 & 2016-01-16 & DB\_H23 & 16x256 & 28.3 & 1.69 & 1.23 & 1.6 & 096.C-0241(G) \\ 
HD45270 & 2016-01-16 & OBS\_YJ & 64x64 & 27.2 & 1.73 & 1.23 & 1.5 & 096.C-0241(G) \\ 
\hline 
\hline 
\caption{Star sample observation logs.}
\label{obs_table_1}
\end{longtable}} \par 
\end{small} 
\vspace{0.2cm}
\textbf{Notes:} $^a$: DIT correspond to the detector integration time per frame, $\Delta$PA is the amplitude of the parallactic rotation, $\tau_0$ corresponds to the coherence time. $^b$: Values extracted from the updated DIMM info and averaged over the sequence. 

\begin{small} 
\onecolumn 
{\setlength\tabcolsep{2pt} 
\begin{longtable}{cccccccccc}
STAR & DATE OBS & FILTER & DIT(s)$\times$Nframe & $\Delta$PA ($\degree$)$^a$ & Seeing (")$^b$ & Airmass$^b$ & $\tau_0$ (ms)$^{a,b}$ & Program ID \\ 
\hline 
\hline 
HD49855 & 2015-12-28 & DB\_H23 & 64x64 & 21.4 & 0.84 & 1.48 & 2.9 & 096.C-0241(C) \\ 
HD49855 & 2015-12-28 & OBS\_YJ & 64x64 & 21.5 & 0.83 & 1.48 & 2.9 & 096.C-0241(C) \\ 
\hline 
HD61005 & 2015-02-03 & DB\_NDH23 & 64x64 & 93.0 & 0.66 & 1.01 & 23.1 & 095.C-0298(H) \\ 
HD61005 & 2015-02-03 & OBS\_YJ & 64x66 & 96.9 & 0.67 & 1.01 & 23.1 & 095.C-0298(H) \\ 
\hline 
HD8558 & 2015-09-24 & DB\_H23 & 64x64 & 28.8 & 2.05 & 1.19 & 0.9 & 095.C-0298(D) \\ 
HD8558 & 2015-09-24 & OBS\_YJ & 64x64 & 28.9 & 2.05 & 1.19 & 0.9 & 095.C-0298(D) \\ 
HD8558 & 2015-10-28 & DB\_H23 & 64x64 & 28.2 & 1.02 & 1.2 & 2 & 096.C-0241(A) \\ 
HD8558 & 2015-10-28 & OBS\_YJ & 64x64 & 28.4 & 1.02 & 1.2 & 2 & 096.C-0241(A) \\ 
HD8558 & 2016-10-14 & DB\_H23 & 64x64 & 28.2 & 0.56 & 1.2 & 2.8 & 198.C-0209(A) \\ 
HD8558 & 2016-10-14 & OBS\_YJ & 64x64 & 28.3 & 0.56 & 1.2 & 2.8 & 198.C-0209(A) \\ 
\hline 
HD90712 & 2016-01-03 & DB\_H23 & 64x64 & 74.8 & 0.67 & 1.02 & 6.1 & 096.C-0241(C) \\ 
HD90712 & 2016-01-03 & OBS\_YJ & 64x64 & 74.9 & 0.65 & 1.02 & 6.1 & 096.C-0241(C) \\ 
\hline 
HD987 & 2016-10-15 & DB\_H23 & 64x90 & 29.7 & 0.71 & 1.56 & 2.2 & 198.C-0209(A) \\ 
HD987 & 2016-10-15 & OBS\_YJ & 64x90 & 29.7 & 0.71 & 1.56 & 2.2 & 198.C-0209(A) \\ 
\hline 
\hline 
\caption{Star sample observation logs (continuation of Table \ref{obs_table_1}).}
\label{obs_table_2}
\end{longtable}} \par 
\end{small} 
\vspace{0.2cm}
\textbf{Notes:} $^a$: DIT corresponds to the detector integration time per frame, $\Delta$PA is the amplitude of the parallactic rotation, $\tau_0$ corresponds to the coherence time. $^b$: Values extracted from the updated DIMM info and averaged over the sequence. 

\begin{table*}[h!]
\vspace{0.4cm}.
\centering
\small 
\makebox[\linewidth]{ 
\begin{tabular}{cccccccc}
NAME & OTHER ID & RA$^a$ & DEC$^a$ & H (mag) & dist (pc)$^b$ & age (Myr)$^c$ & ST \\ 
\hline 
\hline 
Smethells86 & - & 21 44 30.1227 & -60 58 38.894 & 8.09 & 46.4 & $45^{+5}_{-10}$ & M0Ve \\ 
V*CT Tuc & - & 00 25 14.6618 & -61 30 48.252 & 7.94 & 44.1 & $45^{+5}_{-10}$ & M0Ve \\ 
HD108767B & *delCrvB & 12 29 50.8908 & -16 31 15.208 & 6.37 & 26.8 & $180^{+170}_{-80}$ & K1 \\ 
HD16978 & *epsHyi & 02 39 35.3612 & -68 16 01.010 & 4.43 & 46.6 & $45^{+5}_{-10}$ & B9Va \\ 
\hline 
\hline 
\end{tabular}} \par 
\caption{Summary of the main parameters of the test stars}
\vspace{0.2cm}
\textbf{Notes:} $^a$: Coordinates  in J2000 IRCS. $^b$: Distances derived from Simbad parallaxes. $^c$: Age extracted from \cite{2021A&A...651A..70D}. $^d$: Spectral type.
\label{star_table_inj}
\end{table*}

\begin{small}
\onecolumn 
{\setlength\tabcolsep{2pt} 
\begin{longtable}{cccccccccc}
STAR & DATE OBS & FILTER & DIT(s)$\times$Nframe & $\Delta$PA ($\degree$)$^a$ & Seeing (")$^b$ & Airmass$^b$ & $\tau_0$ (ms)$^{a,b}$ & Program ID \\ 
\hline 
\hline 
V*CT Tuc & 2015-07-05 & DB\_H23 & 64x64 & 26.0 & 1.07 & 1.25 &2 & 095.C-0298(C) \\ 
V*CTTuc & 2015-07-05 & OBS\_YJ & 64x64 & 26.2 & 1.07 & 1.25 & 2 & 095.C-0298(C) \\ 
\hline
Smethells 86 & 2015-11-29 & DB\_H23 & 64x64 & 25.8 & 1.57 & 1.26 & 7.2 & 096.C-0241(B) \\ 
Smethells 86 & 2015-11-29 & OBS\_YJ & 64x64 & 25.8 & 1.57 & 1.26 & 7.2 & 096.C-0241(B) \\ 
\hline 
HD108767B & 2018-01-25 & DB\_H23 & 64x72 & 94.4 & 0.59 & 1.02 & 8.3 & 1100.C-0481(D) \\ 
HD108767B & 2018-01-25 & OBS\_YJ & 64x72 & 94.3 & 0.59 & 1.02 &8.3 & 1100.C-0481(D) \\ 
\hline 
HD16978 & 2016-09-16 & DB\_H23 & 32x160 & 29.1 & 0.42 & 1.38 & 9.2 & 097.C-0865(D) \\ 
HD16978 & 2016-09-16 & OBS\_YJ & 32x144 & 26.8 & 0.42 & 1.38 & 9.2 & 097.C-0865(D) \\ 
\hline 
\hline 
\caption{Test targets observation logs.}
\label{obs_table_inj}
\end{longtable}} \par 
\end{small} 
\vspace{0.2cm}
\textbf{Notes:} $^a$: DIT corresponds to the detector integration time per frame, $\Delta$PA is the amplitude of the parallactic rotation, $\tau_0$ corresponds to the coherence time. $^b$: Values extracted from the updated DIMM info and averaged over the sequence. 

\clearpage

\section{Compared detection and contrast maps between PACO ASDI, TLOCI and PCA for 5$\sigma$ injected sources from PACO detection limits}
\label{appendix_inj}

\begin{table}[h]
\footnotesize
\begin{center}
\begin{tabular}{ccc|ccc|ccc}
SEP (mas) & PA ($\degree$) & CONTRAST & SEP (mas) & PA ($\degree$) & CONTRAST & SEP (mas) & PA ($\degree$) & CONTRAST  \\ 
\hline
\hline
250 & 100 & $4.81\times10^{-6}$ & 250 & 220 & $4.18\times10^{-6}$ & 250 & 340 & $4.92\times10^{-6}$ \\
500 & 85 & $2.09\times10^{-6}$ & 500 & 205 & $1.24\times10^{-6}$ & 500 & 325 & $1.38\times10^{-6}$ \\
750 & 70 & $1.45\times10^{-6}$ & 750 & 190 & $7.12\times10^{-7}$ & 750 & 310 & $9.73\times10^{-7}$ \\
1000 & 55 & $8.46\times10^{-7}$ & 1000 & 175 & $8.10\times10^{-7}$ & 1000 & 295 & $7.78\times10^{-7}$ \\
1250 & 40 & $4.44\times10^{-7}$ & 12500 & 160 & $4.05\times10^{-7}$ & 1250 & 280 & $4.68\times10^{-7}$ \\
\hline
\hline
\end{tabular}
\caption{IRDIS injected fake planets parameters.}
\label{table_inj_ird_close}
\end{center}
\end{table}

\begin{table}[h]
\footnotesize
\begin{center}
\begin{tabular}{ccc|ccc|ccc}
SEP (mas) & PA ($\degree$) & CONTRAST & SEP (mas) & PA ($\degree$) & CONTRAST & SEP (mas) & PA ($\degree$) & CONTRAST  \\ 
\hline
\hline
1500 & 100 & $3.07\times10^{-7}$ & 1500 & 160 & $3.20\times10^{-7}$ & 1500 & 220 & $3.08\times10^{-7}$ \\ 
1500 & 280 & $2.87\times10^{-7}$ & 1500 & 340 & $3.18\times10^{-7}$ & 1750 & 85 & $2.51\times10^{-7}$ \\ 
1750 & 145 & $2.57\times10^{-7}$ & 1750 & 205 & $2.44\times10^{-7}$ & 1750 & 265 & $2.43\times10^{-7}$ \\ 
1750 & 325 & $2.50\times10^{-7}$ & 2000 & 70 & $2.24\times10^{-7}$ & 2000 & 130 & $2.07\times10^{-7}$ \\ 
2000 & 190 & $2.19\times10^{-7}$ & 2000 & 250 & $2.15\times10^{-7}$ & 2000 & 310 & $2.19\times10^{-7}$ \\ 
2250 & 55 & $2.10\times10^{-7}$ & 2250 & 115 & $1.93\times10^{-7}$ & 2250 & 175 & $2.12\times10^{-7}$ \\ 
2250 & 235 & $2.16\times10^{-7}$ & 2250 & 295 & $1.96\times10^{-7}$ & 2500 & 40 & $1.94\times10^{-7}$ \\ 
2500 & 100 & $1.89\times10^{-7}$ & 2500 & 160 & $1.92\times10^{-7}$ & 2500 & 220 & $1.96\times10^{-7}$ \\ 
2500 & 280 & $1.88\times10^{-7}$ & 2750 & 25 & $1.91\times10^{-7}$ & 2750 & 85 & $1.89\times10^{-7}$ \\ 
2750 & 145 & $1.73\times10^{-7}$ & 2750 & 205 & $1.91\times10^{-7}$ & 2750 & 265 & $1.87\times10^{-7}$ \\ 
3000 & 10 & $1.72\times10^{-7}$ & 3000 & 70 & $1.81\times10^{-7}$ & 3000 & 130 & $1.66\times10^{-7}$ \\ 
3000 & 190 & $1.78\times10^{-7}$ & 3000 & 250 & $1.77\times10^{-7}$ & 3250 & 355 & $1.71\times10^{-7}$ \\ 
3250 & 55 & $1.68\times10^{-7}$ & 3250 & 115 & $1.87\times10^{-7}$ & 3250 & 175 & $1.79\times10^{-7}$ \\ 
3250 & 235 & $1.76\times10^{-7}$ & 3500 & 340 & $1.66\times10^{-7}$ & 3500 & 40 & $1.71\times10^{-7}$ \\ 
3500 & 100 & $1.68\times10^{-7}$ & 3500 & 160 & $1.76\times10^{-7}$ & 3500 & 220 & $1.72\times10^{-7}$ \\ 
3750 & 325 & $1.68\times10^{-7}$ & 3750 & 25 & $1.76\times10^{-7}$ & 3750 & 85 & $1.67\times10^{-7}$ \\ 
3750 & 145 & $1.72\times10^{-7}$ & 3750 & 205 & $1.85\times10^{-7}$ & 4000 & 310 & $1.74\times10^{-7}$ \\ 
4000 & 10 & $1.76\times10^{-7}$ & 4000 & 70 & $1.80\times10^{-7}$ & 4000 & 130 & $1.71\times10^{-7}$ \\ 
4000 & 190 & $1.74\times10^{-7}$ & 4250 & 295 & $1.72\times10^{-7}$ & 4250 & 355 & $1.63\times10^{-7}$ \\ 
4250 & 55 & $1.73\times10^{-7}$ & 4250 & 115 & $1.64\times10^{-7}$ & 4250 & 175 & $1.80\times10^{-7}$ \\ 
4500 & 280 & $1.69\times10^{-7}$ & 4500 & 340 & $1.74\times10^{-7}$ & 4500 & 40 & $1.63\times10^{-7}$ \\ 
4500 & 100 & $1.71\times10^{-7}$ & 4500 & 160 & $1.69\times10^{-7}$ & 4750 & 265 & $1.69\times10^{-7}$ \\ 
4750 & 325 & $1.65\times10^{-7}$ & 4750 & 25 & $1.63\times10^{-7}$ & 4750 & 85 & $1.66\times10^{-7}$ \\ 
4750 & 145 & $1.65\times10^{-7}$ & 5000 & 190 & $1.65\times10^{-7}$ & 5000 & 250 & $1.87\times10^{-7}$ \\ 
5000 & 10 & $1.69\times10^{-7}$ & 5000 & 70 & $1.73\times10^{-7}$ & 5000 & 130 & $1.64\times10^{-7}$ \\ 
\hline
\hline
\end{tabular}
\caption{IRDIS injected fake planets parameters.}
\label{table_inj_ird_far}
\end{center}
\end{table}

\begin{figure}[h!]
\centering
\includegraphics[width=.42\linewidth]{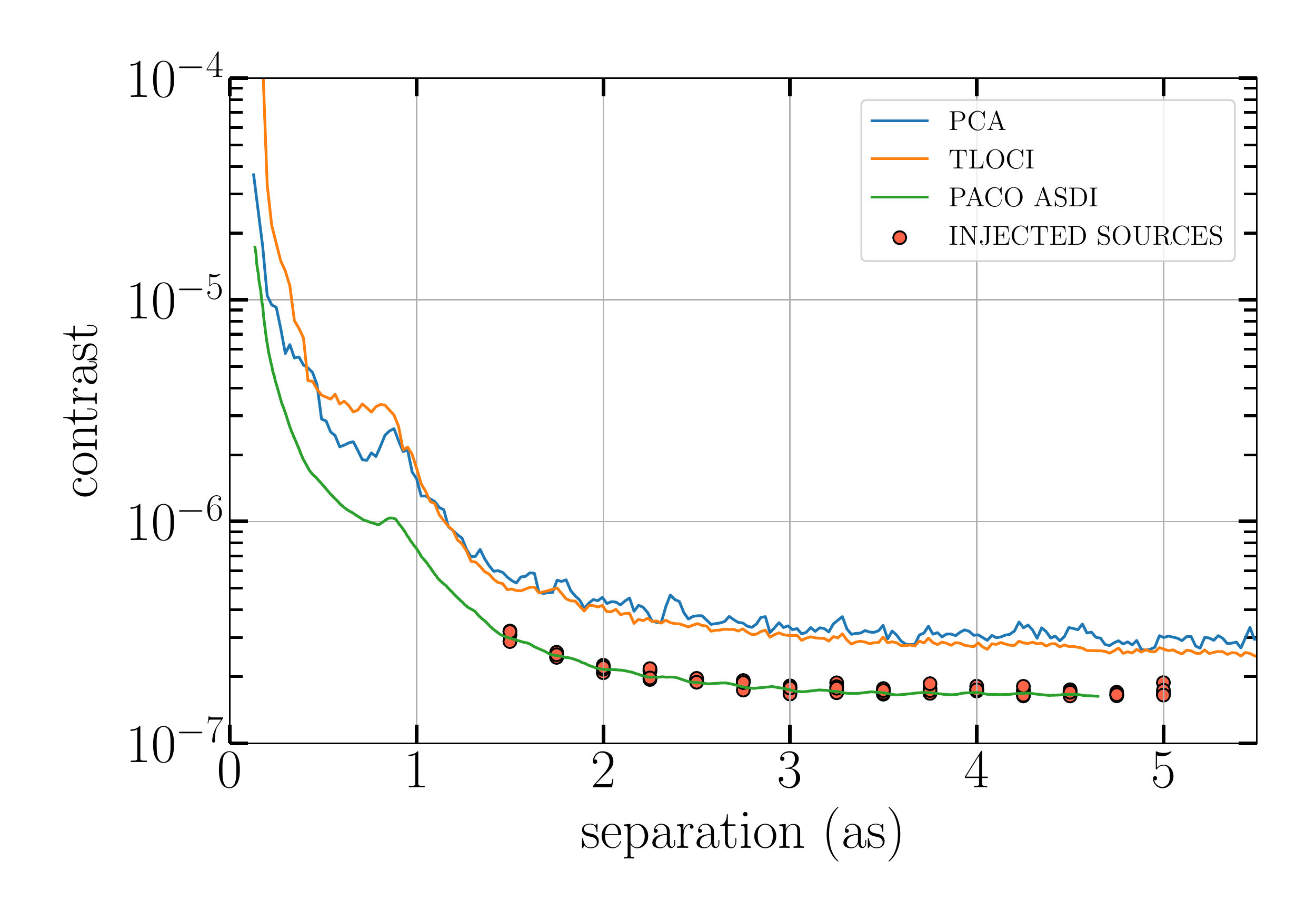}
\caption{Contrast comparison between PCA (10 modes), TLOCI and PACO. The contrast of the injected fake planets were computed using 2-D contrast maps, hence the differences with the $5\sigma$ curve: local variations of the achieved contrast are averaged azimuthally.}
\label{contrast_curve_comp_ird_far}
\end{figure}

\comment{
\begin{figure}[ht!]
\centering
    \includegraphics[angle=90,width=0.72\textwidth]{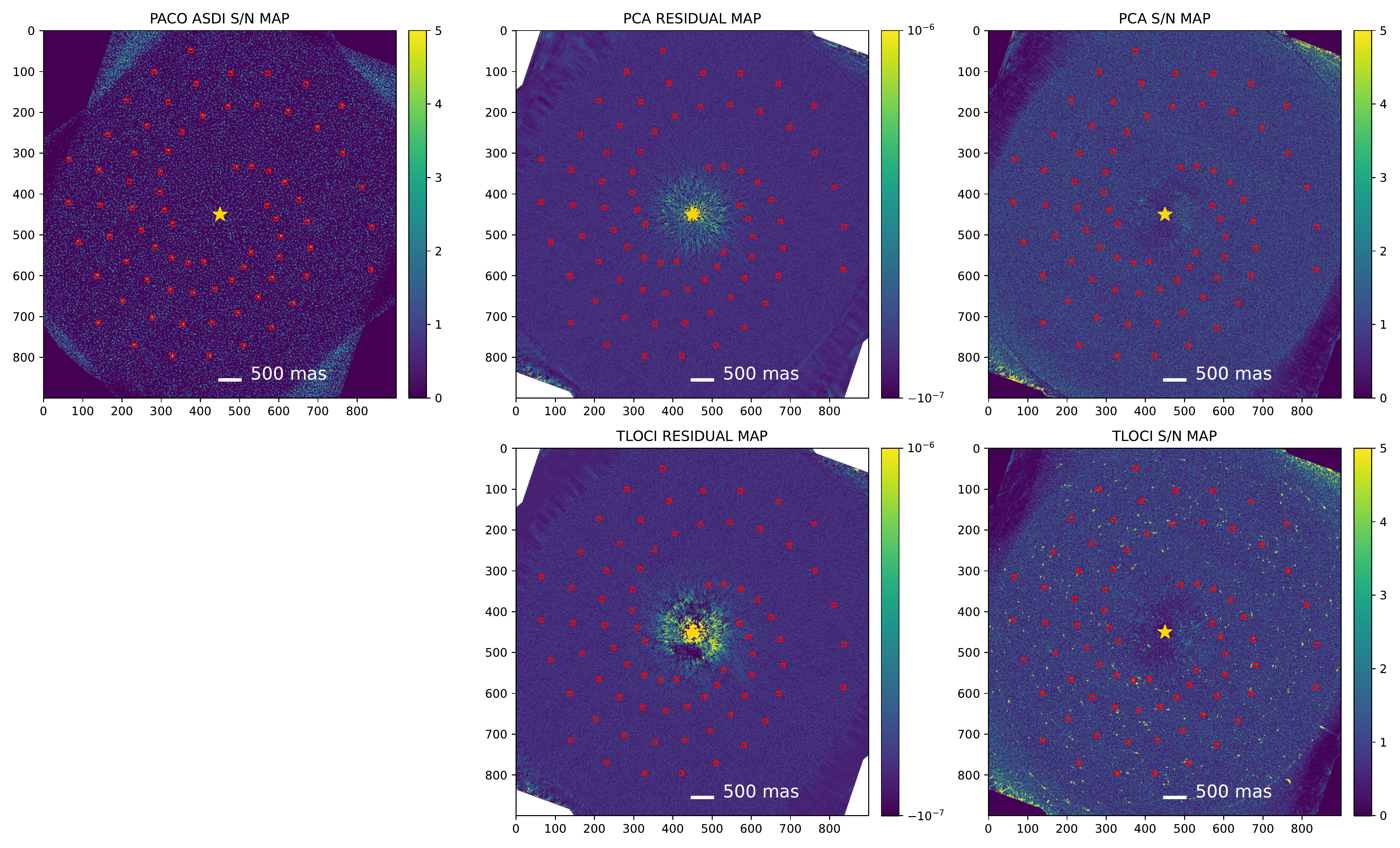}
    \caption{S/N map provided by PACO (top left), residual map provided by SpeCal PCA using 10 modes (top middle), associated S/N map (top left), residual map from TLOCI (bottom middle) and associated S/N map (bottom right). Injections are clearly visible on the S/N maps from PACO. The stationarity of the S/N maps from PACO is also better than for PCA or TLOCI. None of the detection peaks at high S/N on the PCA/TLOCI maps  corresponds to injected sources.}
    \label{inj_comp_ird_far}
\end{figure}
}

\begin{figure}[htp]
\centering

\includegraphics[width=.95\textwidth]{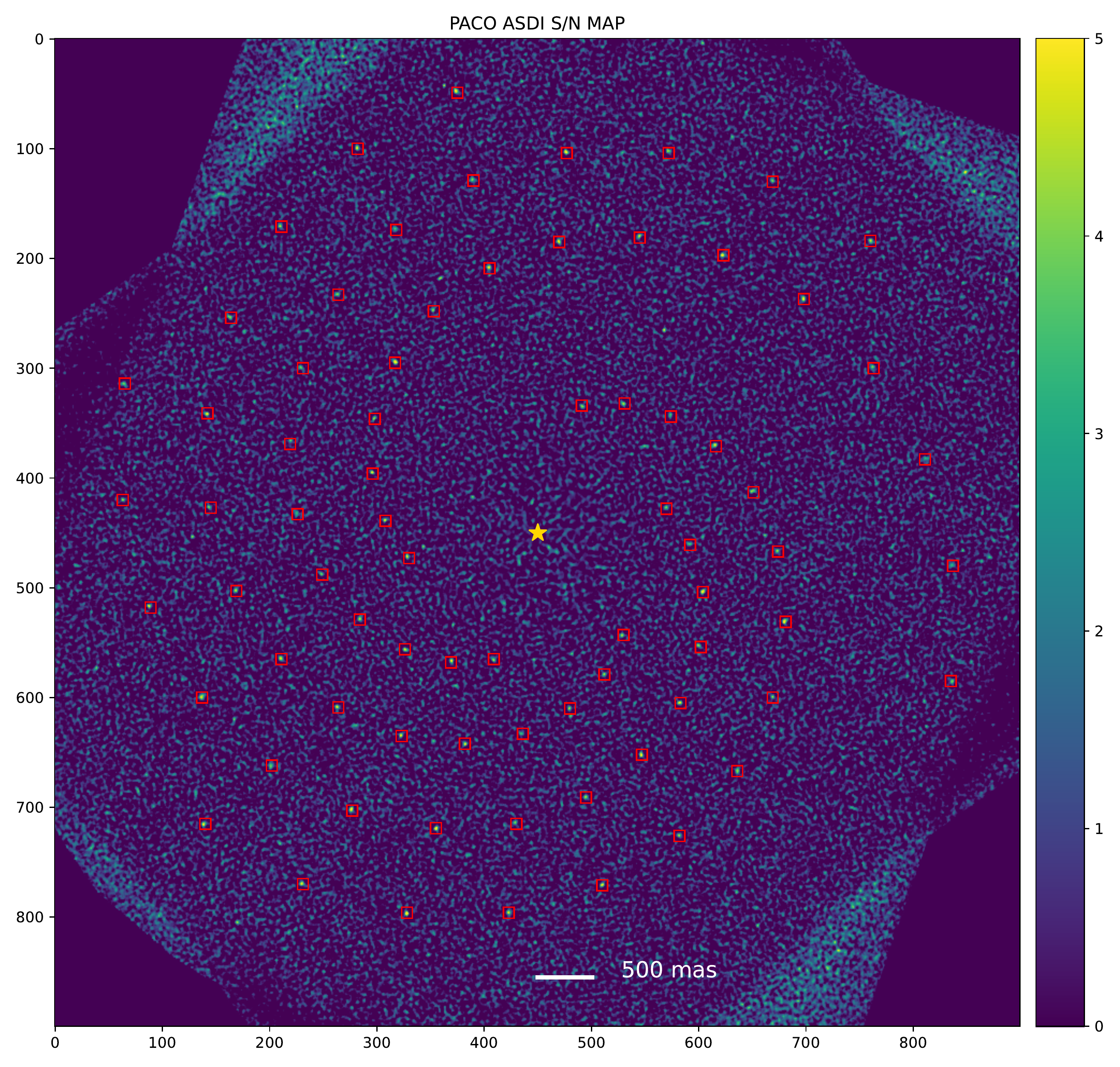}

\caption{PACO ASDI S/N map assuming the SED of sought sources is flat. The locations of injected sources are highlighted by the red boxes.}
\label{paco_snr_map_inj_far}
\end{figure}
\clearpage
\begin{figure}[htp]
\centering
\includegraphics[width=.95\textwidth]{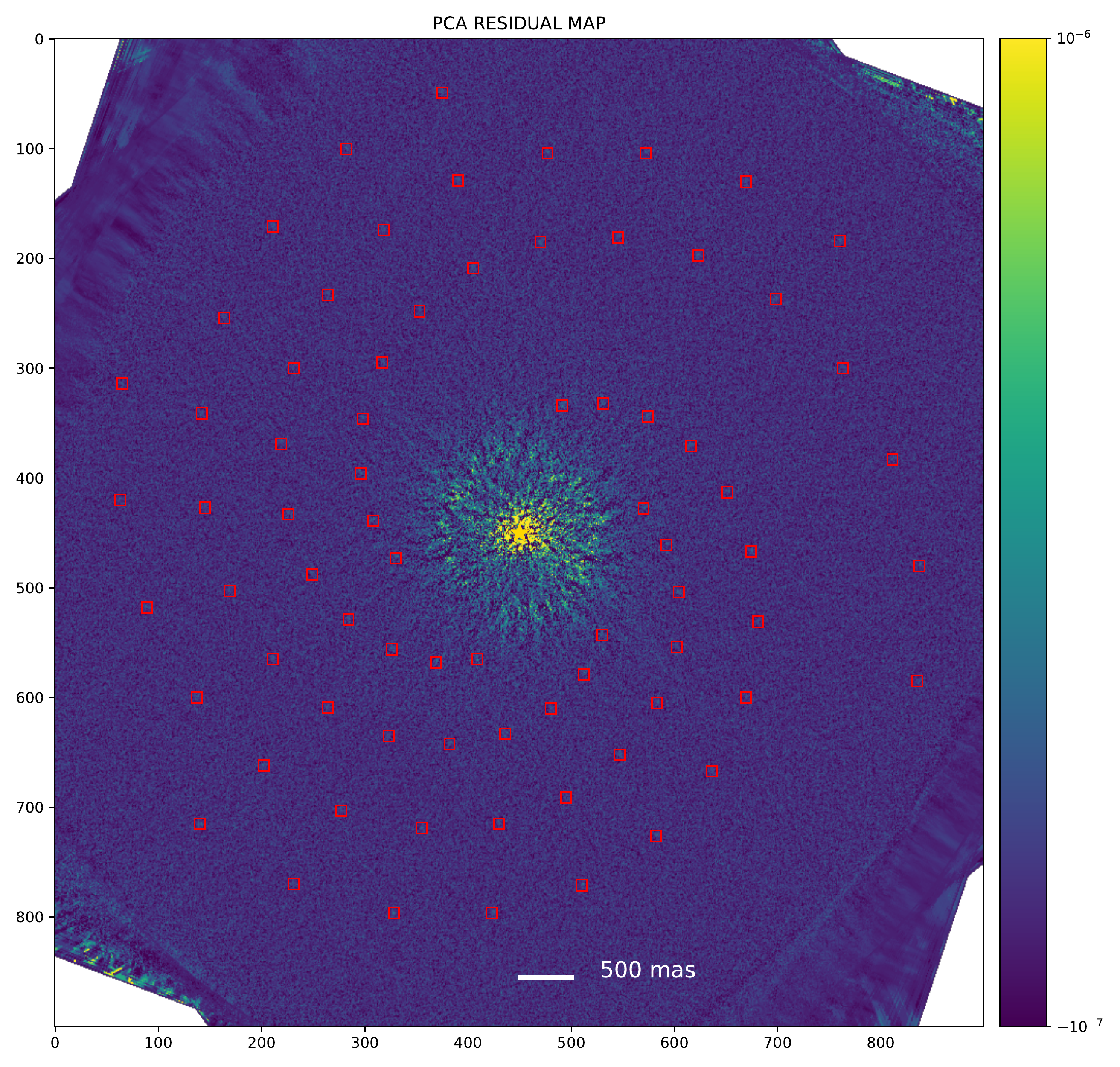}
\caption{PCA residual map. The locations of injected sources are highlighted by  red boxes.}
\label{pca_residual_map_inj_far}
\end{figure}
\clearpage
\begin{figure}[htp]
\centering

\includegraphics[width=.95\textwidth]{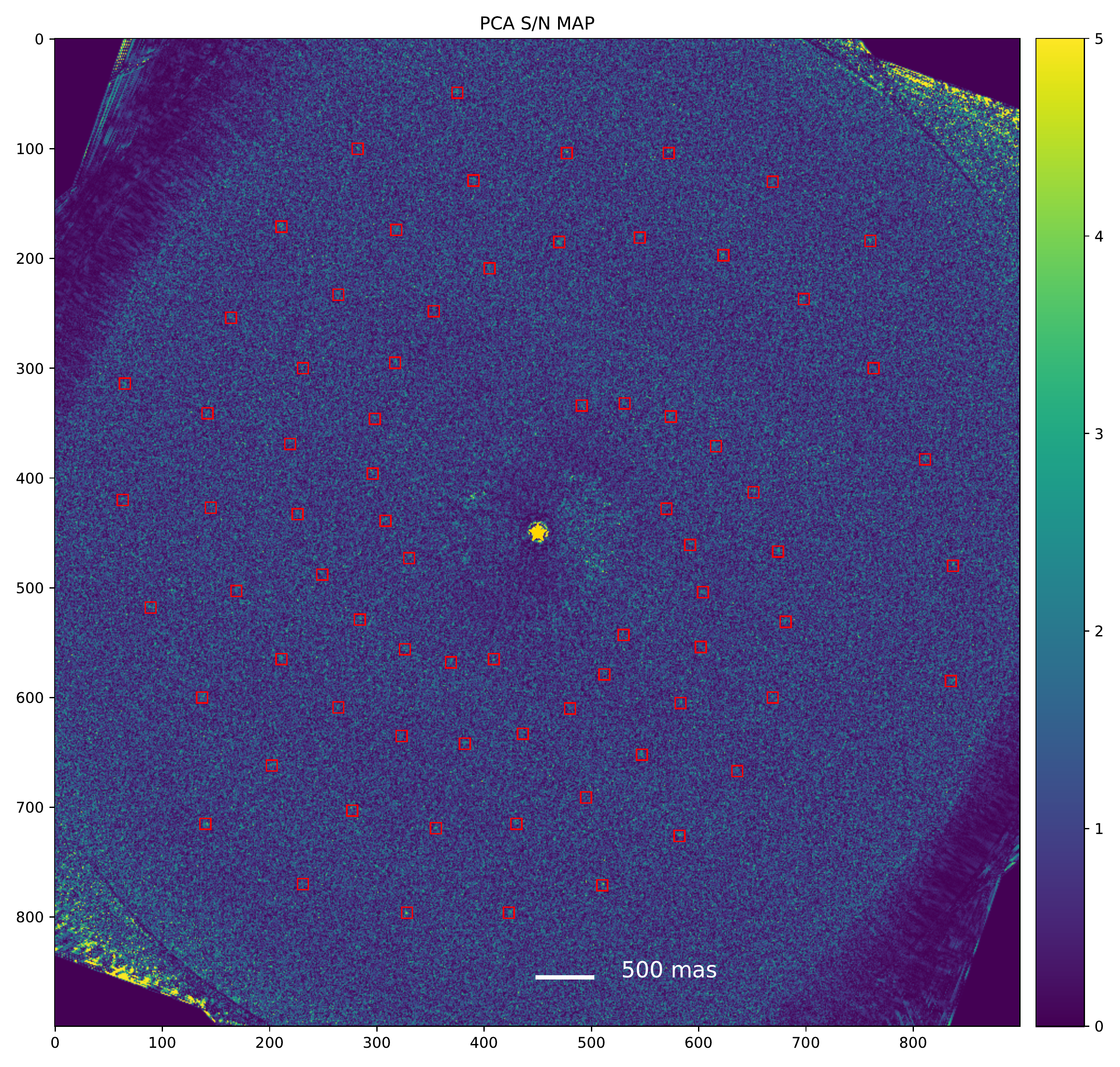}

\caption{PCA S/N map. The locations of injected sources are highlighted by  red boxes.}
\label{pca_snr_map_inj_far}
\end{figure}
\clearpage
\begin{figure}[htp]
\centering
\includegraphics[width=.95\textwidth]{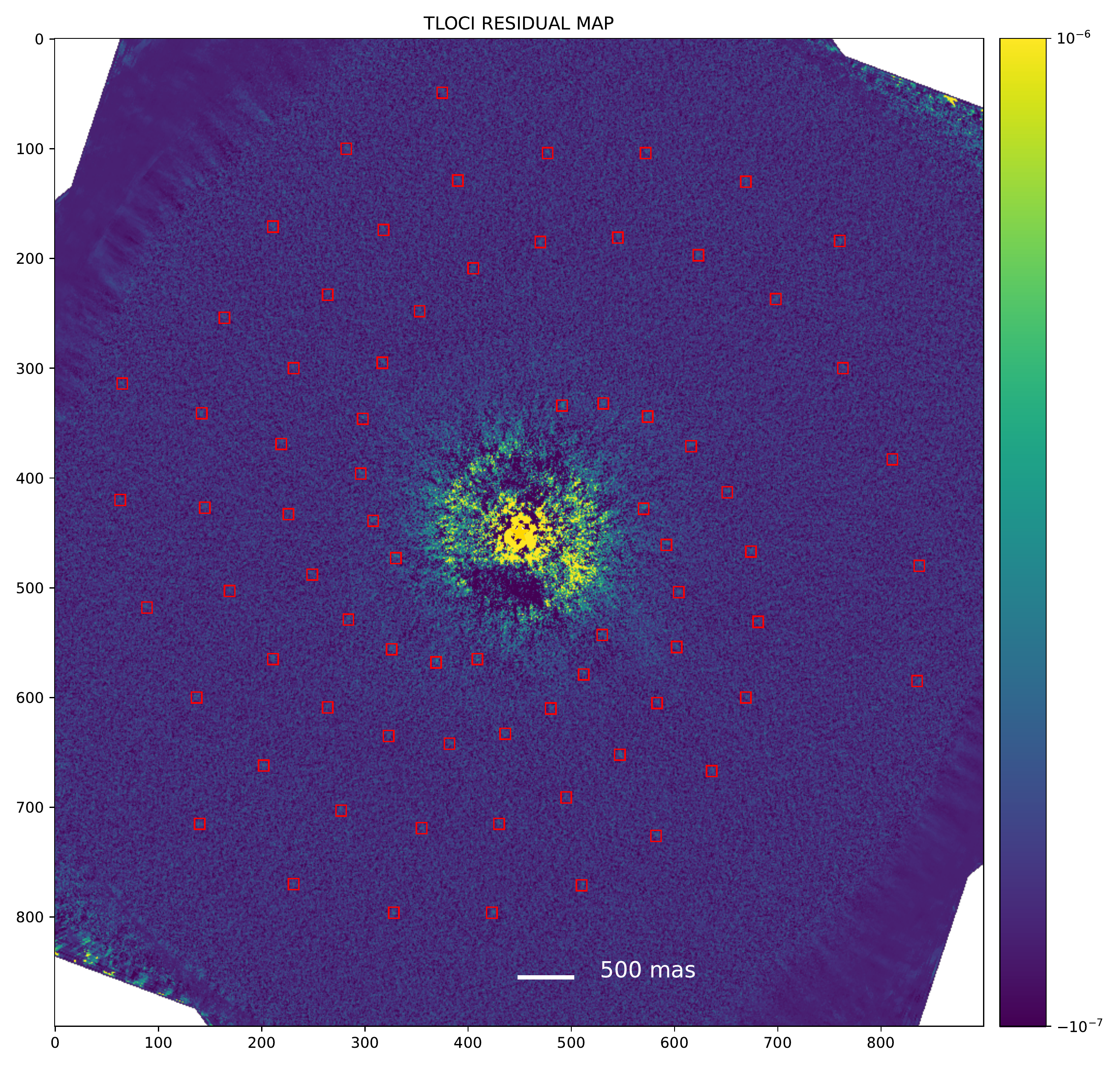}
\caption{TLOCI residual map. The locations of injected sources are highlighted by  red boxes.}
\label{tloci_residual_map_inj_far}
\end{figure}
\clearpage
\begin{figure}[htp]
\centering
\includegraphics[width=.95\textwidth]{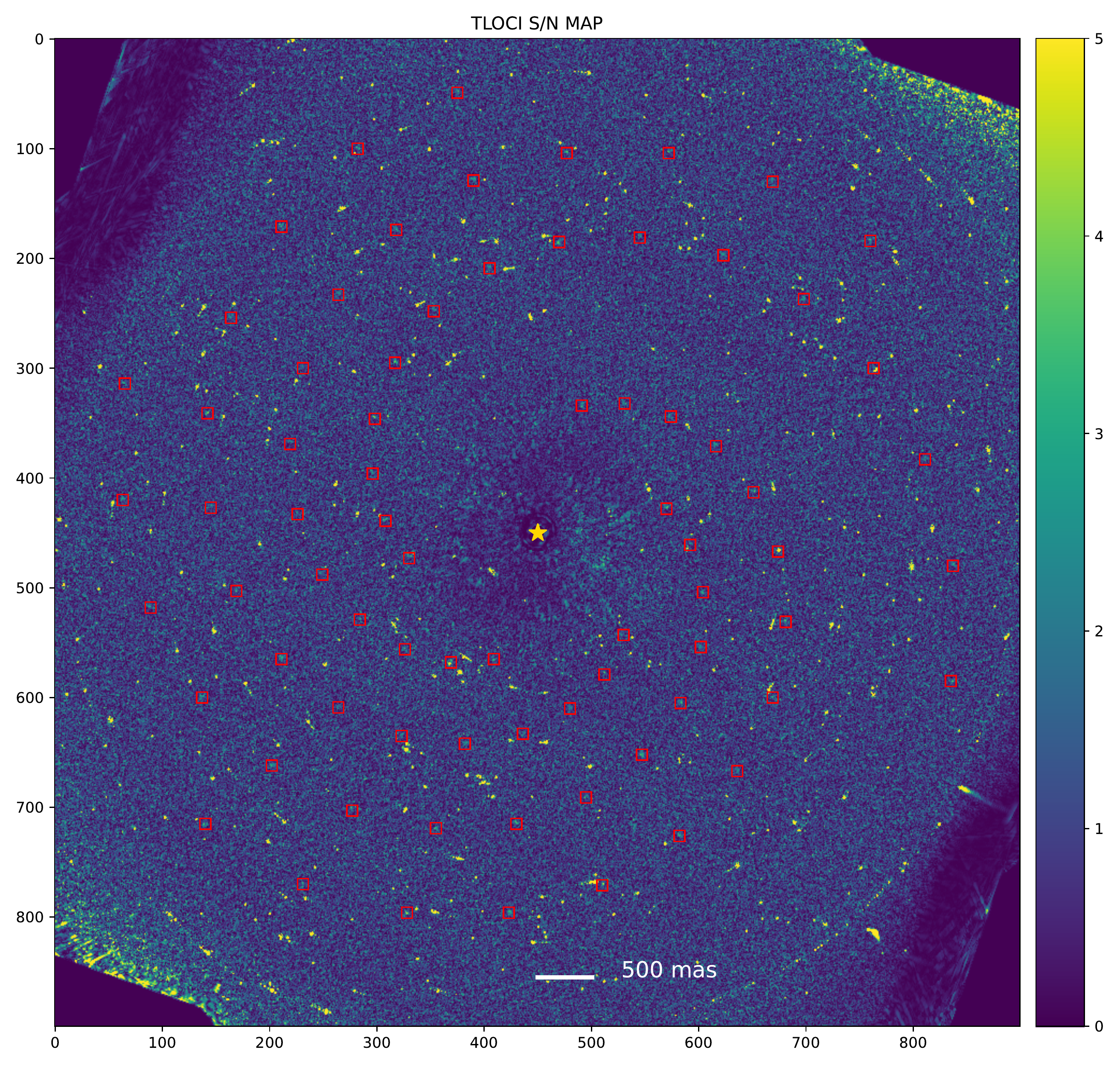}
\caption{TLOCI S/N map. The locations of injected sources are highlighted by  red boxes.}
\label{tloci_snr_map_inj_far}
\end{figure}
\clearpage
\begin{figure*}[t!]
\centering
\includegraphics[width=.5\textwidth]{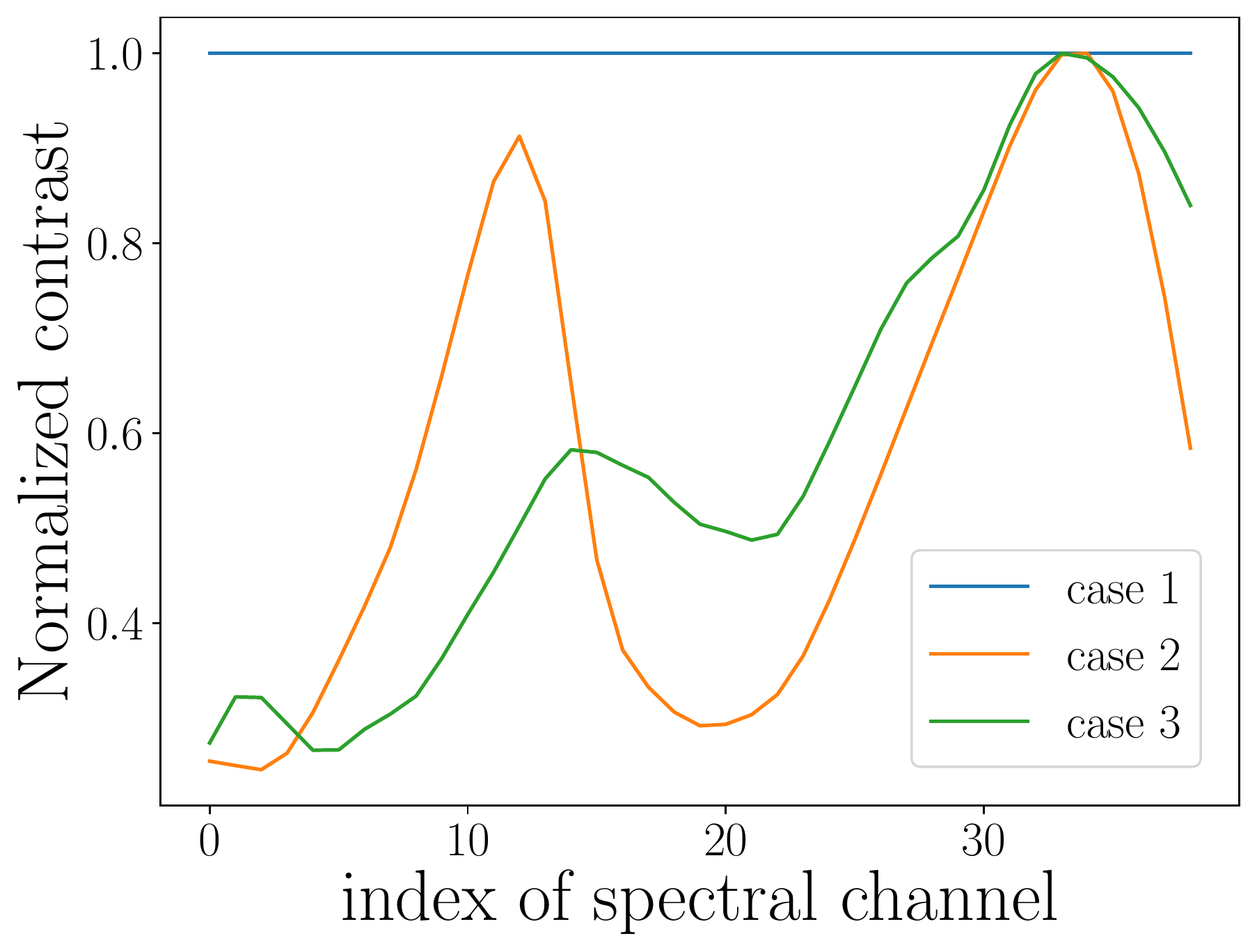}\quad
\caption{Normalized injected contrasts for IFS for the three cases considered.}
\label{normalized_spectra}
\end{figure*}

\begin{table}[h!]
\footnotesize
\begin{center}
\begin{tabular}{c|ccc|ccc}

CASE & SEP (mas) & PA ($\degree$) & MEAN CONTRAST & SEP (mas) & PA ($\degree$) & MEAN CONTRAST  \\ 
\hline
\hline
1 & 200 & 240 & $6.18\times10^{-6}$ & 400 & 225 & $2.27\times10^{-6}$\\ 
1 & 600 & 210 & $1.43\times10^{-6}$ & 800 & 195 & $1.06\times10^{-6}$\\ 
2 & 200 & 120 & $3.58\times10^{-6}$& 400 & 105 & $1.61\times10^{-6}$\\ 
2 & 600 & 90 & $1.13\times10^{-6}$& 800 & 75 & $1.48\times10^{-6}$\\ 
3 & 200 & 0 & $2.81\times10^{-6}$& 400 & 345 & $1.83\times10^{-6}$\\ 
3 & 600 & 330 & $1.14\times10^{-6}$& 800 & 305 & $1.16\times10^{-6}$\\ 

\hline
\hline
\end{tabular}
\caption{IFS injected fake planets parameters.}
\label{table_inj_ifs}
\end{center}
\end{table}

\clearpage
\section{MESS2 results}
\label{MESS2}

\begin{figure}[htp]
\centering

\includegraphics[width=.45\textwidth]{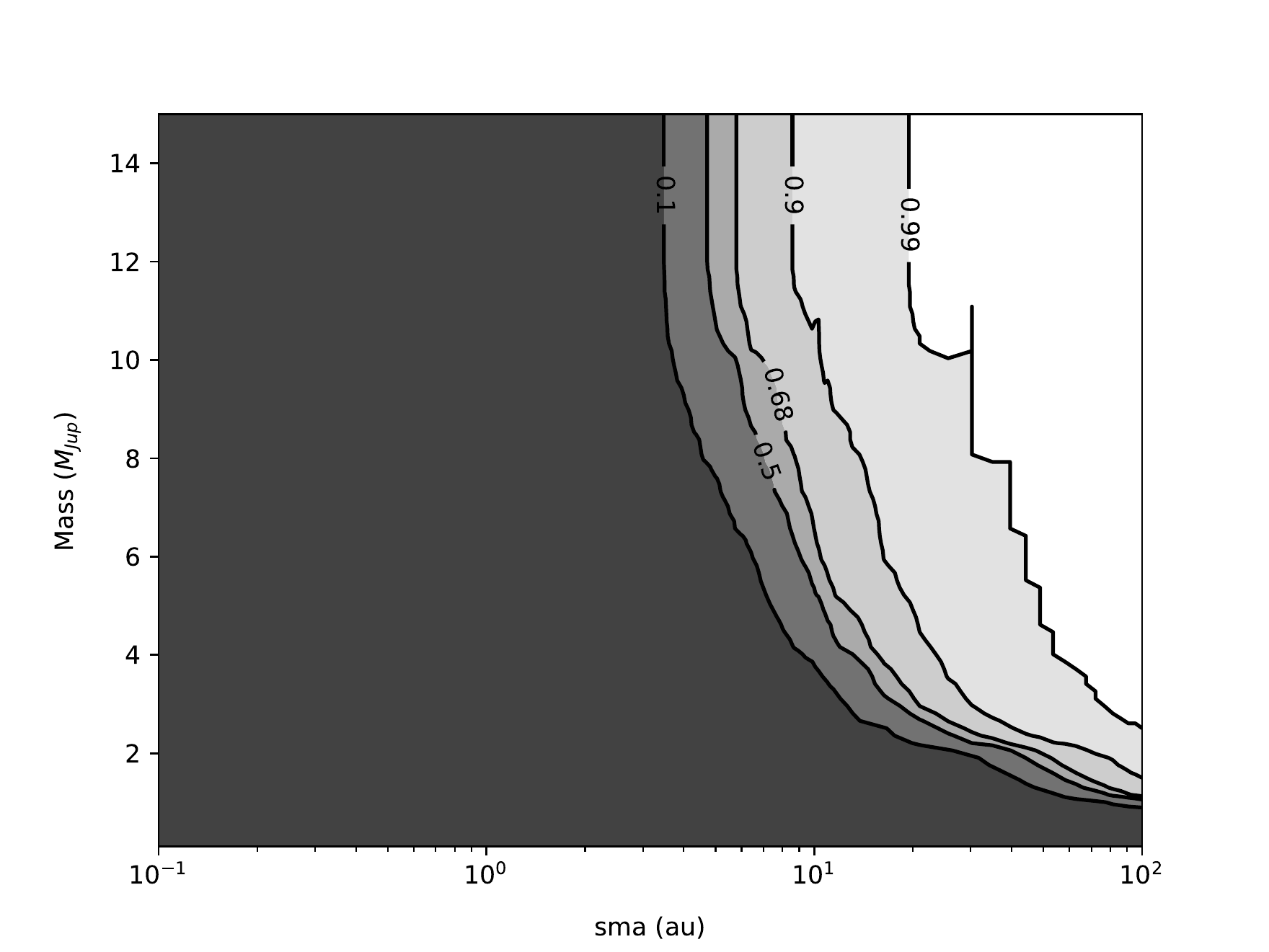}\quad
\includegraphics[width=.45\textwidth]{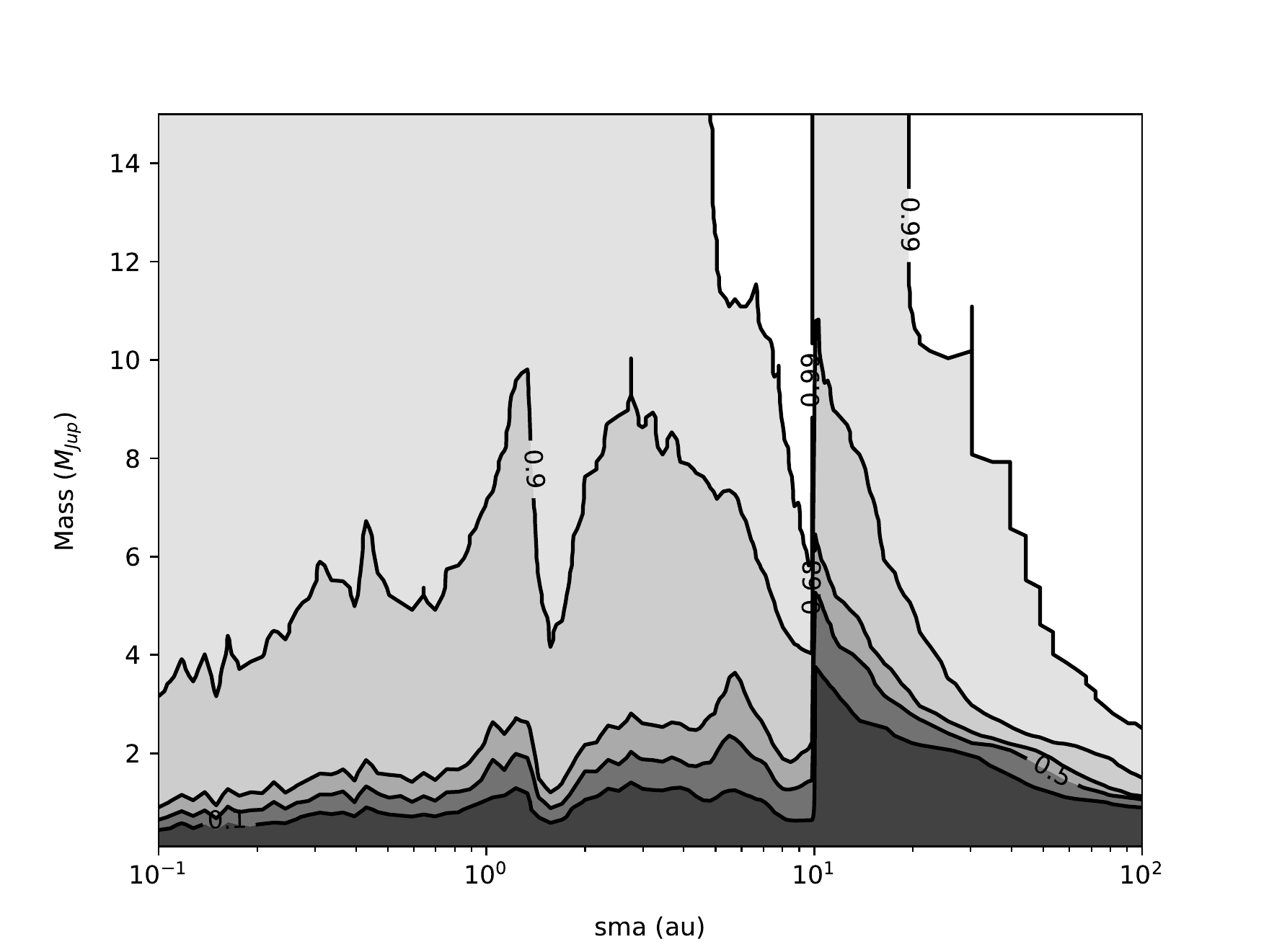}
\caption{Results for HD 105,  DI only (left), DI+RV (right). One epoch was available for DI.}
\end{figure}

\begin{figure}[htp]
\centering

\includegraphics[width=.45\textwidth]{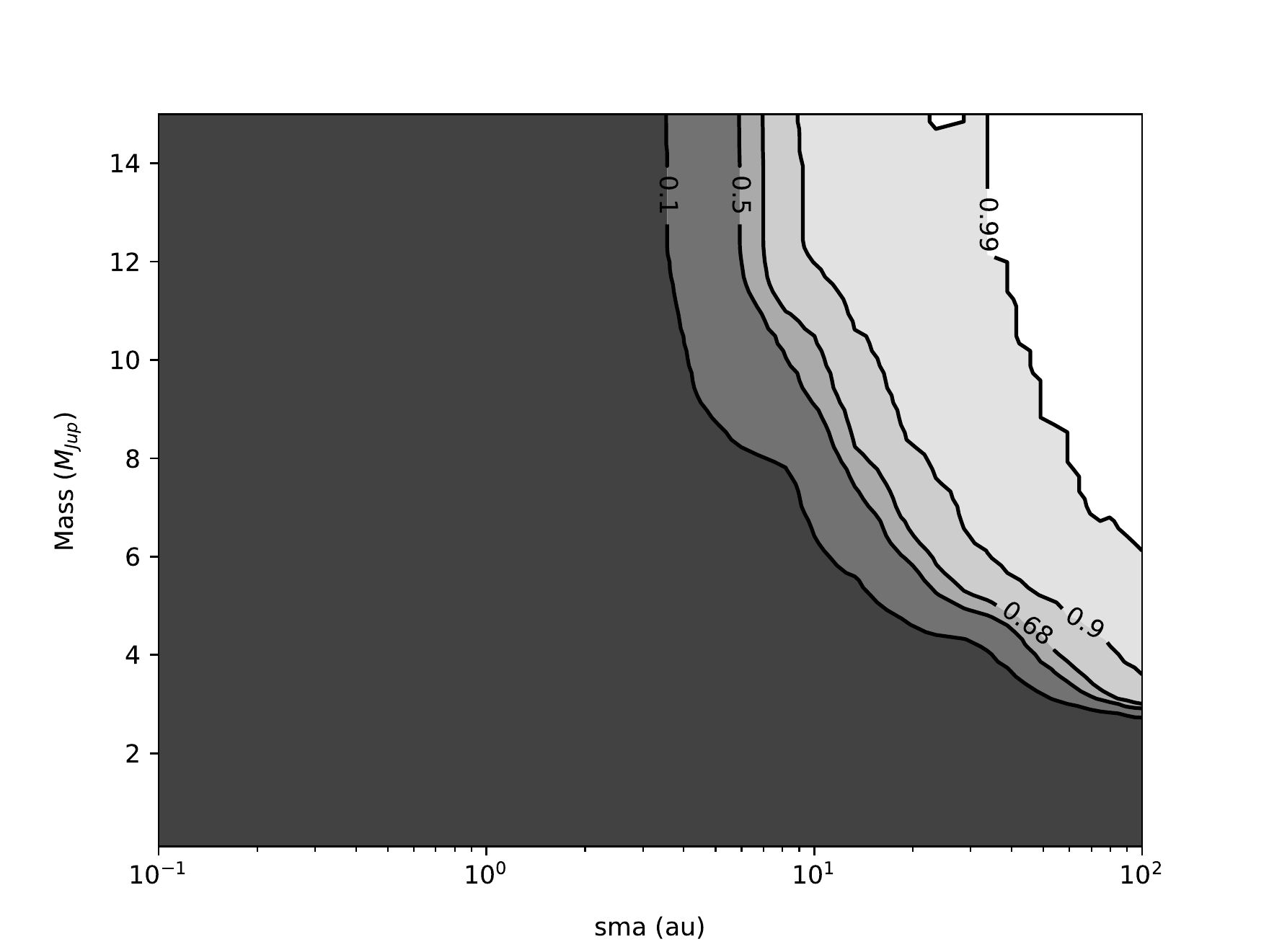}\quad
\includegraphics[width=.45\textwidth]{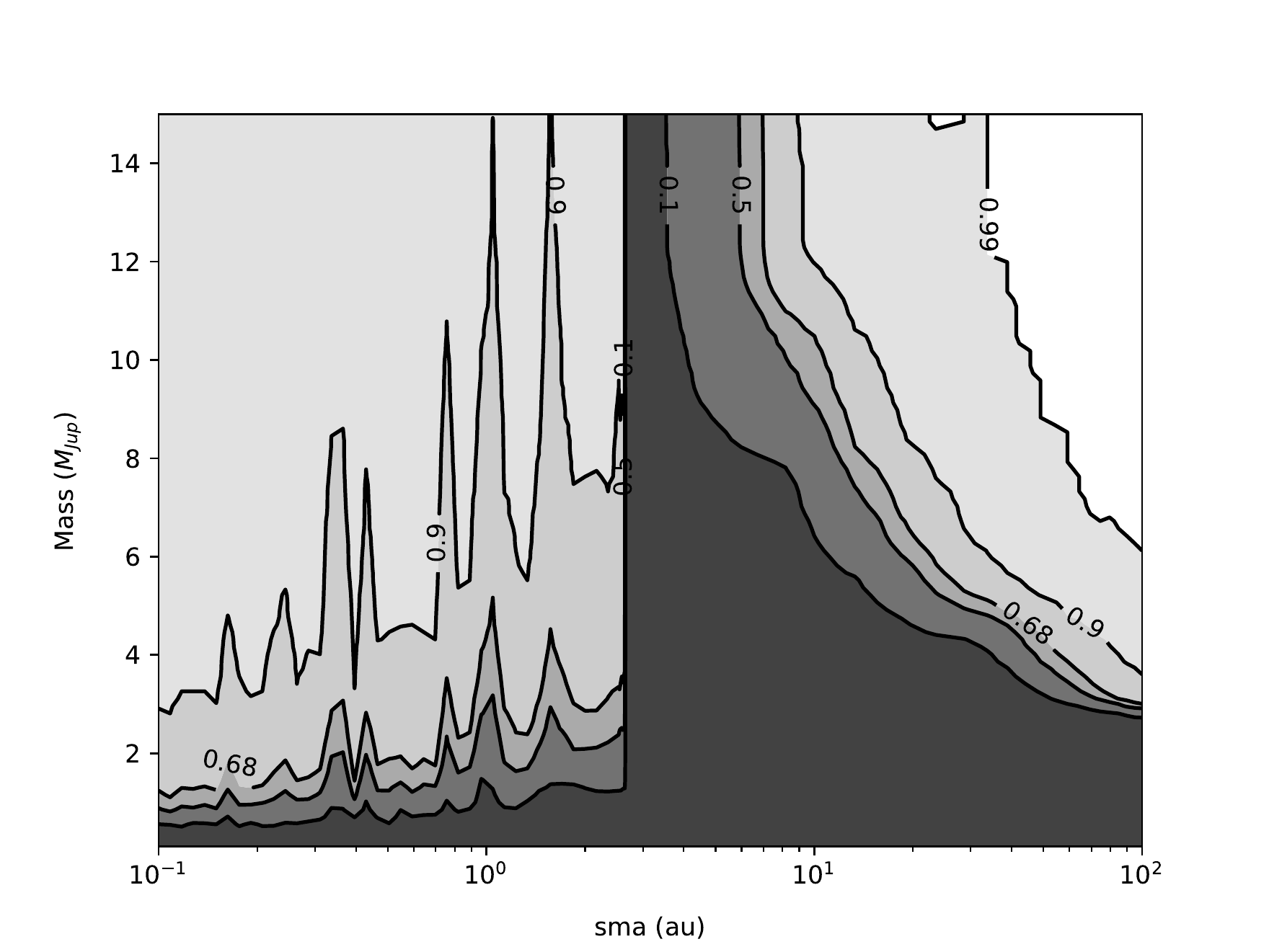}
\caption{Results for HD 377,  DI only (left), DI+RV (right). One epoch was available for DI.}
\end{figure}

\begin{figure}[htp]
\centering

\includegraphics[width=.45\textwidth]{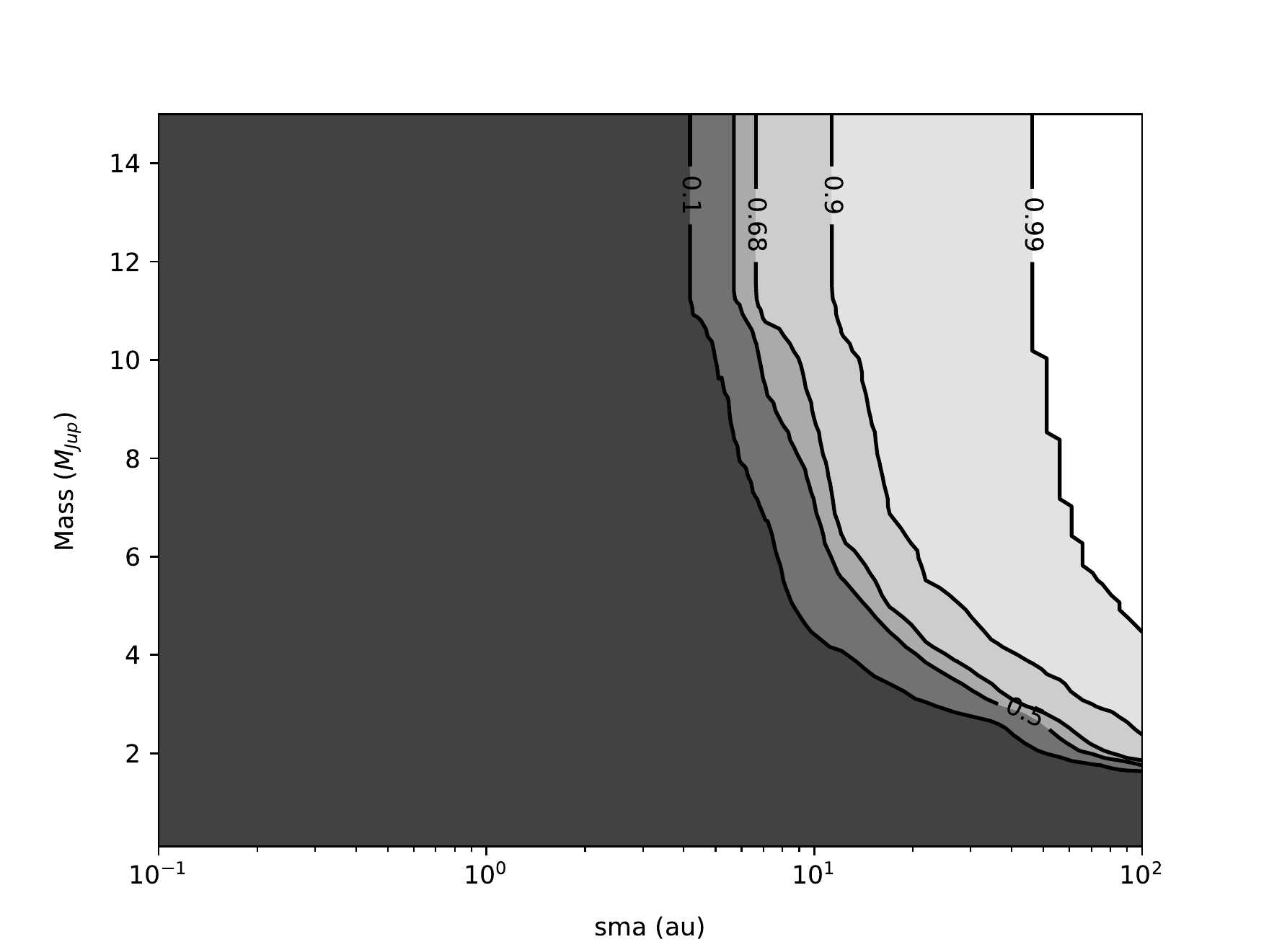}\quad
\includegraphics[width=.45\textwidth]{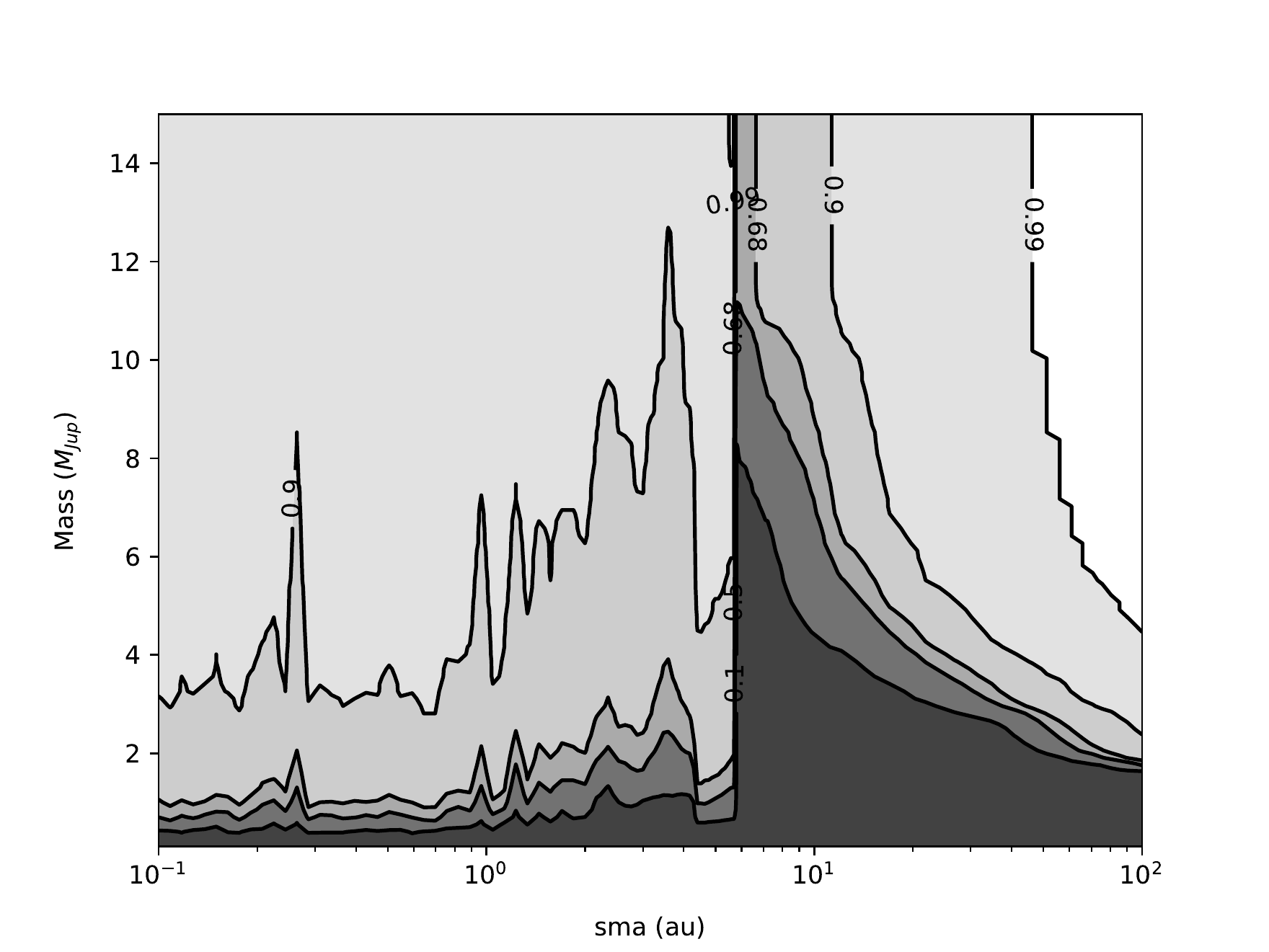}
\caption{Results for HD 987,  DI only (left), DI+RV (right). One epoch was available for DI.}
\end{figure}

\begin{figure}[htp]
\centering

\includegraphics[width=.45\textwidth]{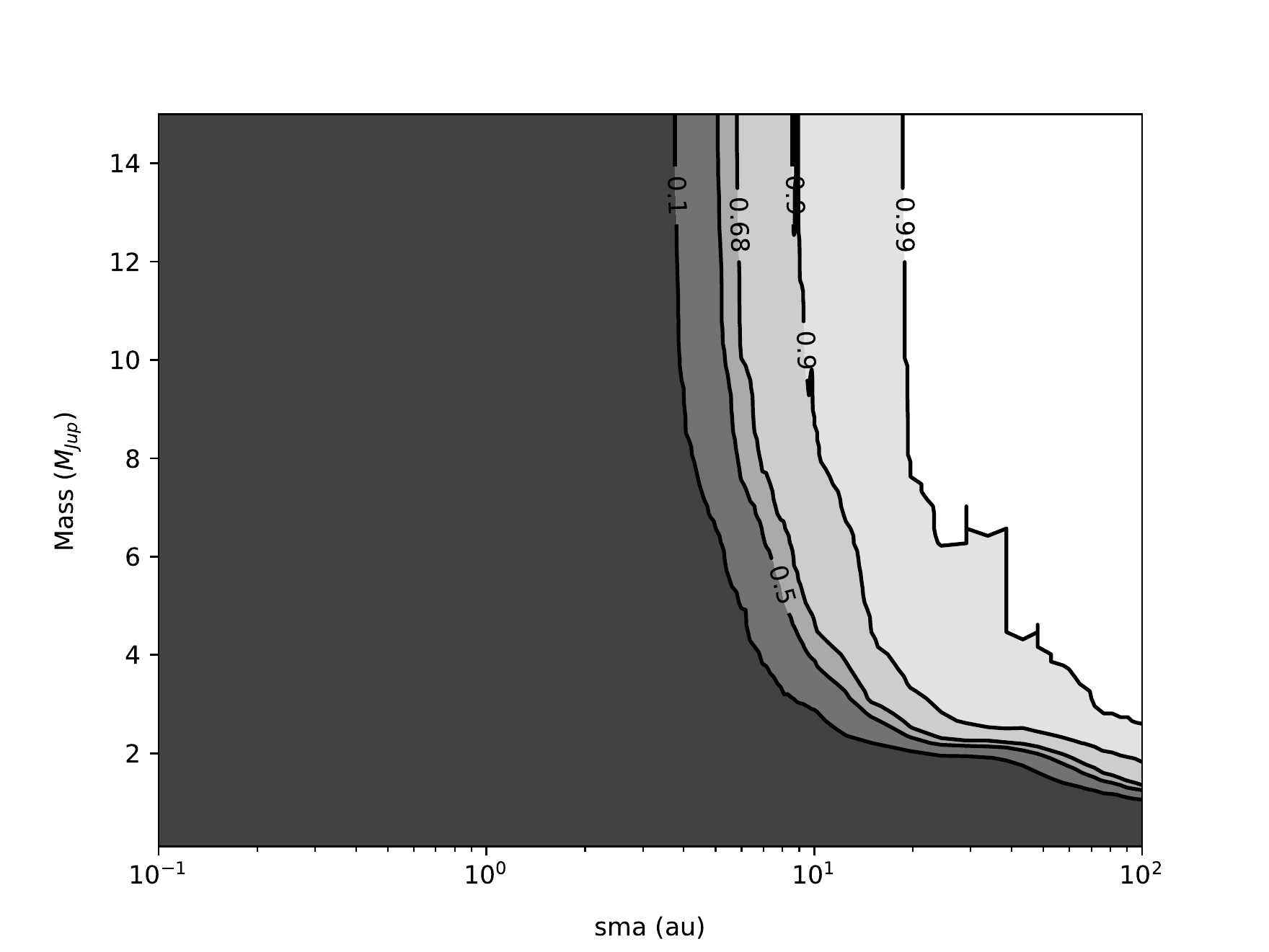}\quad
\includegraphics[width=.45\textwidth]{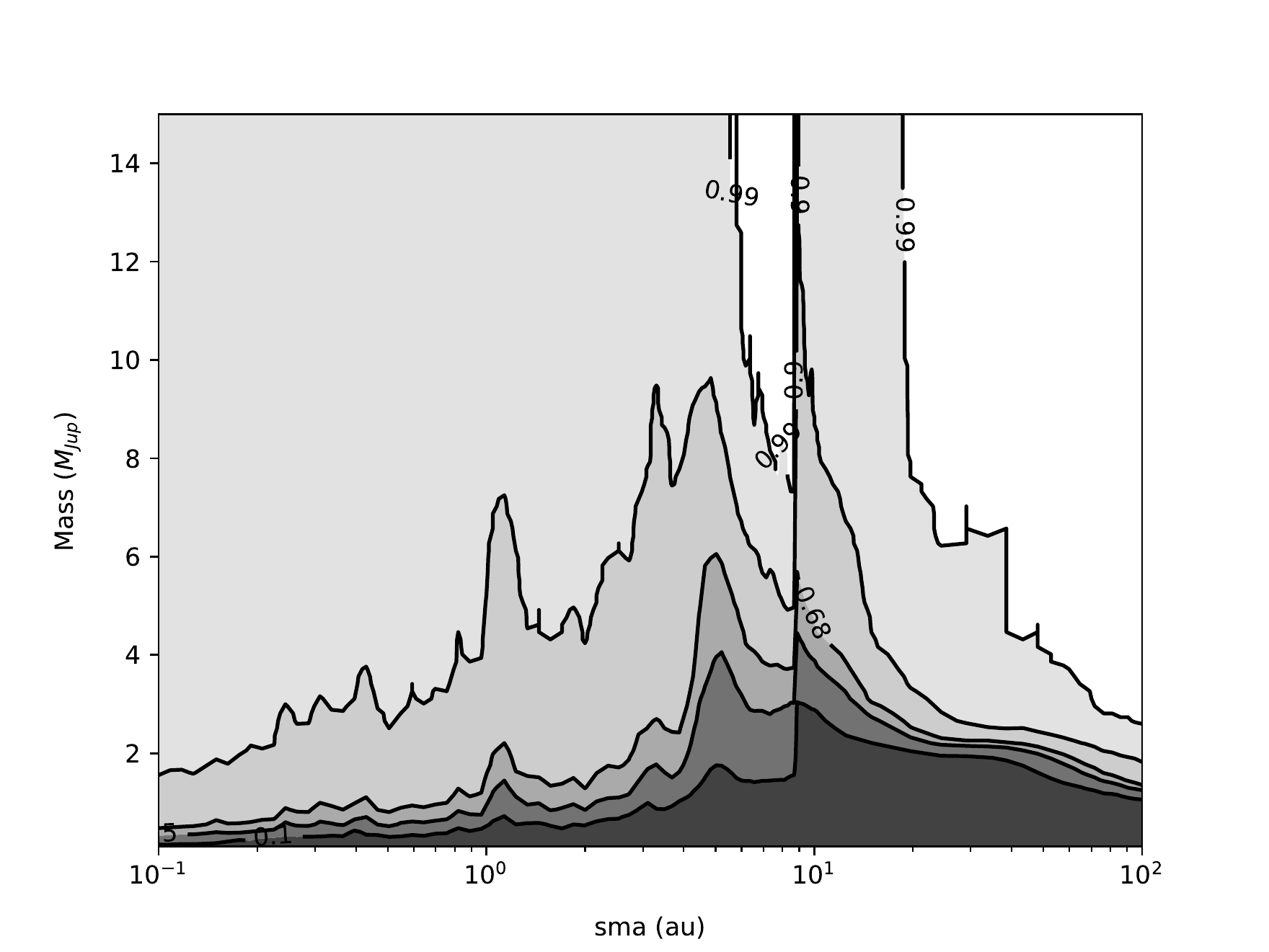}
\caption{Results for HD 1466,  DI only (left), DI+RV (right). Two  epochs were available for DI.}
\end{figure}

\begin{figure}[htp]
\centering

\includegraphics[width=.45\textwidth]{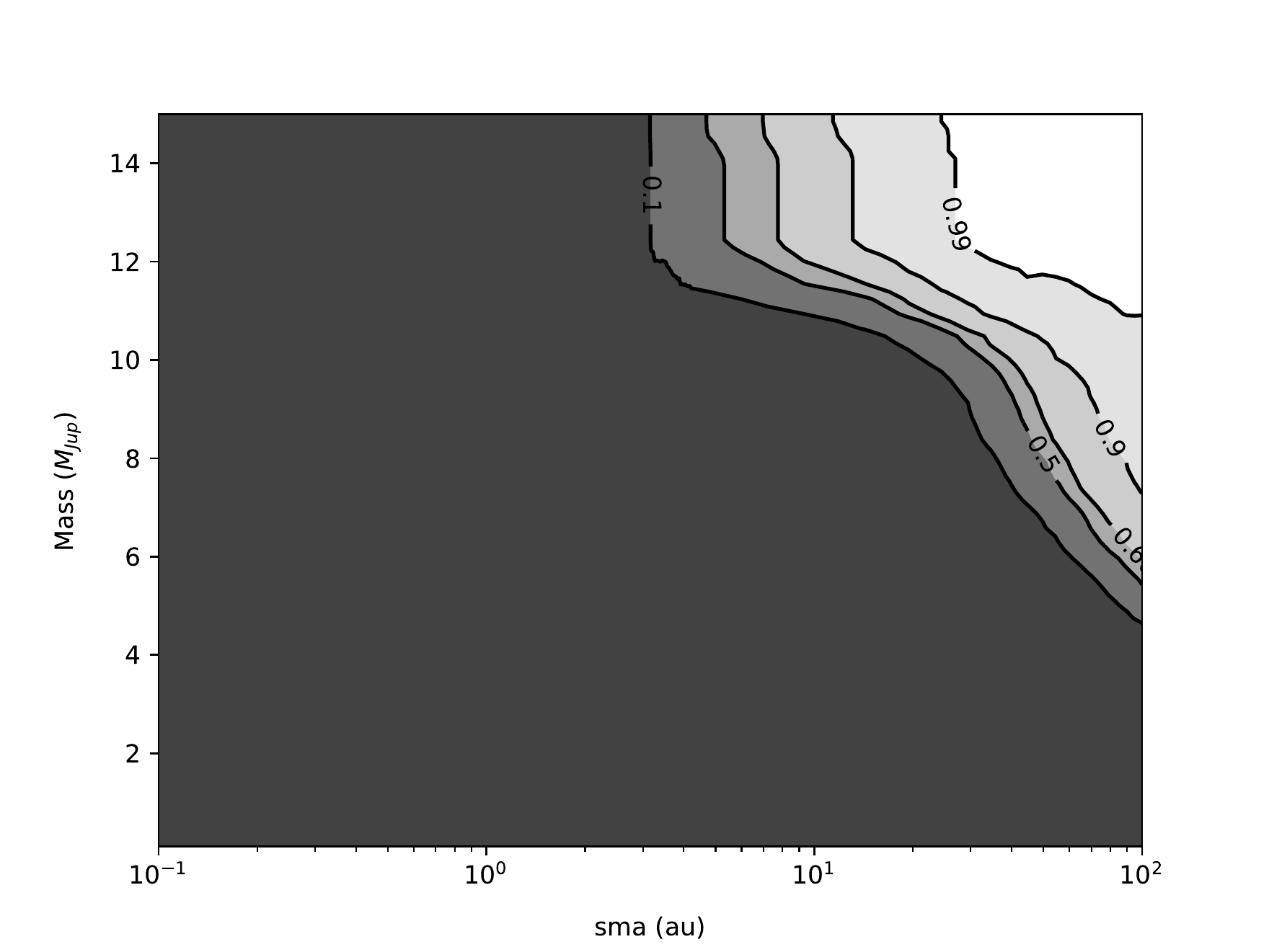}\quad
\includegraphics[width=.45\textwidth]{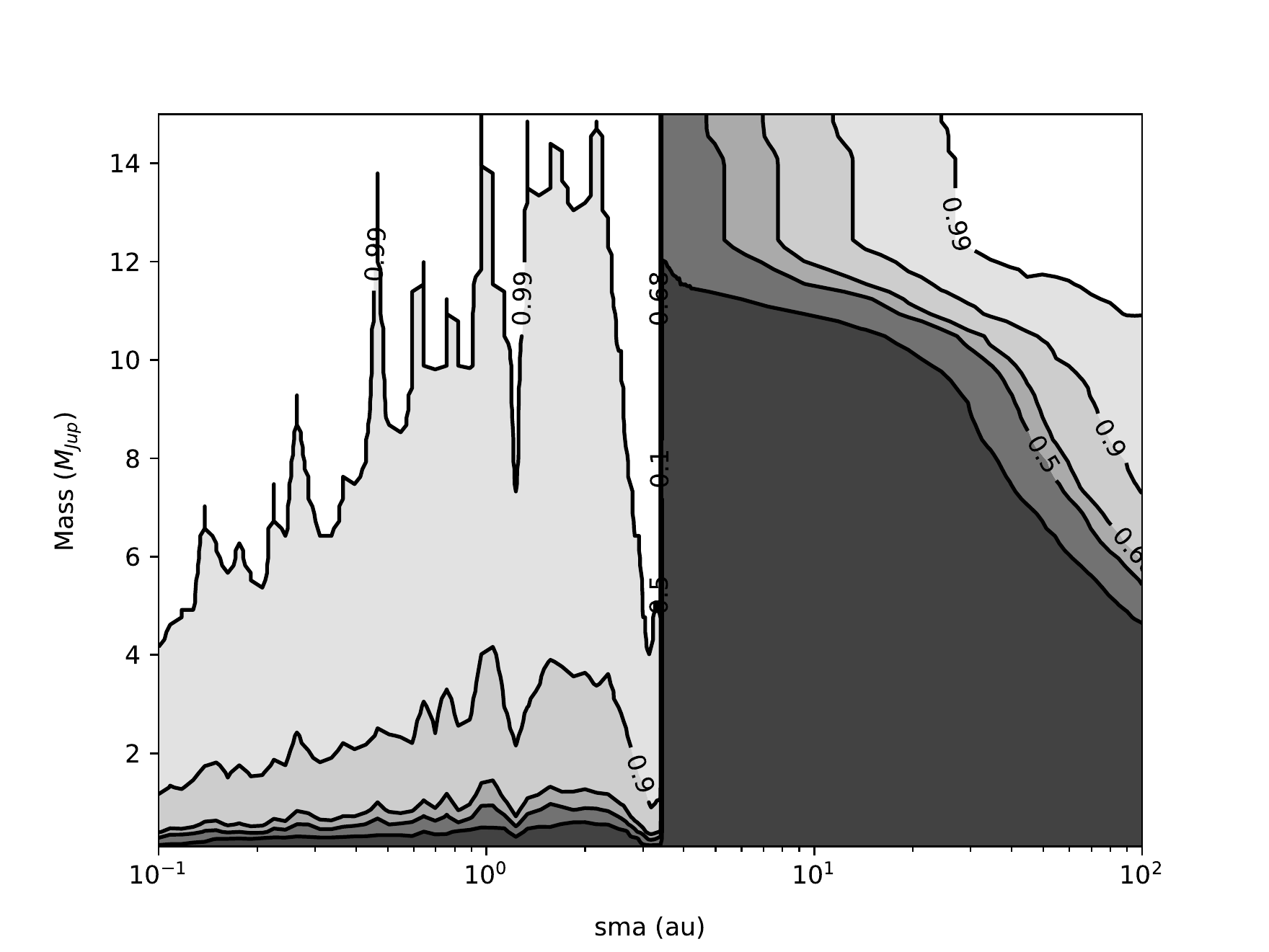}
\caption{Results for BD-12 243, DI only (left), DI+RV (right). One epoch was available for DI.}
\end{figure}

\begin{figure}[htp]
\centering

\includegraphics[width=.45\textwidth]{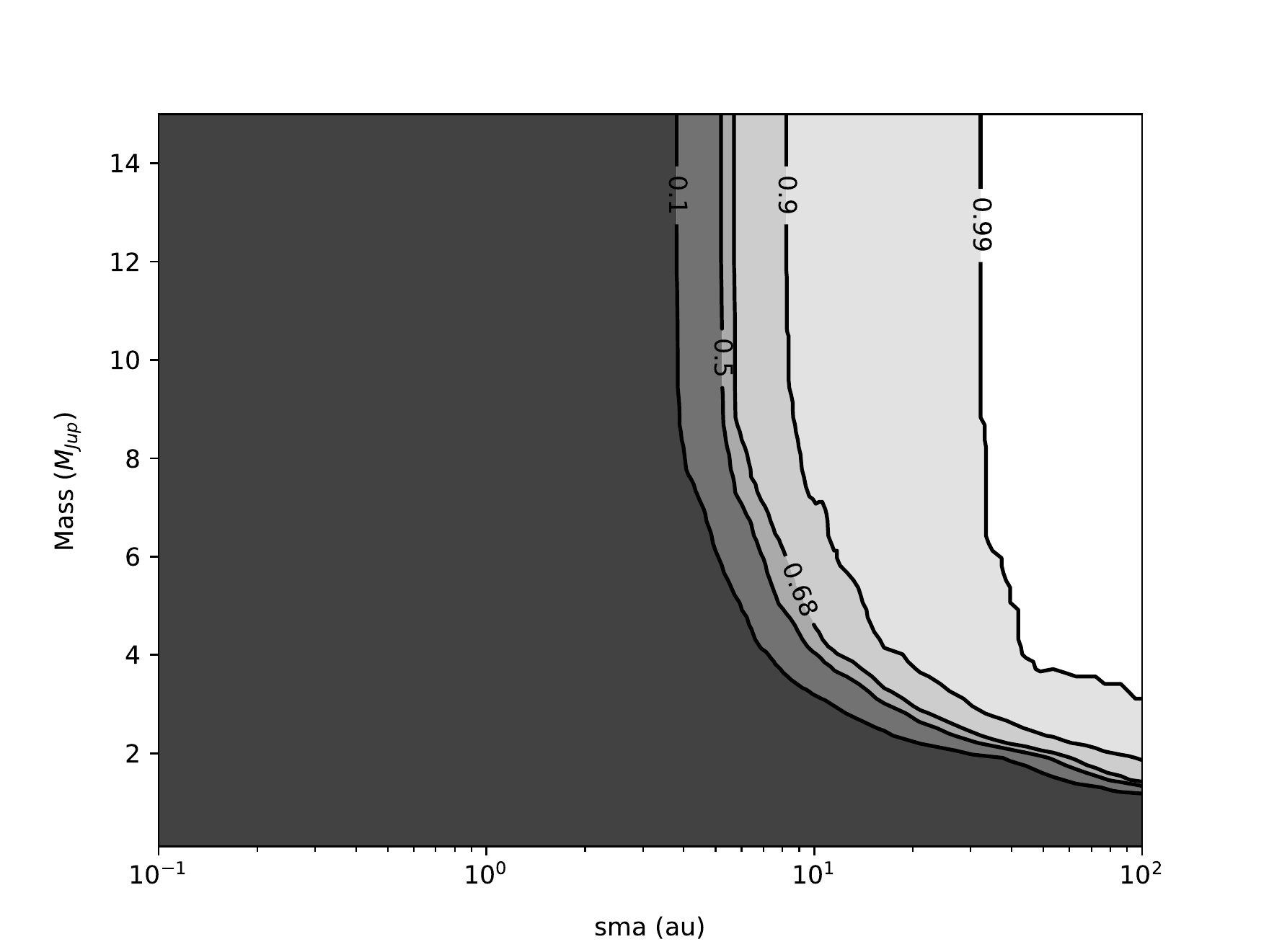}\quad
\includegraphics[width=.45\textwidth]{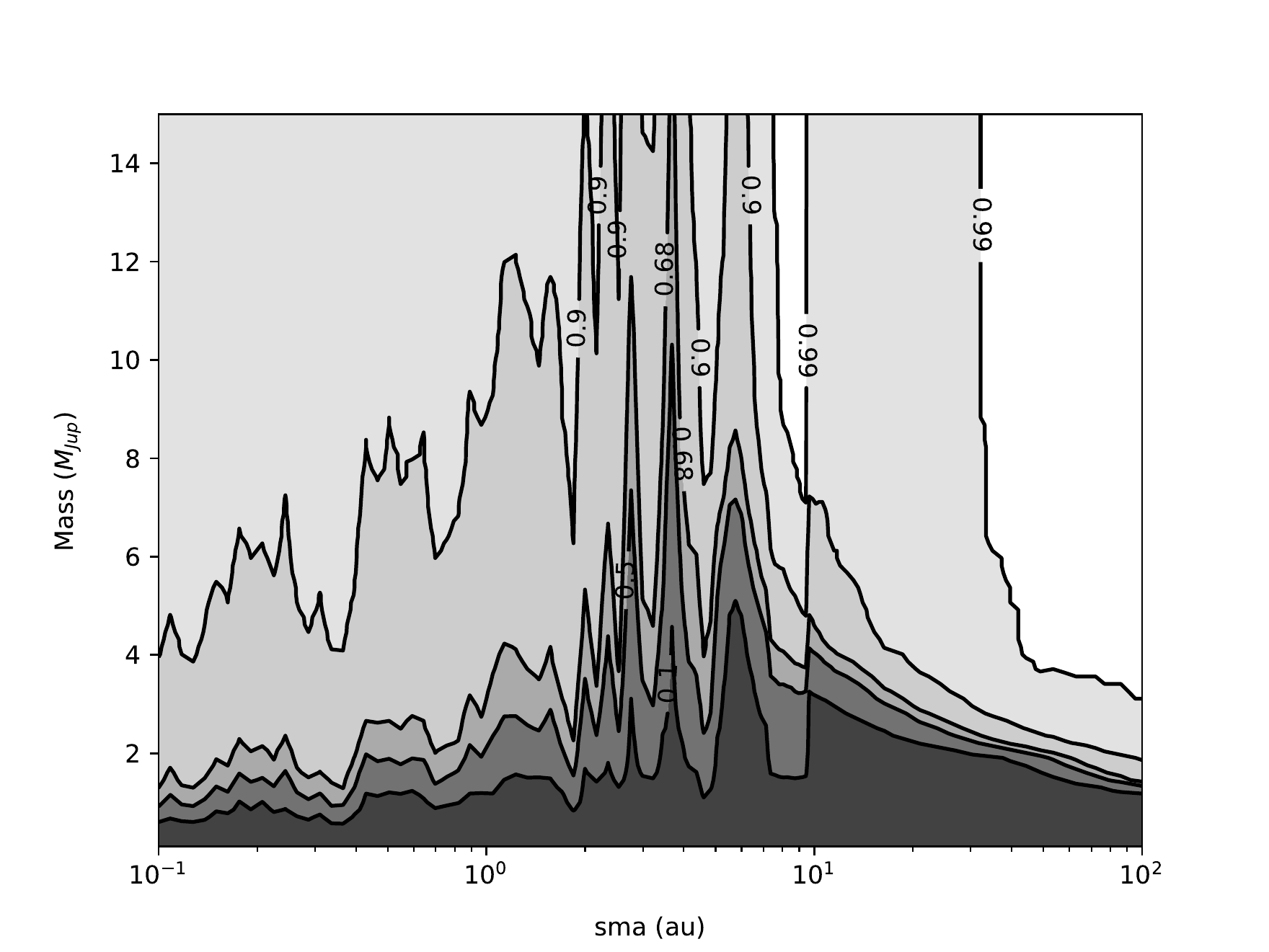}
\caption{Results for HD 8558, DI only (left), DI+RV (right). Three epochs were available for DI.}
\end{figure}

\begin{figure}[htp]
\centering

\includegraphics[width=.45\textwidth]{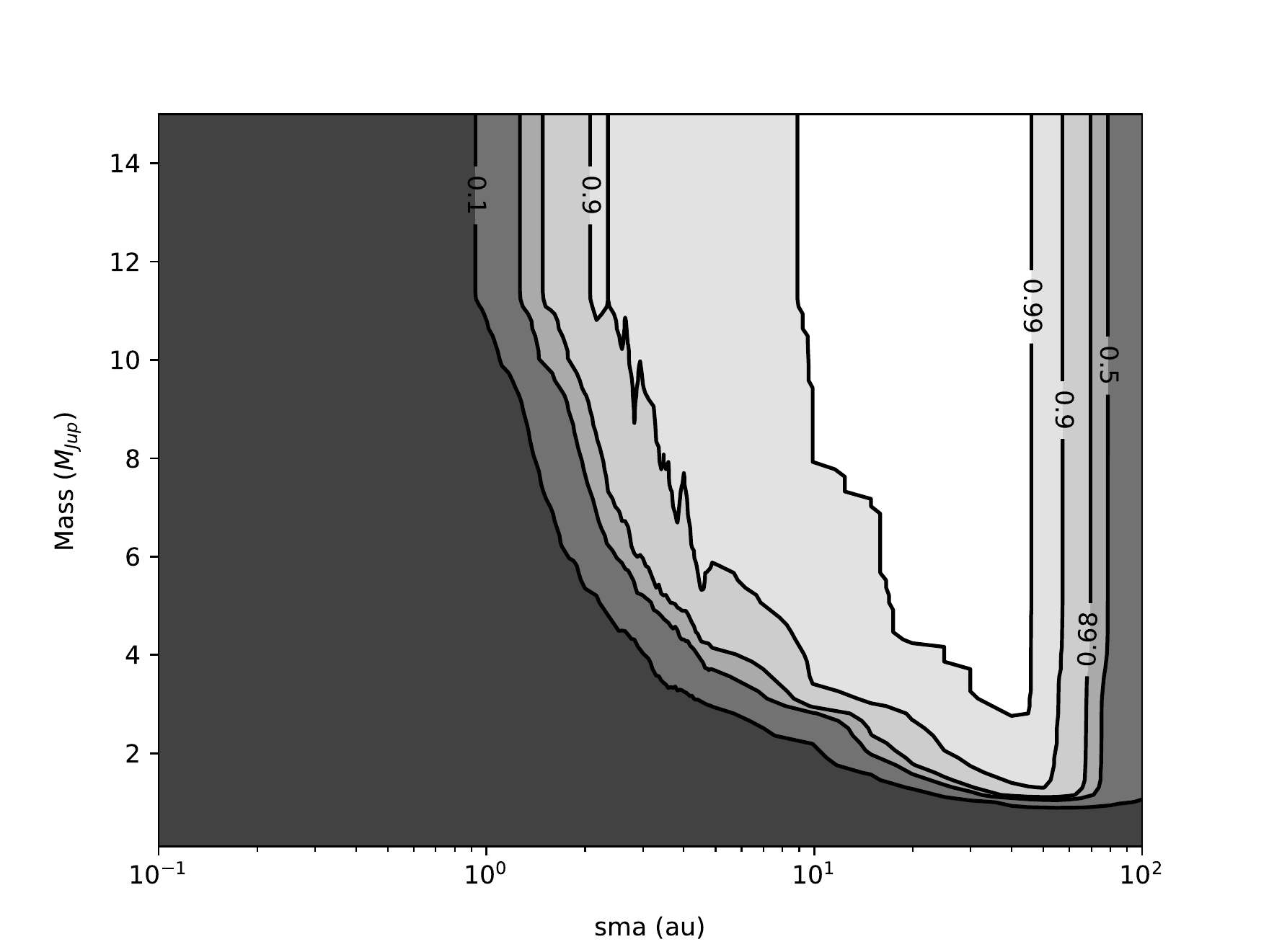}\quad
\includegraphics[width=.45\textwidth]{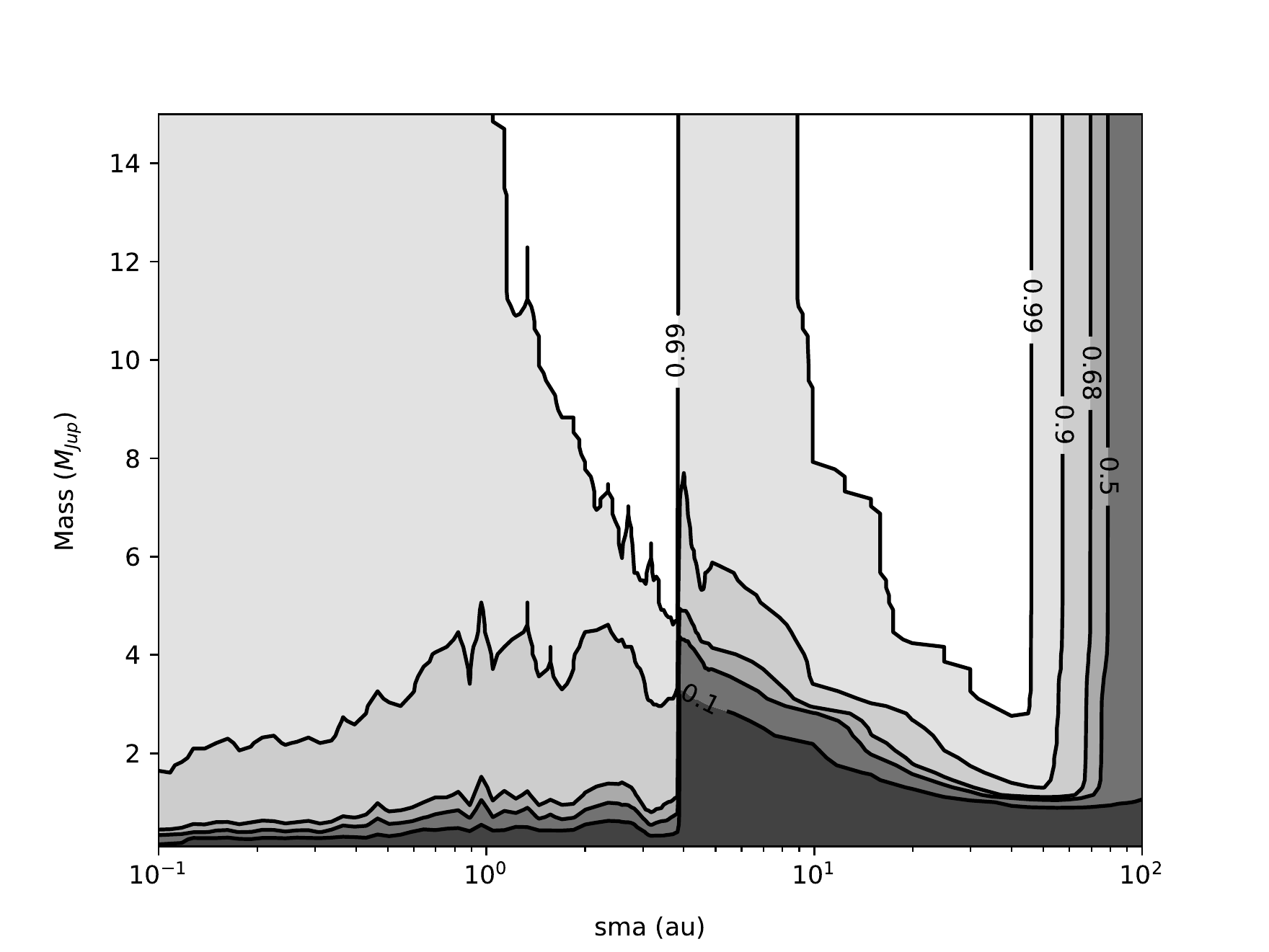}
\caption{Results for HD 17925, DI only (left), DI+RV (right). One epoch was available for DI.}
\end{figure}

\begin{figure}[htp]
\centering

\includegraphics[width=.45\textwidth]{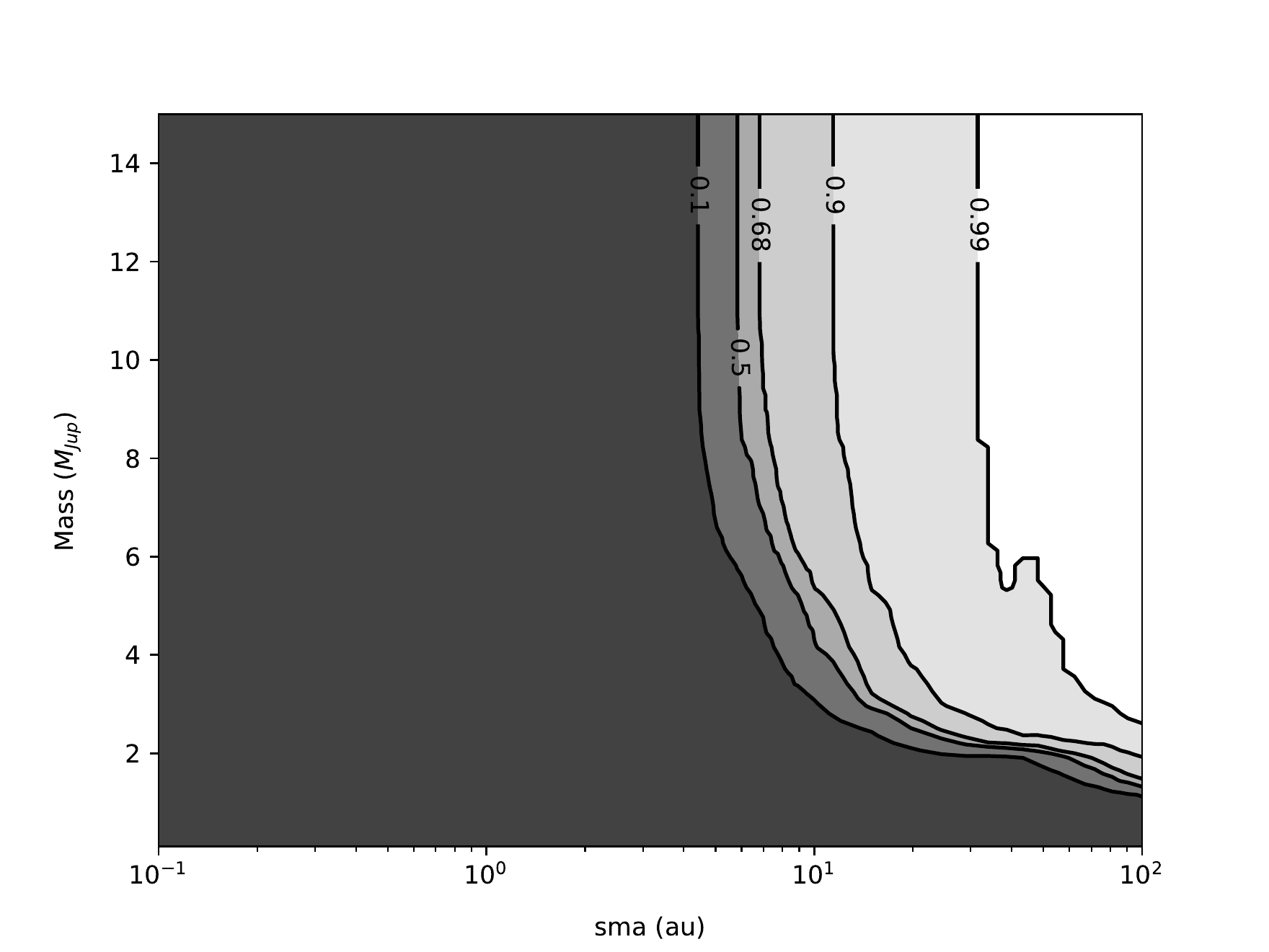}\quad
\includegraphics[width=.45\textwidth]{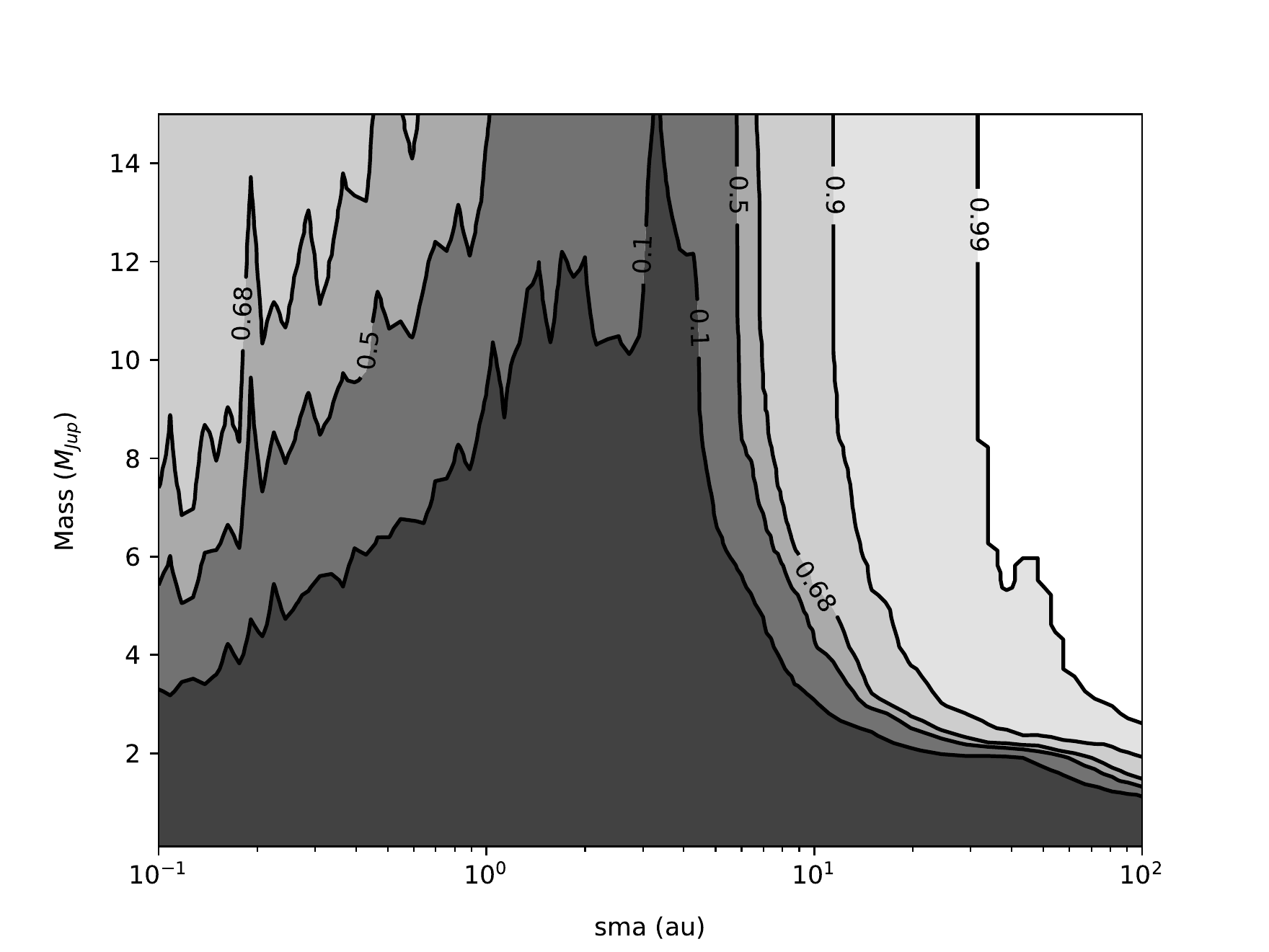}
\caption{Results for HD 43989, DI only (left), DI+RV (right). Two  epochs were available for DI.}
\end{figure}

\begin{figure}[htp]
\centering

\includegraphics[width=.45\textwidth]{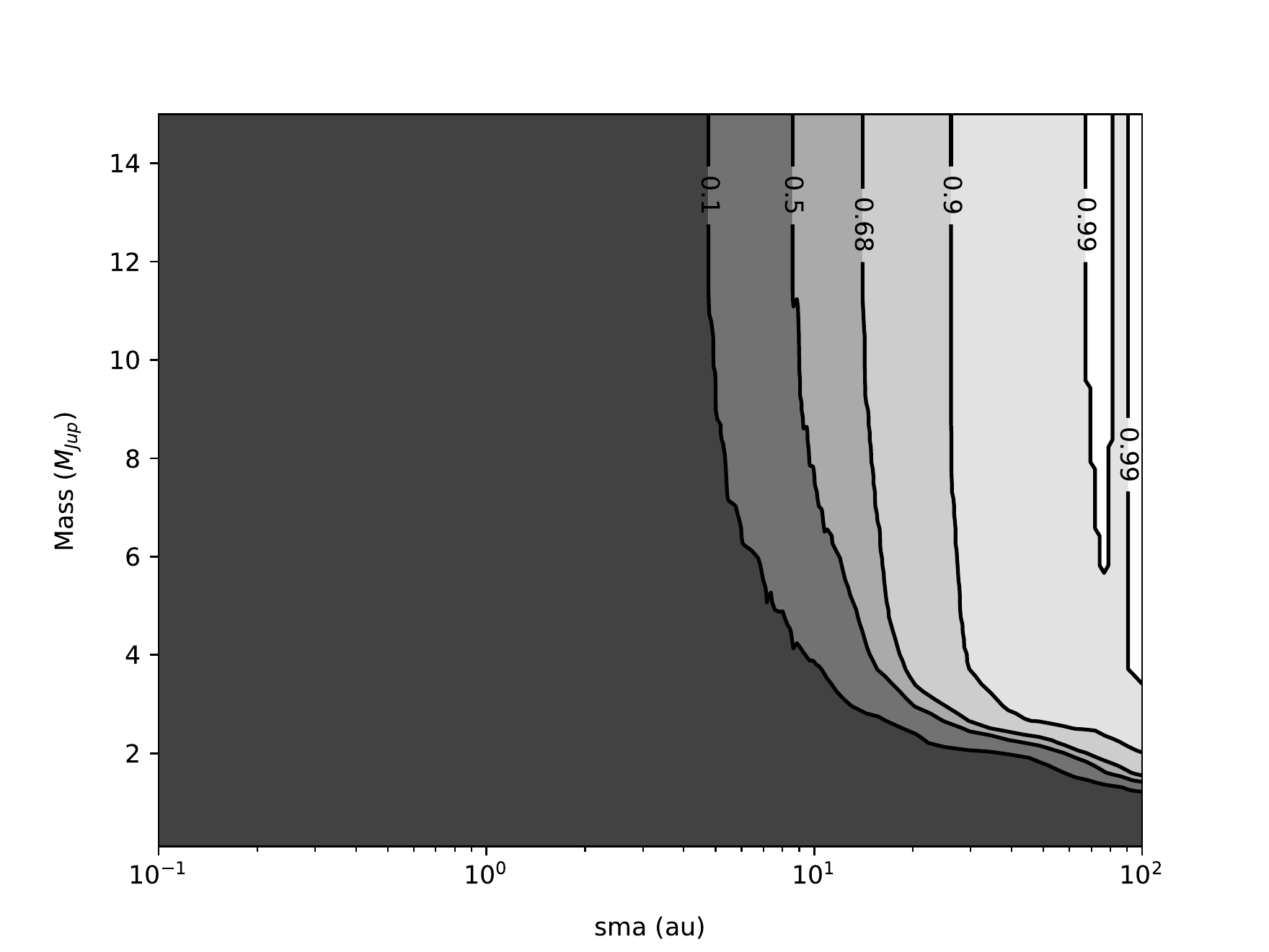}\quad
\includegraphics[width=.45\textwidth]{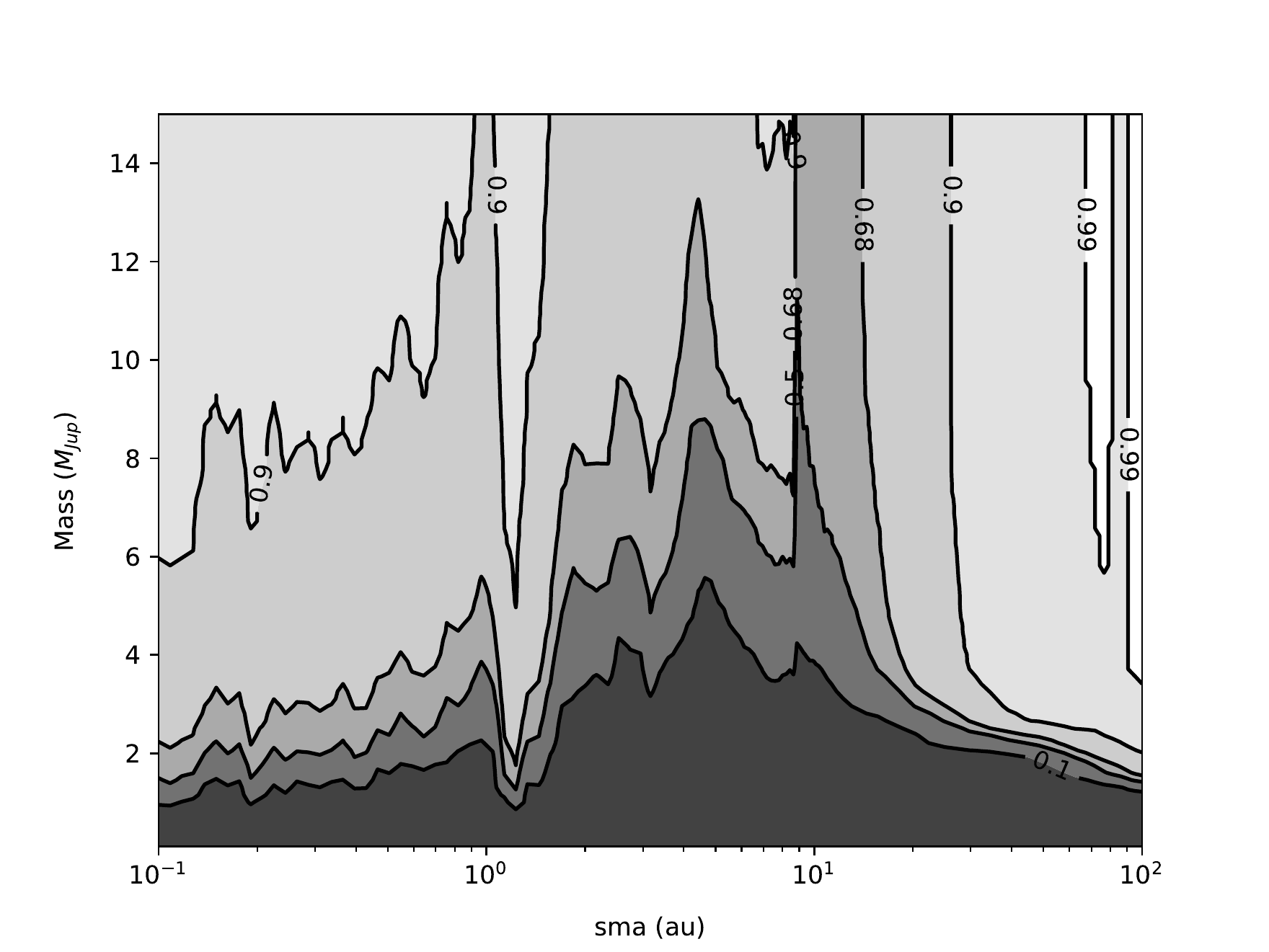}
\caption{Results for HD 44627, DI only (left), DI+RV (right). One epoch was available for DI.}
\end{figure}

\begin{figure}[htp]
\centering

\includegraphics[width=.45\textwidth]{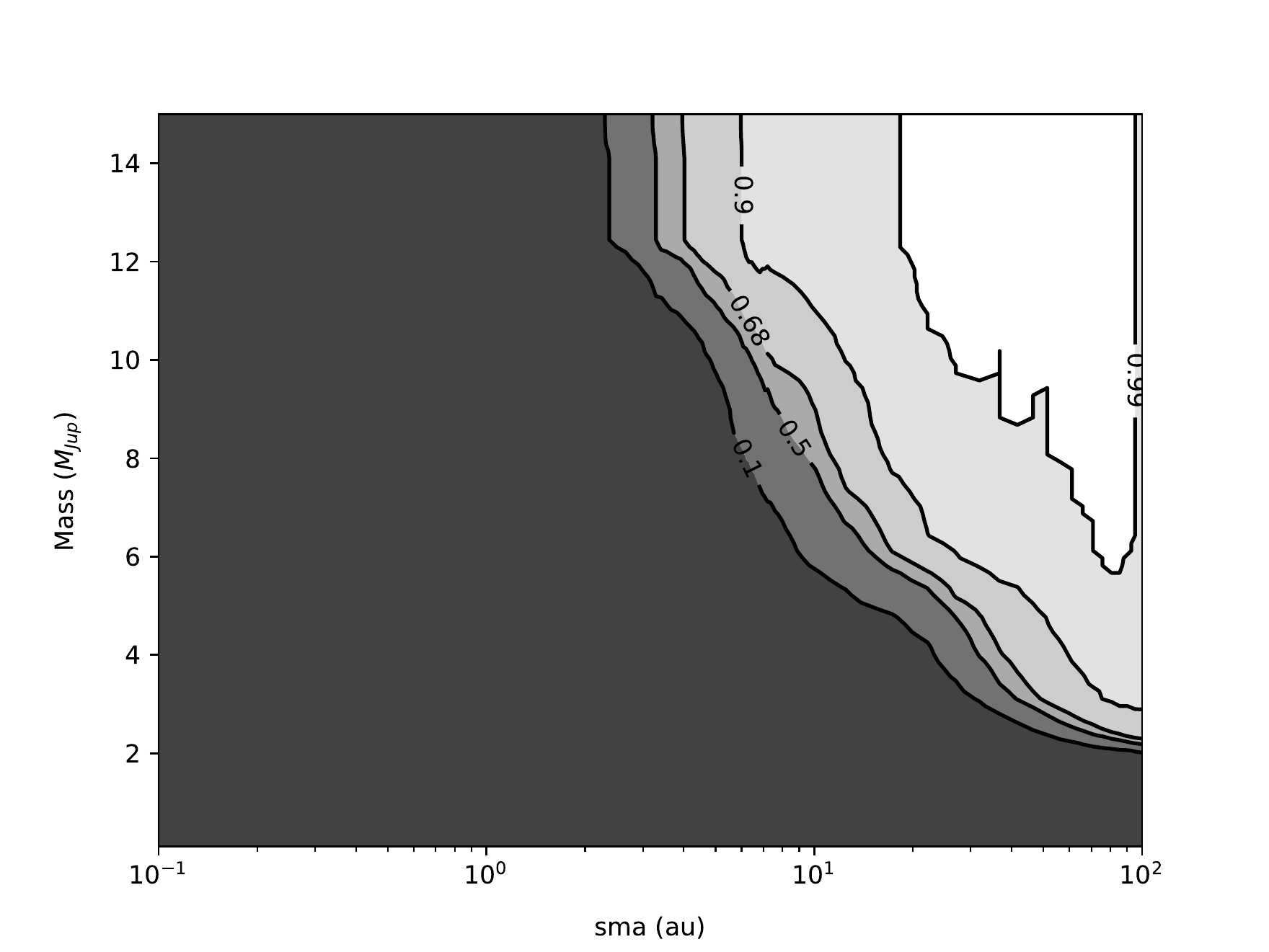}\quad
\includegraphics[width=.45\textwidth]{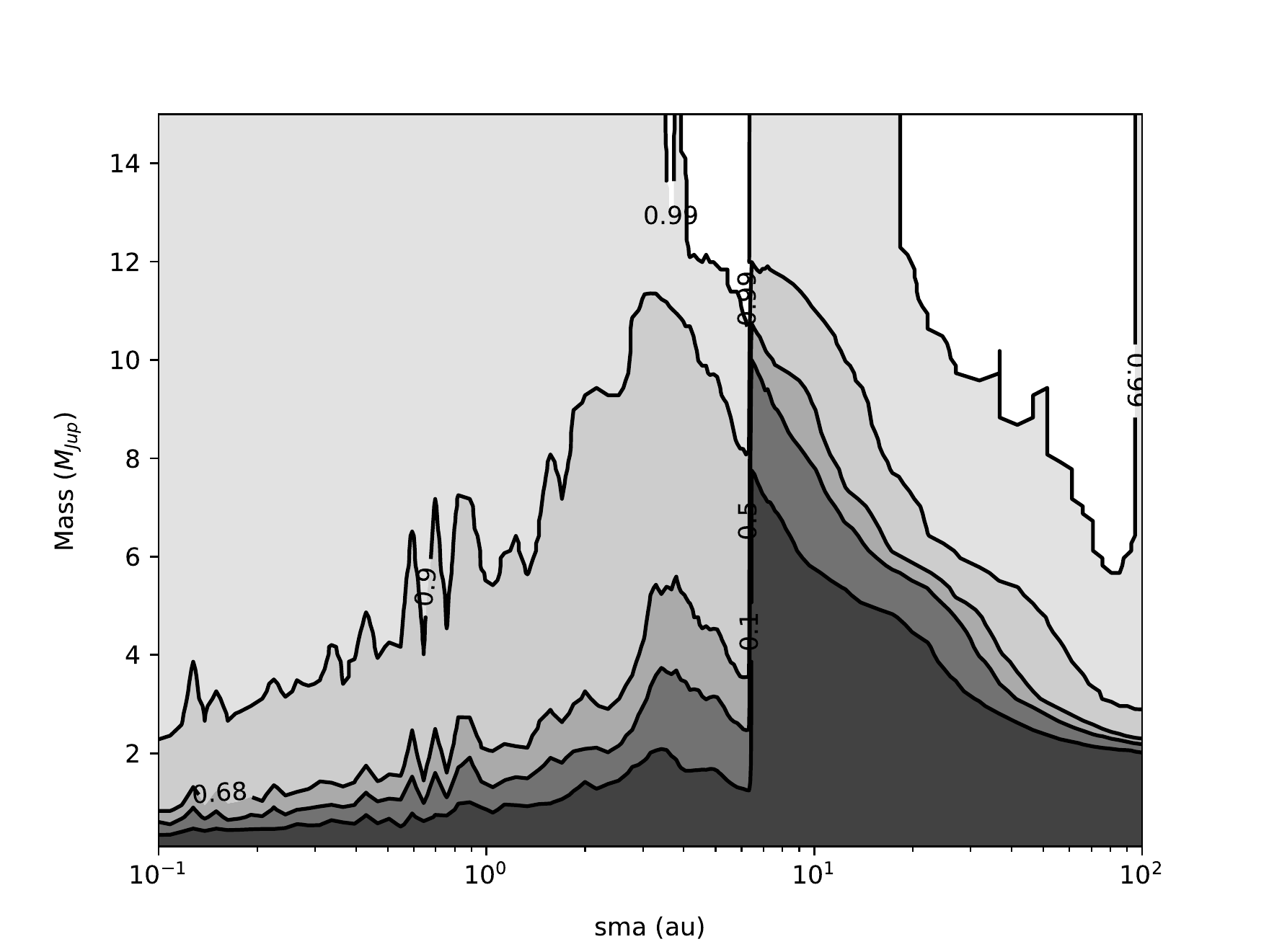}
\caption{Results for HD 45270, DI only (left), DI+RV (right). One epoch was available for DI.}
\end{figure}

\begin{figure}[htp]
\centering

\includegraphics[width=.45\textwidth]{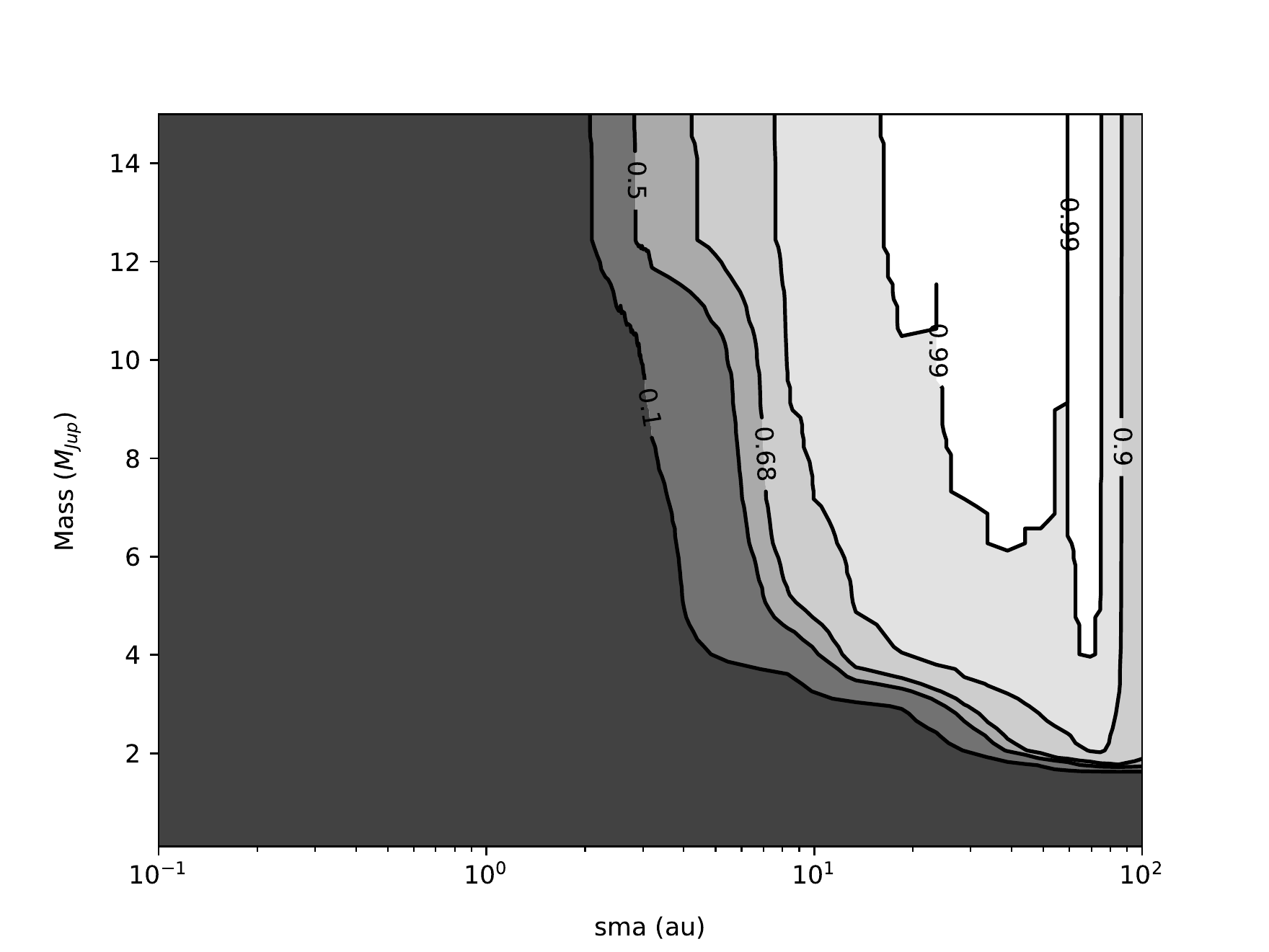}\quad
\includegraphics[width=.45\textwidth]{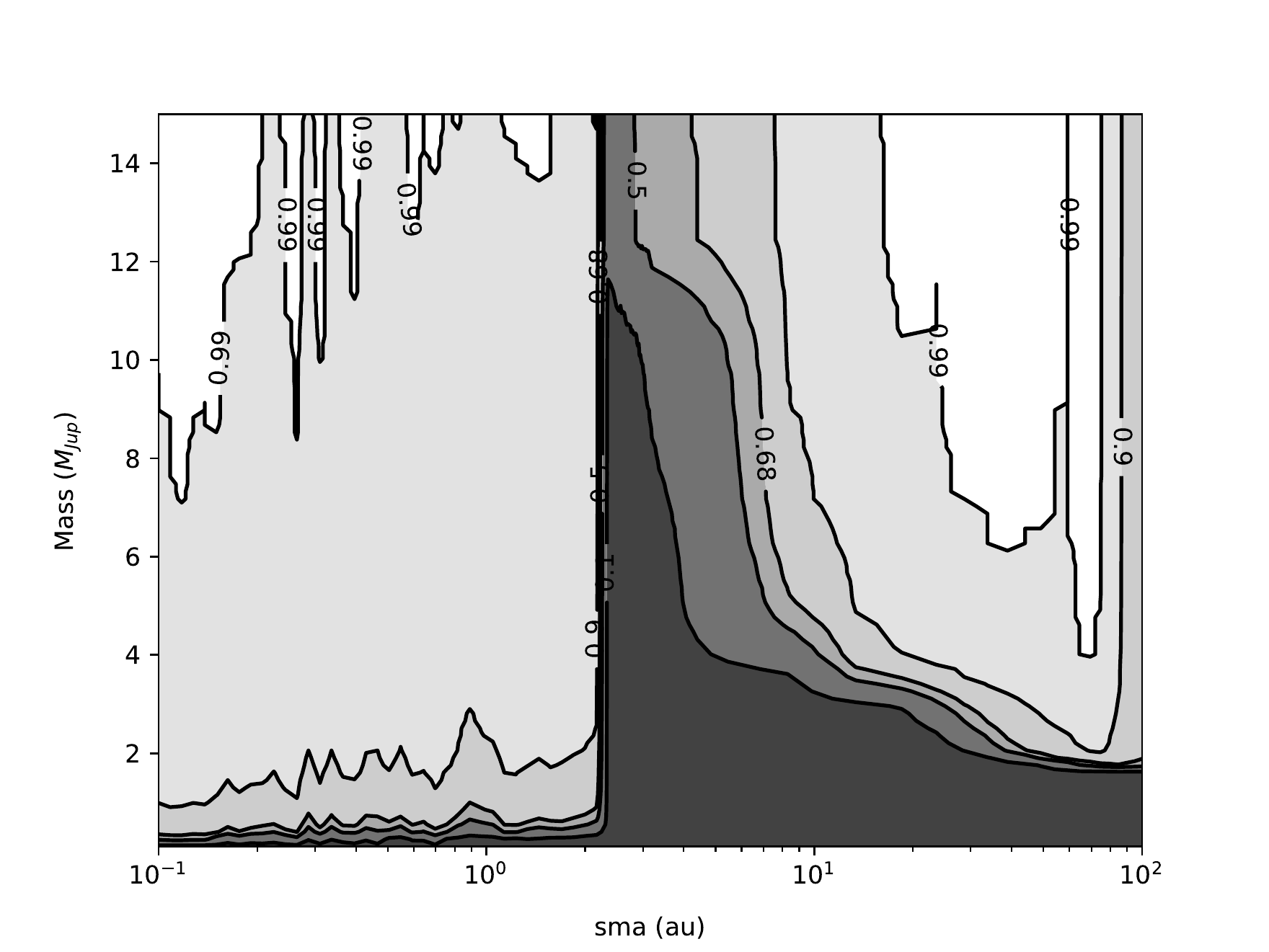}
\caption{Results for CD-61 1439, DI only (left), DI+RV (right). One epoch was available for DI.}
\end{figure}

\begin{figure}[htp]
\centering

\includegraphics[width=.45\textwidth]{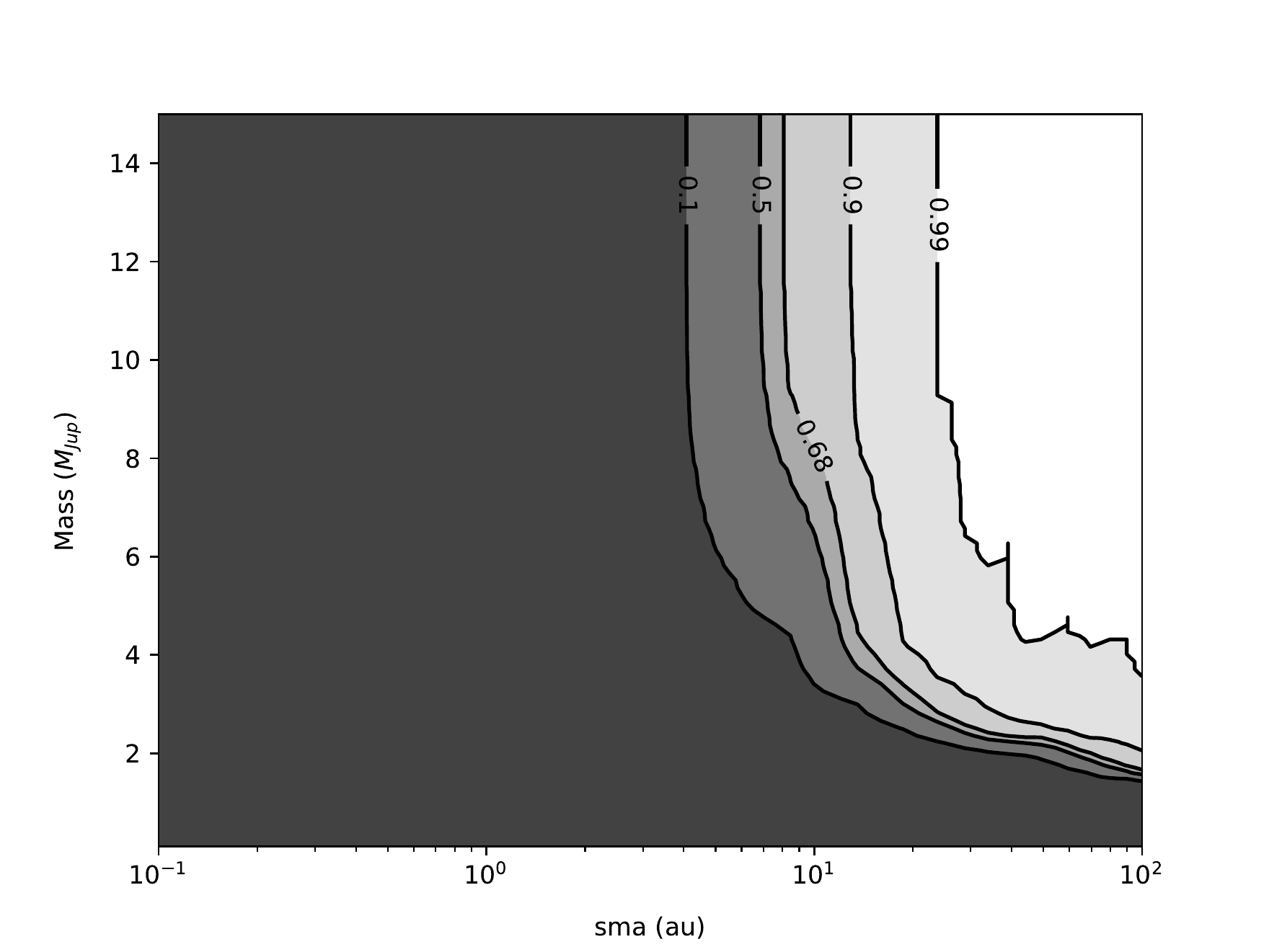}\quad
\includegraphics[width=.45\textwidth]{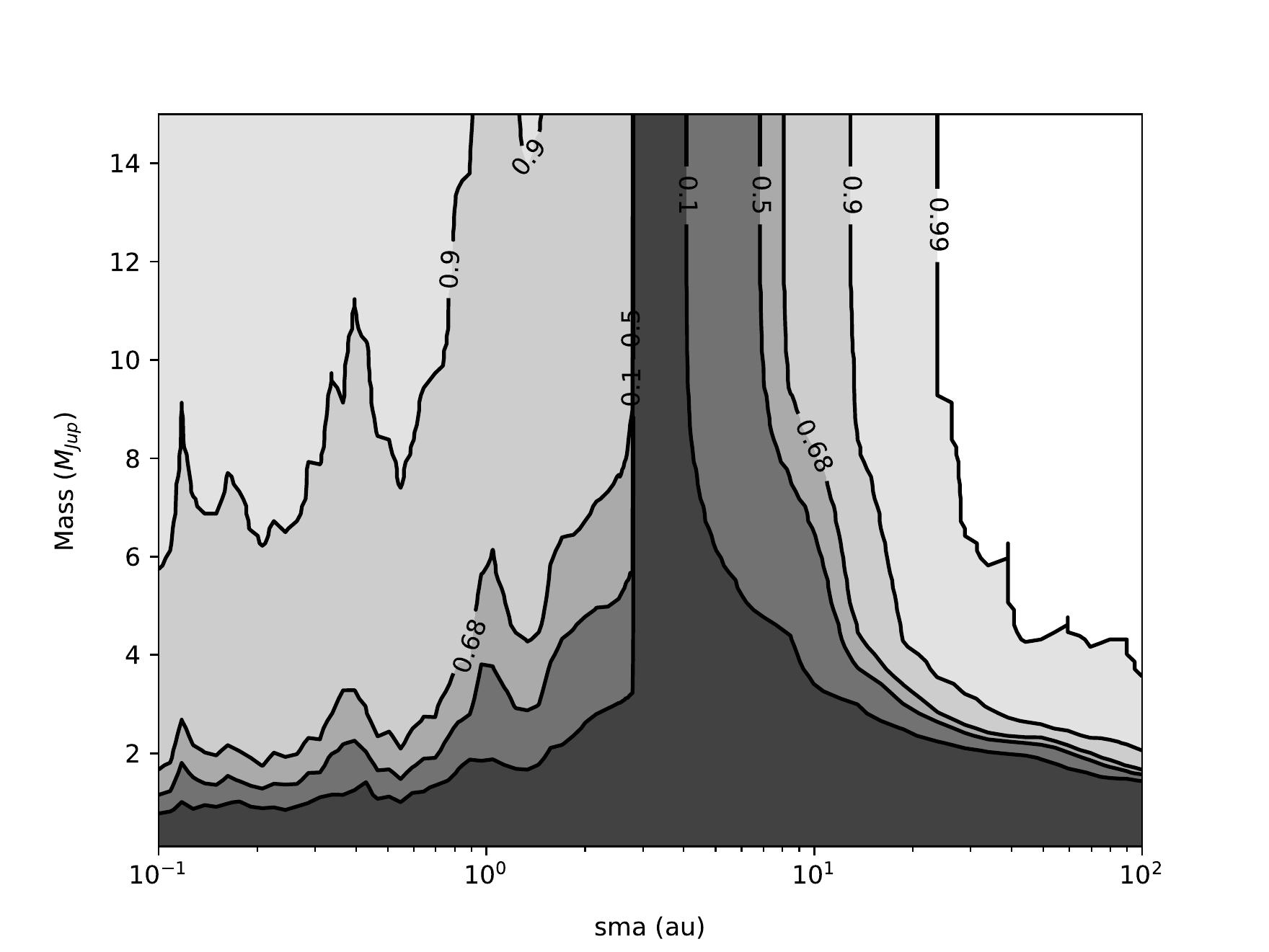}
\caption{Results for HD 49855, DI only (left), DI+RV (right). One epoch was available for DI.}
\end{figure}

\begin{figure}[htp]
\centering

\includegraphics[width=.45\textwidth]{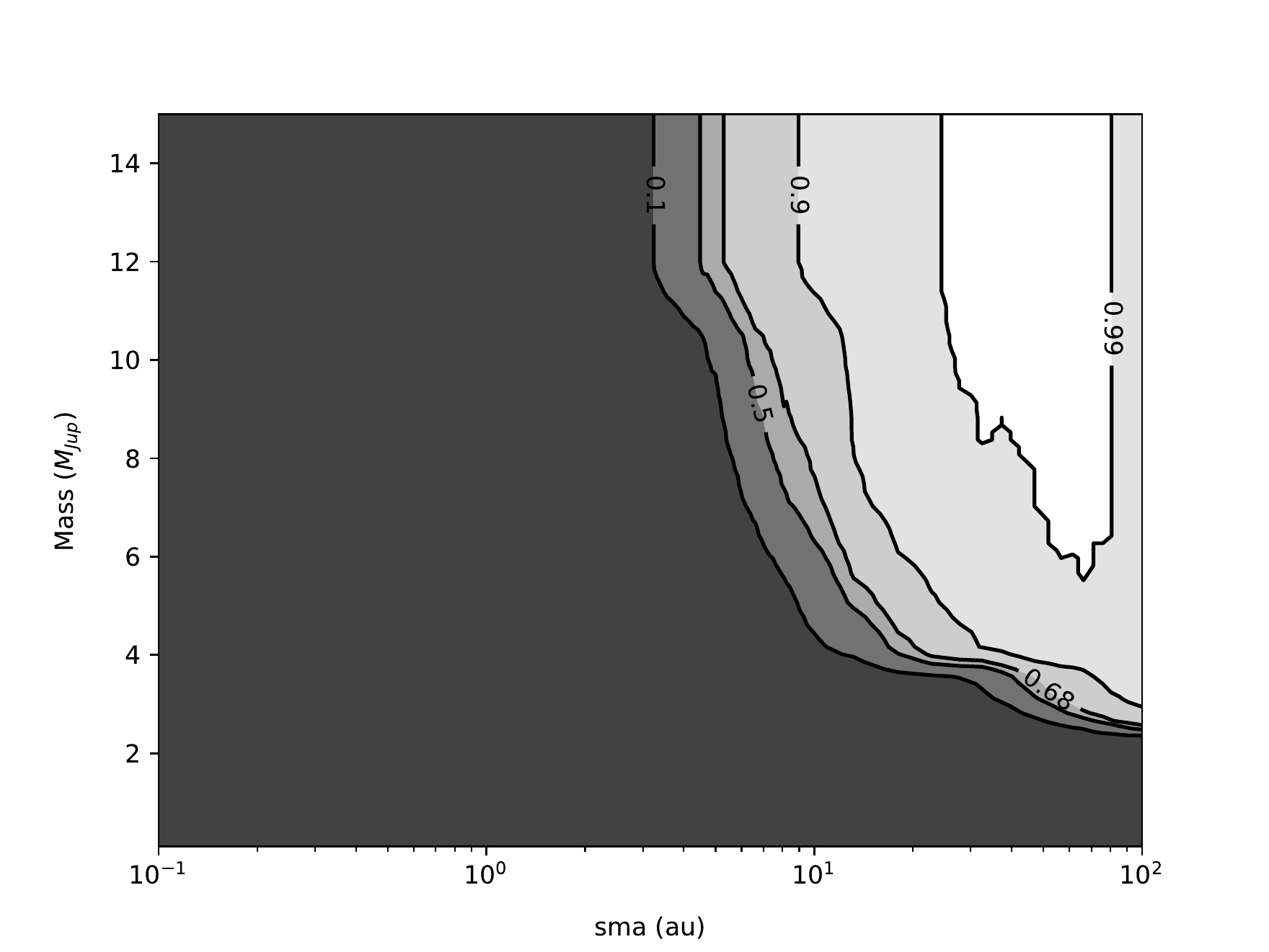}\quad
\includegraphics[width=.45\textwidth]{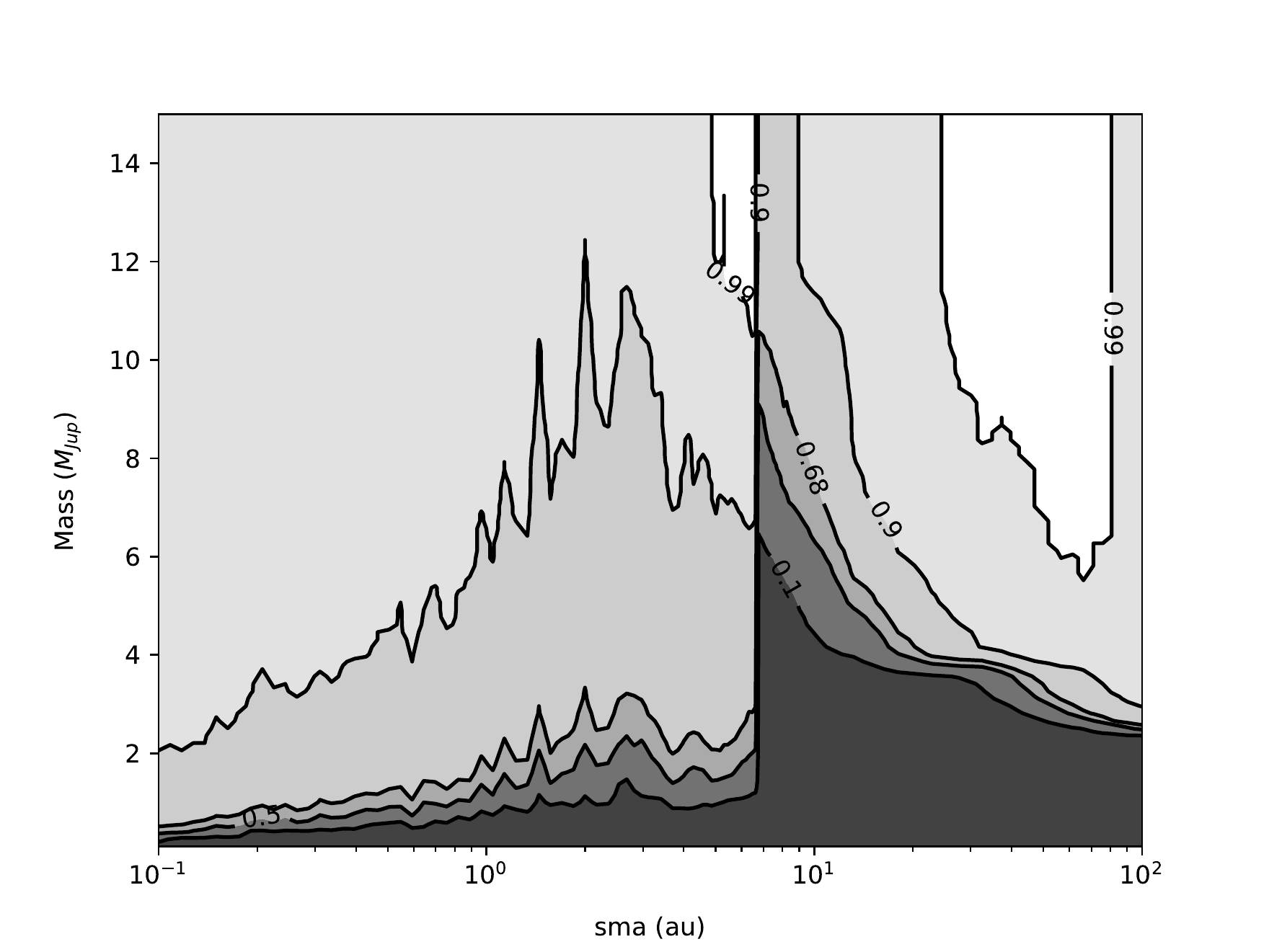}
\caption{Results for HD 61005, DI only (left), DI+RV (right). One epoch was available for DI.}
\end{figure}

\begin{figure}[htp]
\centering

\includegraphics[width=.45\textwidth]{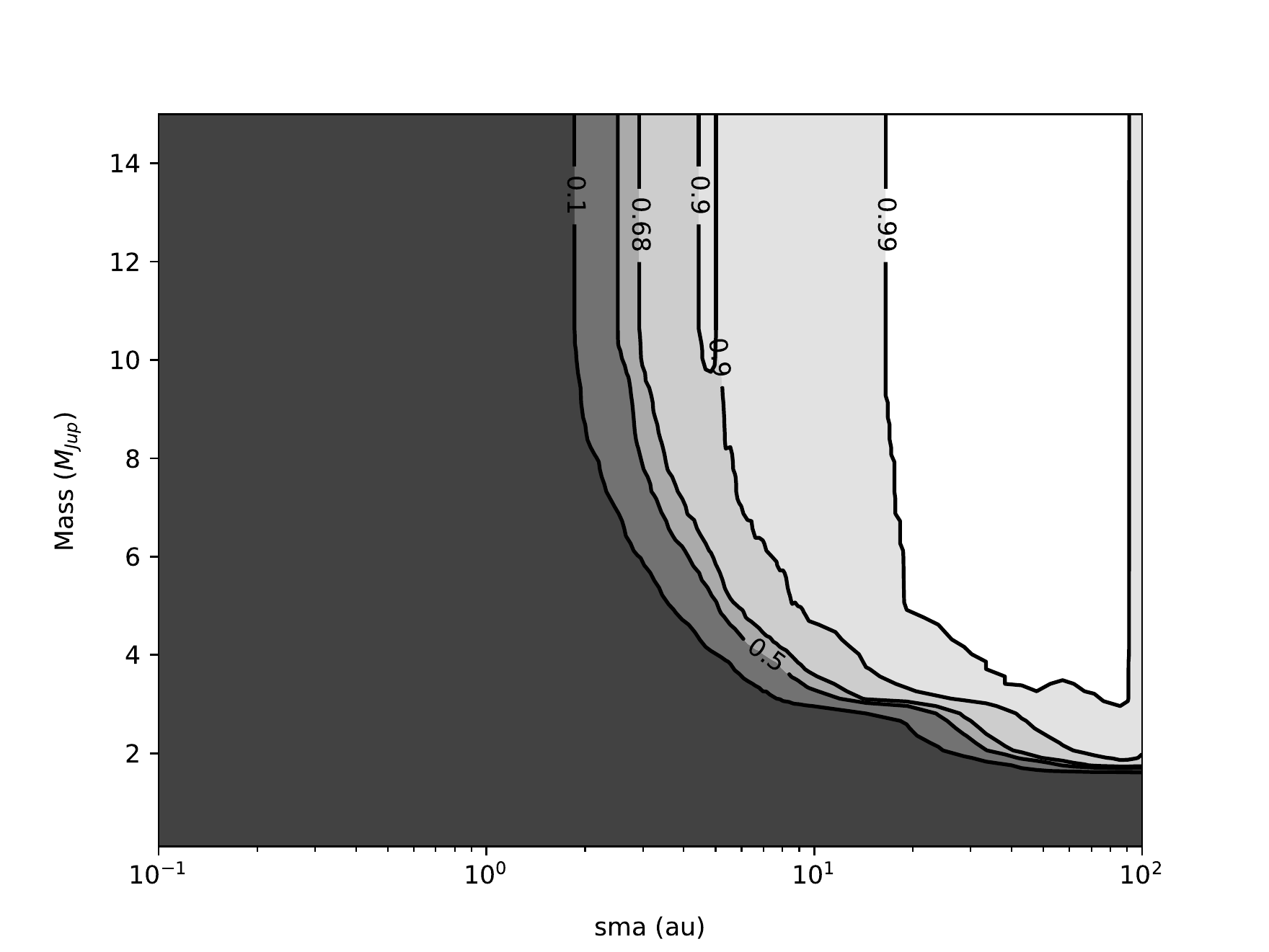}\quad
\includegraphics[width=.45\textwidth]{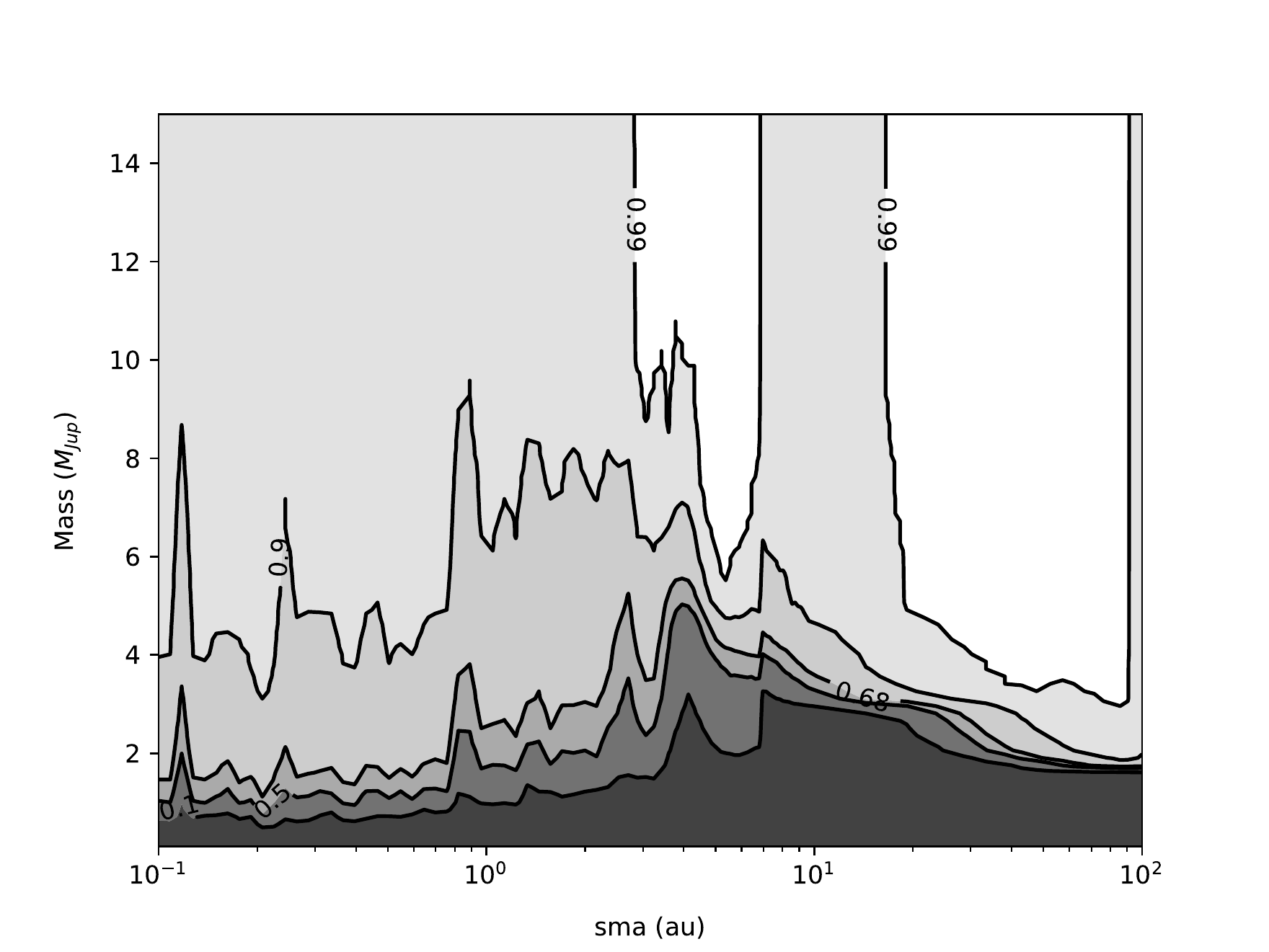}
\caption{Results for HD 118100, DI only (left), DI+RV (right). One epoch was available for DI.}
\end{figure}

\begin{figure}[htp]
\centering

\includegraphics[width=.45\textwidth]{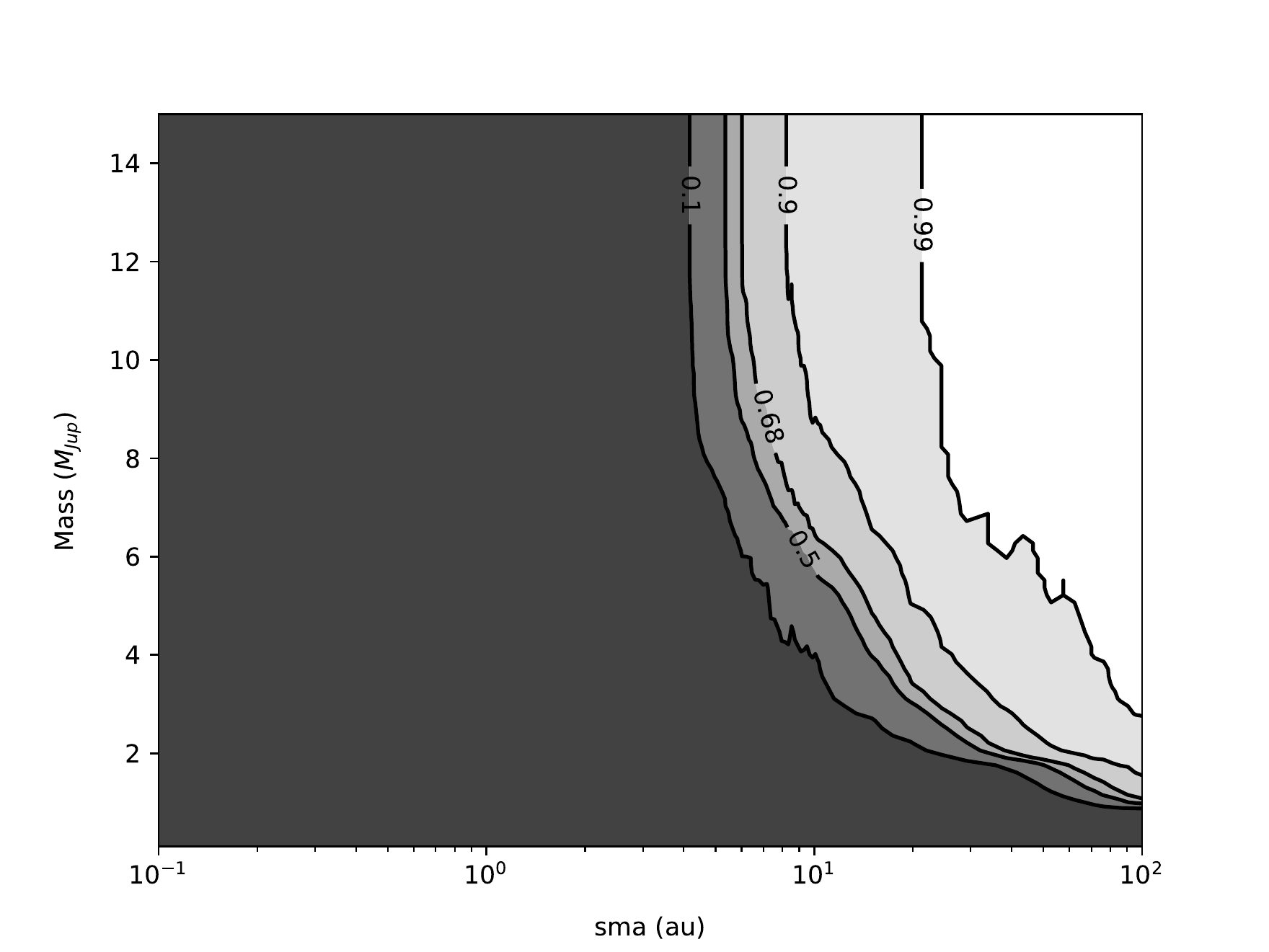}\quad
\includegraphics[width=.45\textwidth]{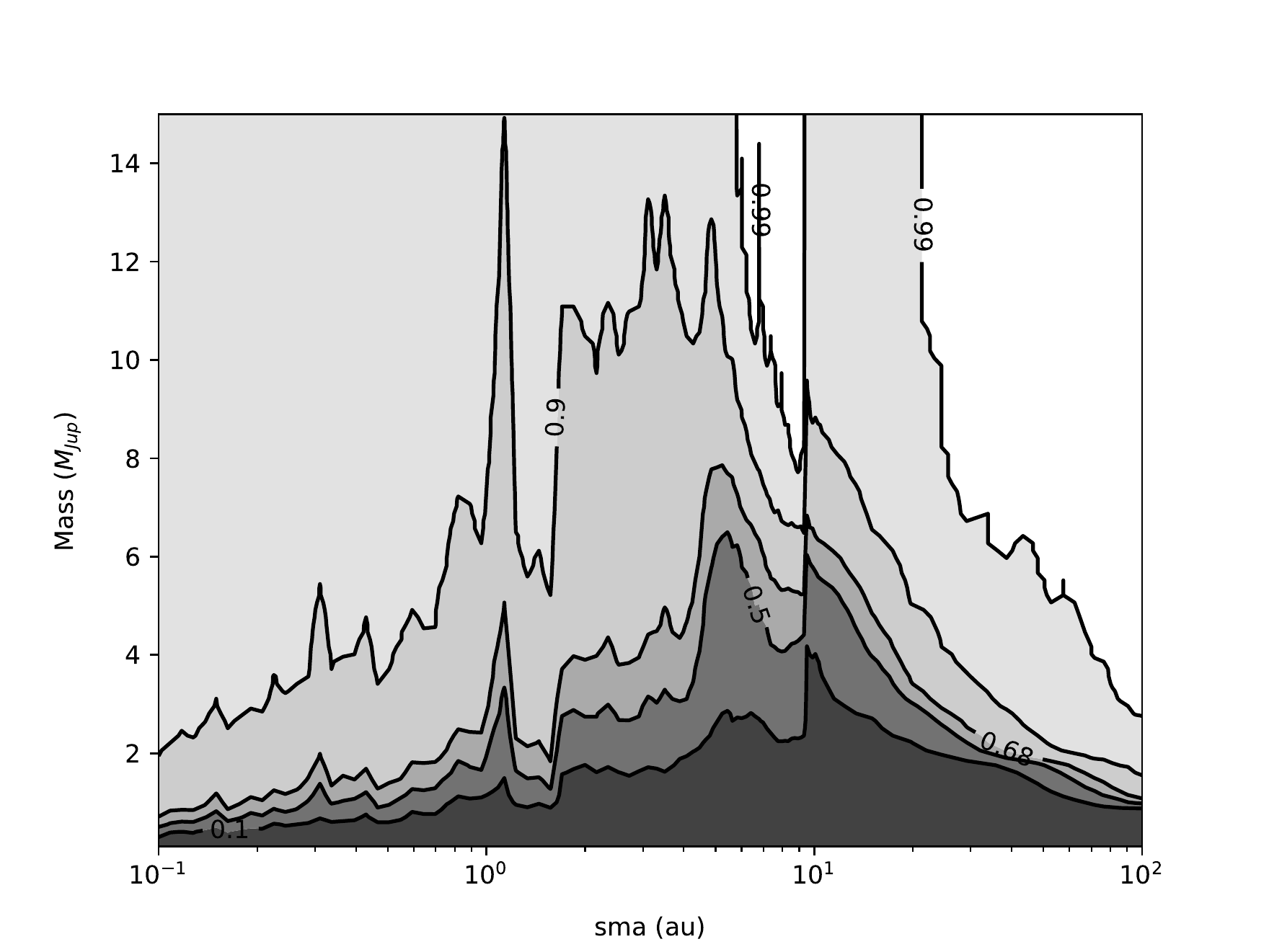}
\caption{Results for HD 1642490, DI only (left), DI+RV (right). Four epochs were available for DI.}
\end{figure}

\begin{figure}[htp]
\centering

\includegraphics[width=.45\textwidth]{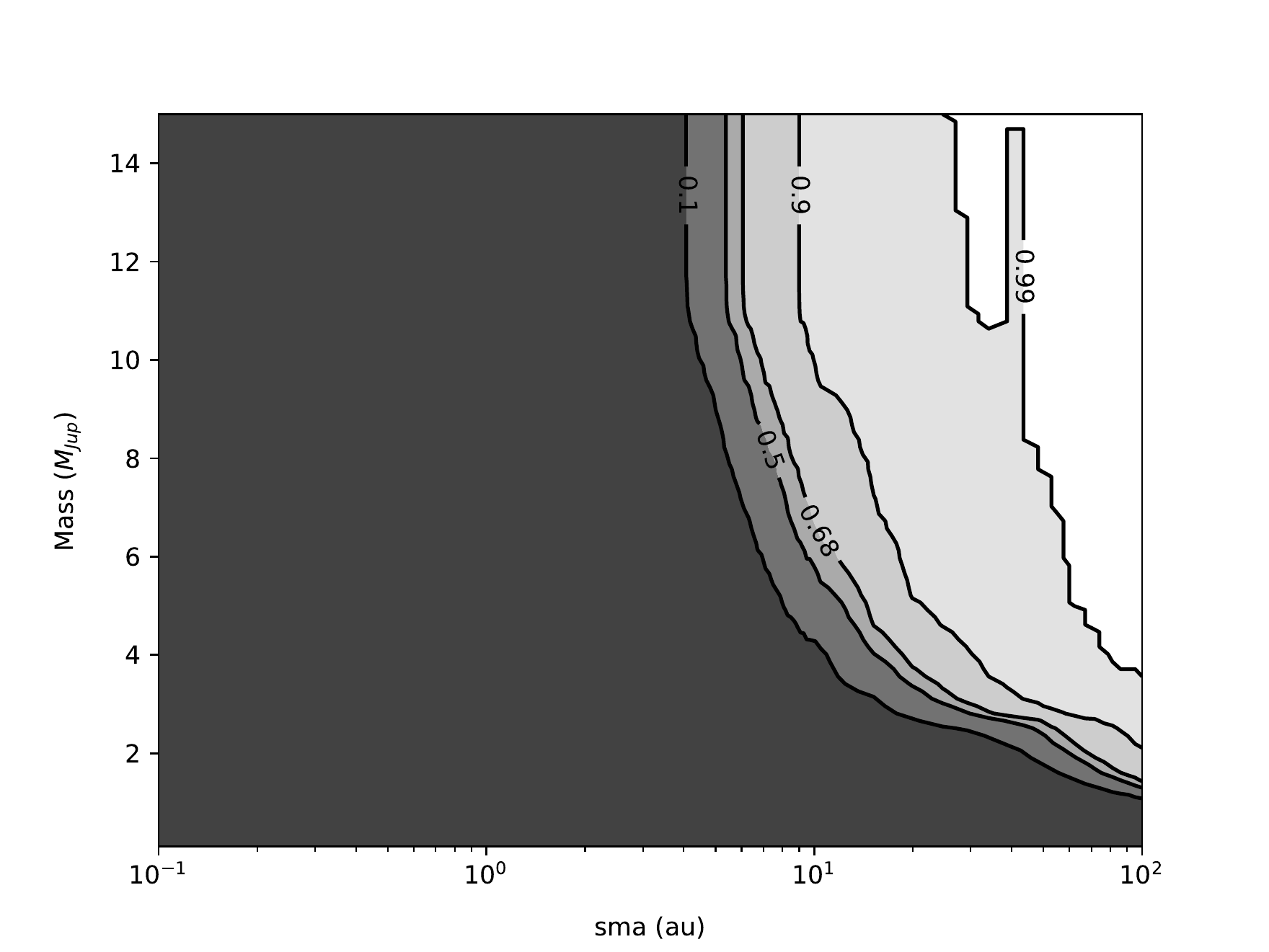}\quad
\includegraphics[width=.45\textwidth]{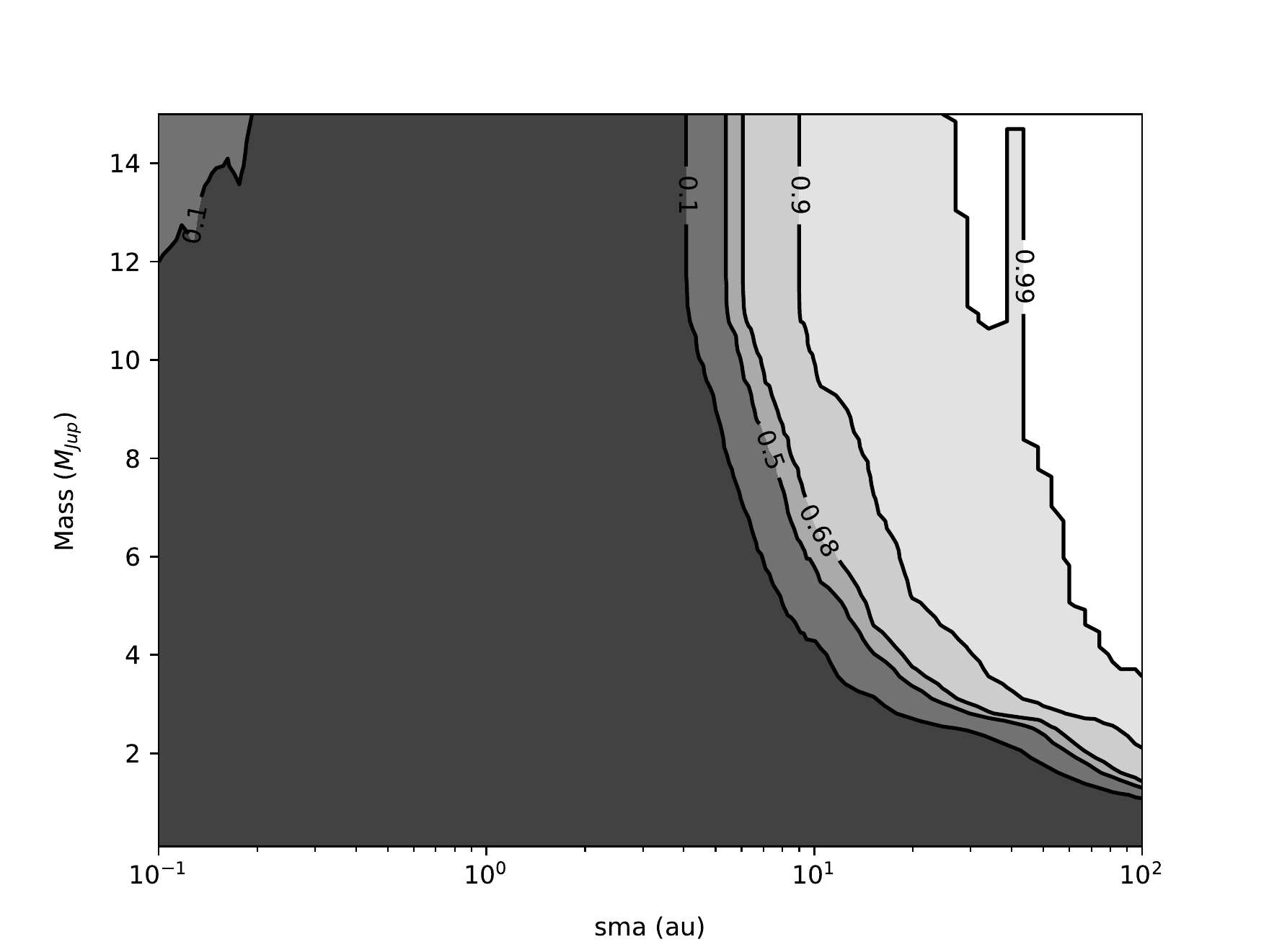}
\caption{Results for HD 174429, DI only (left), DI+RV (right). Five epochs were available for DI.}
\end{figure}

\begin{figure}[htp]
\centering

\includegraphics[width=.45\textwidth]{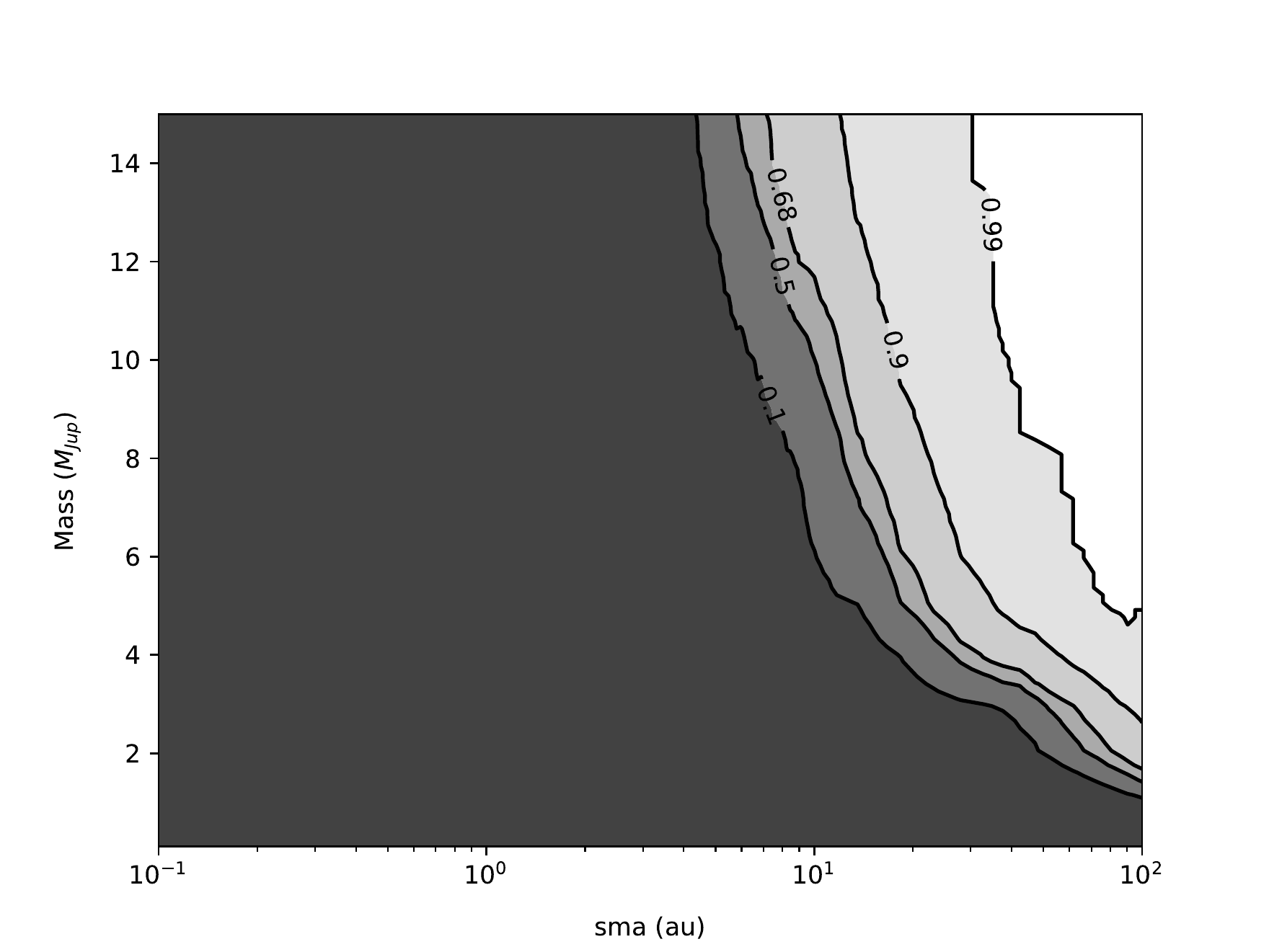}\quad
\includegraphics[width=.45\textwidth]{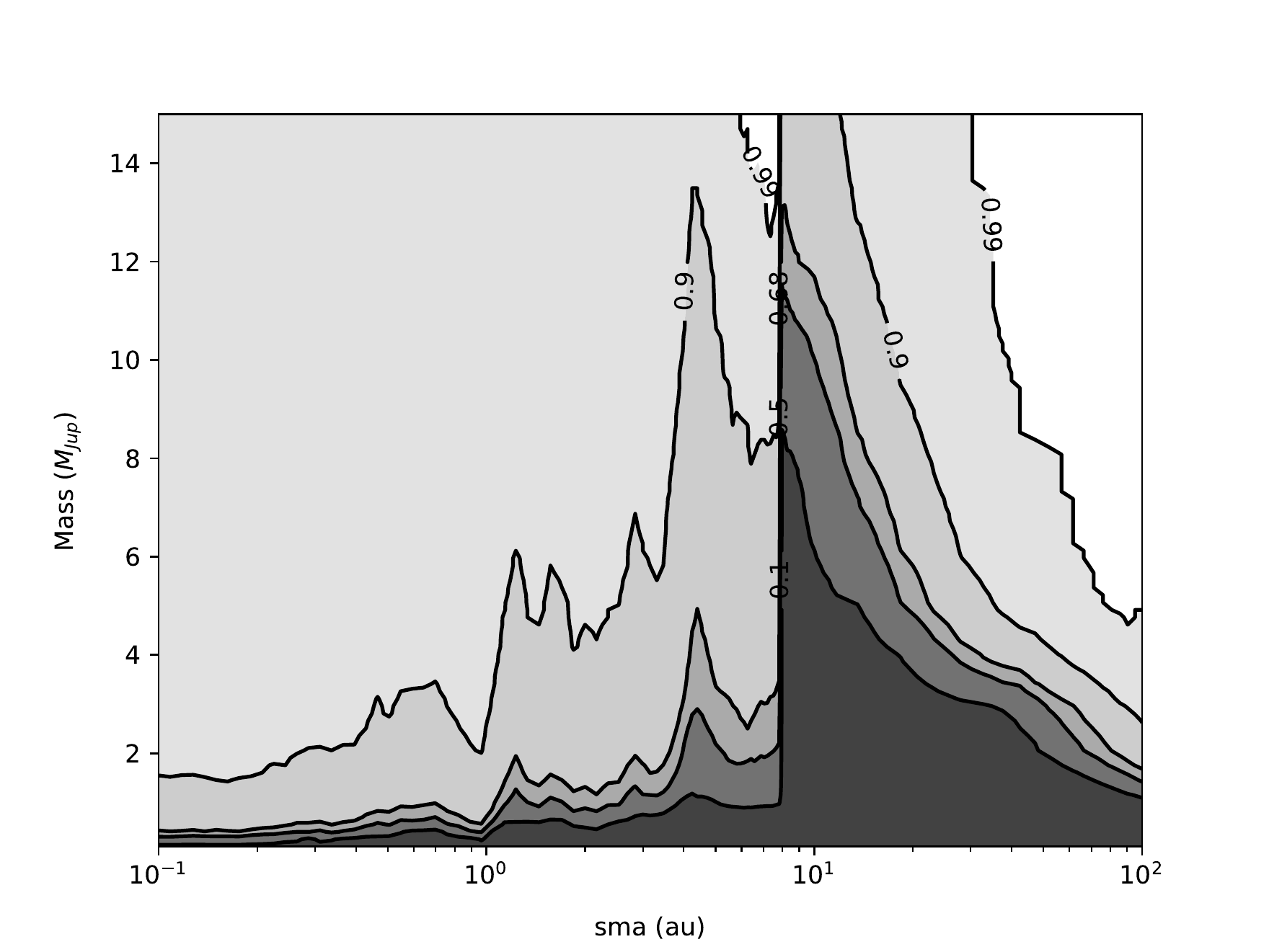}
\caption{Results for HD 181327, DI only (left), DI+RV (right). One epoch was available for DI.}
\end{figure}

\begin{figure}[htp]
\centering

\includegraphics[width=.45\textwidth]{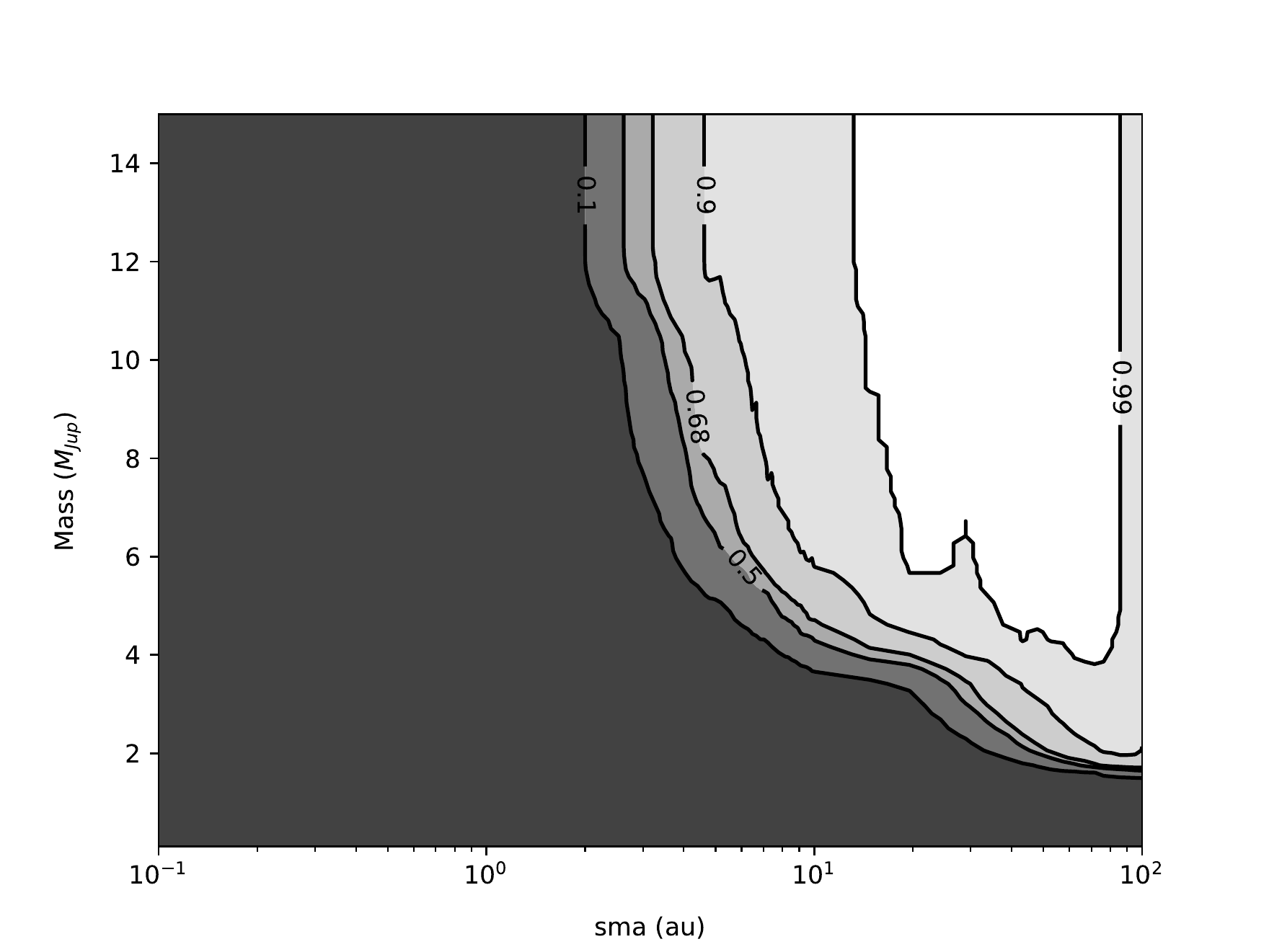}\quad
\includegraphics[width=.45\textwidth]{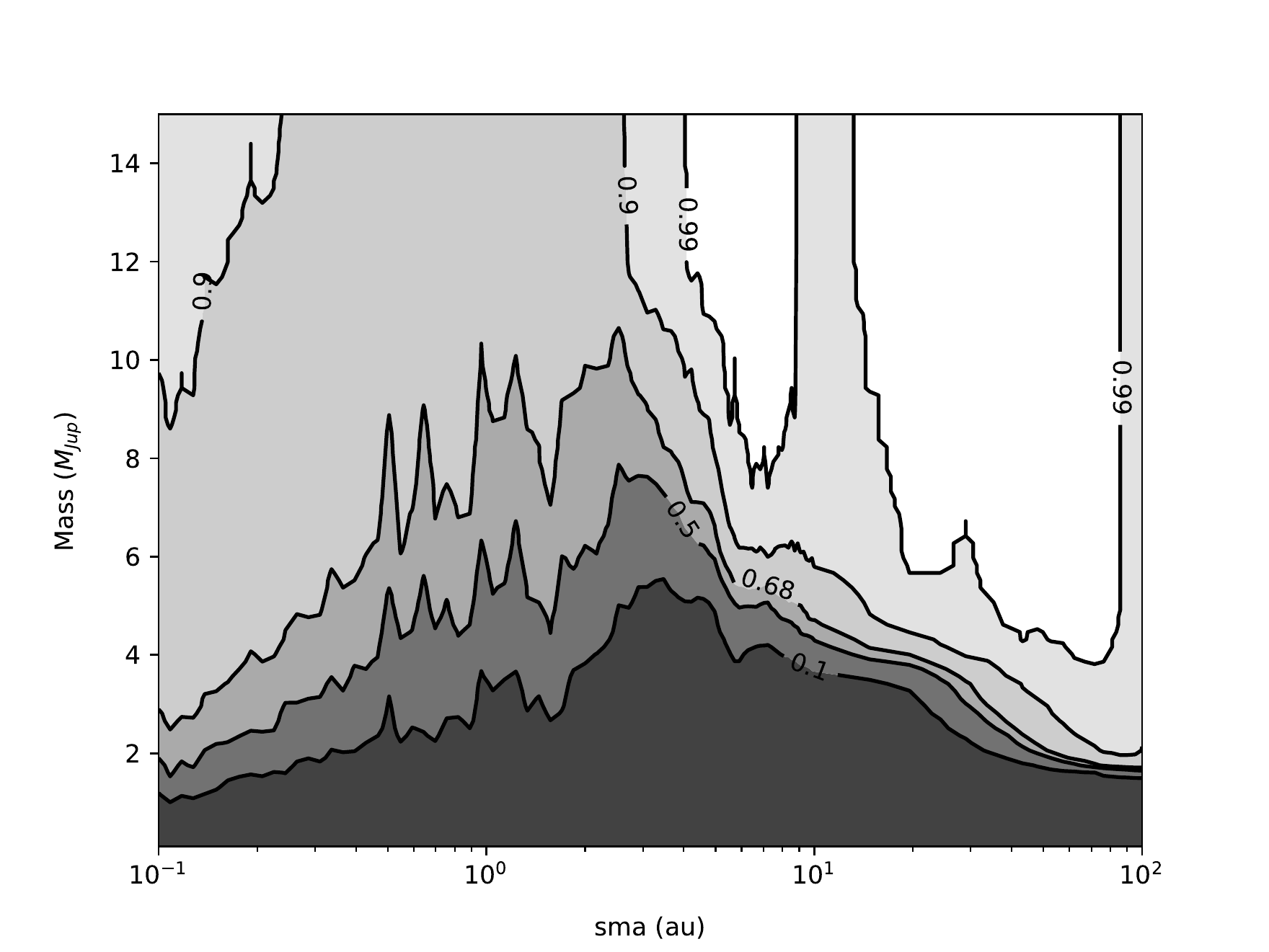}
\caption{Results for HD 189245, DI only (left), DI+RV (right). One epoch was available for DI.}
\end{figure}

\begin{figure}[htp]
\centering

\includegraphics[width=.45\textwidth]{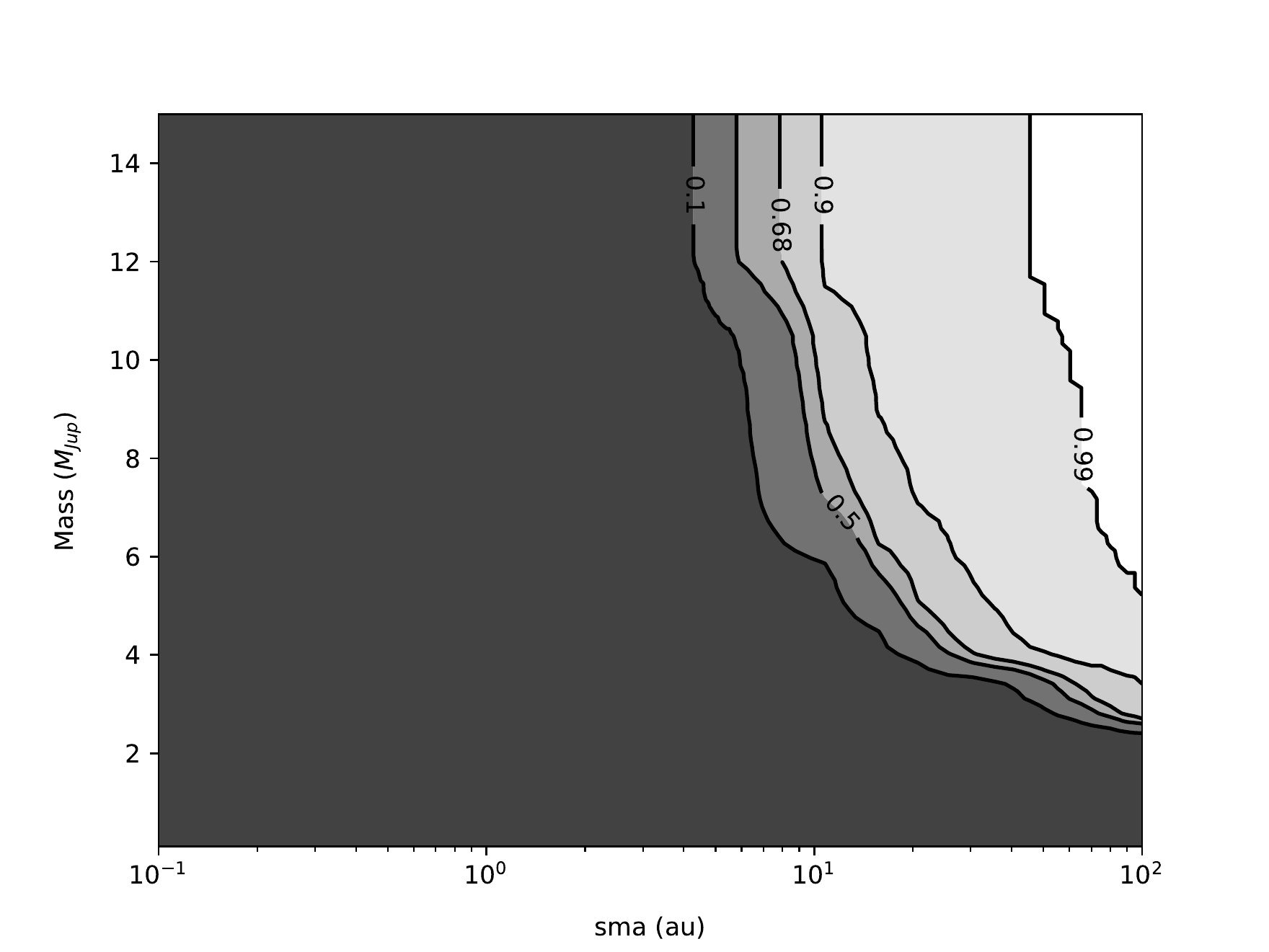}\quad
\includegraphics[width=.45\textwidth]{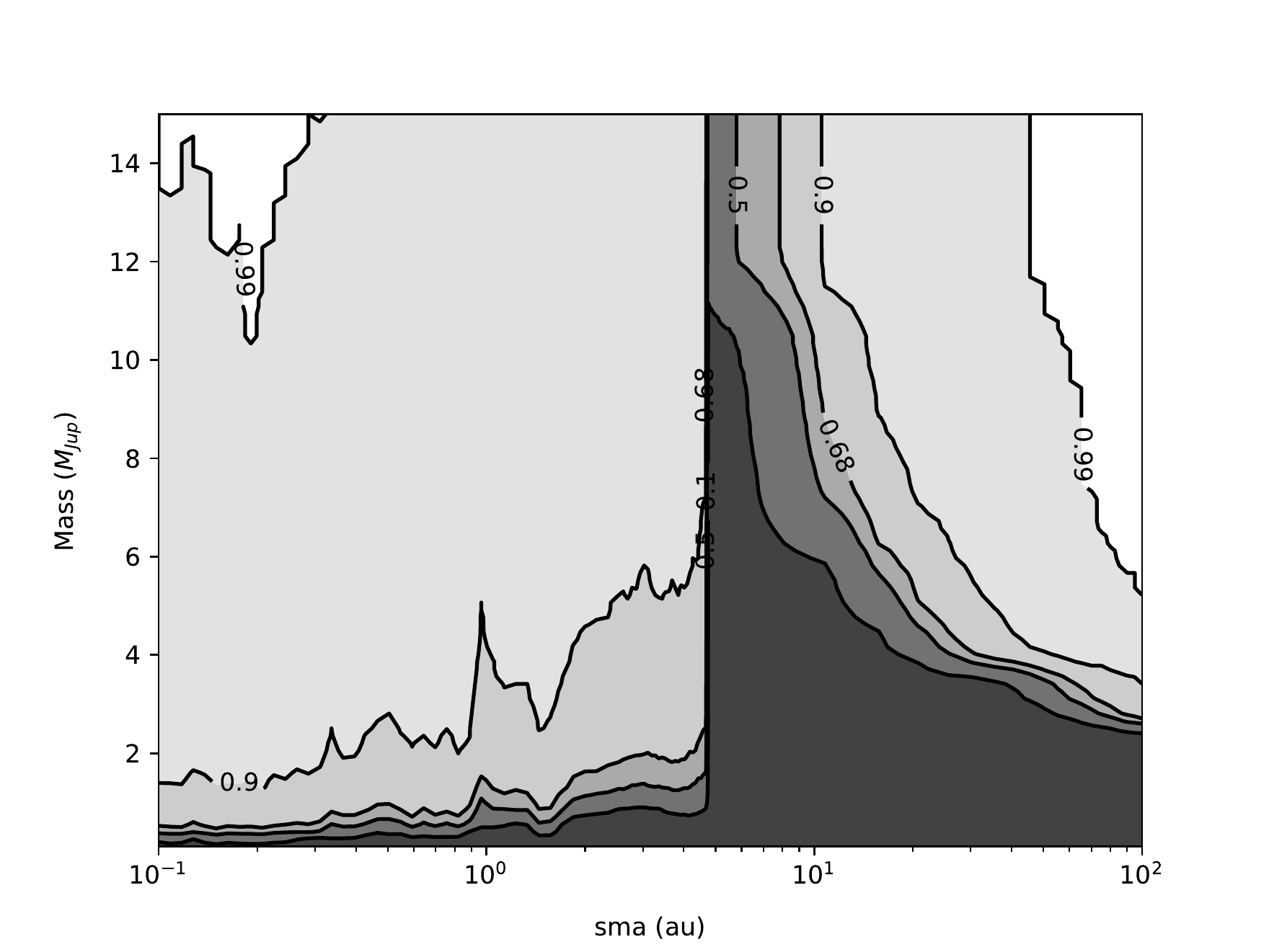}
\caption{Results for HD 218860, DI only (left), DI+RV (right). One epoch was available for DI.}
\end{figure}

\begin{figure}[htp]
\centering

\includegraphics[width=.45\textwidth]{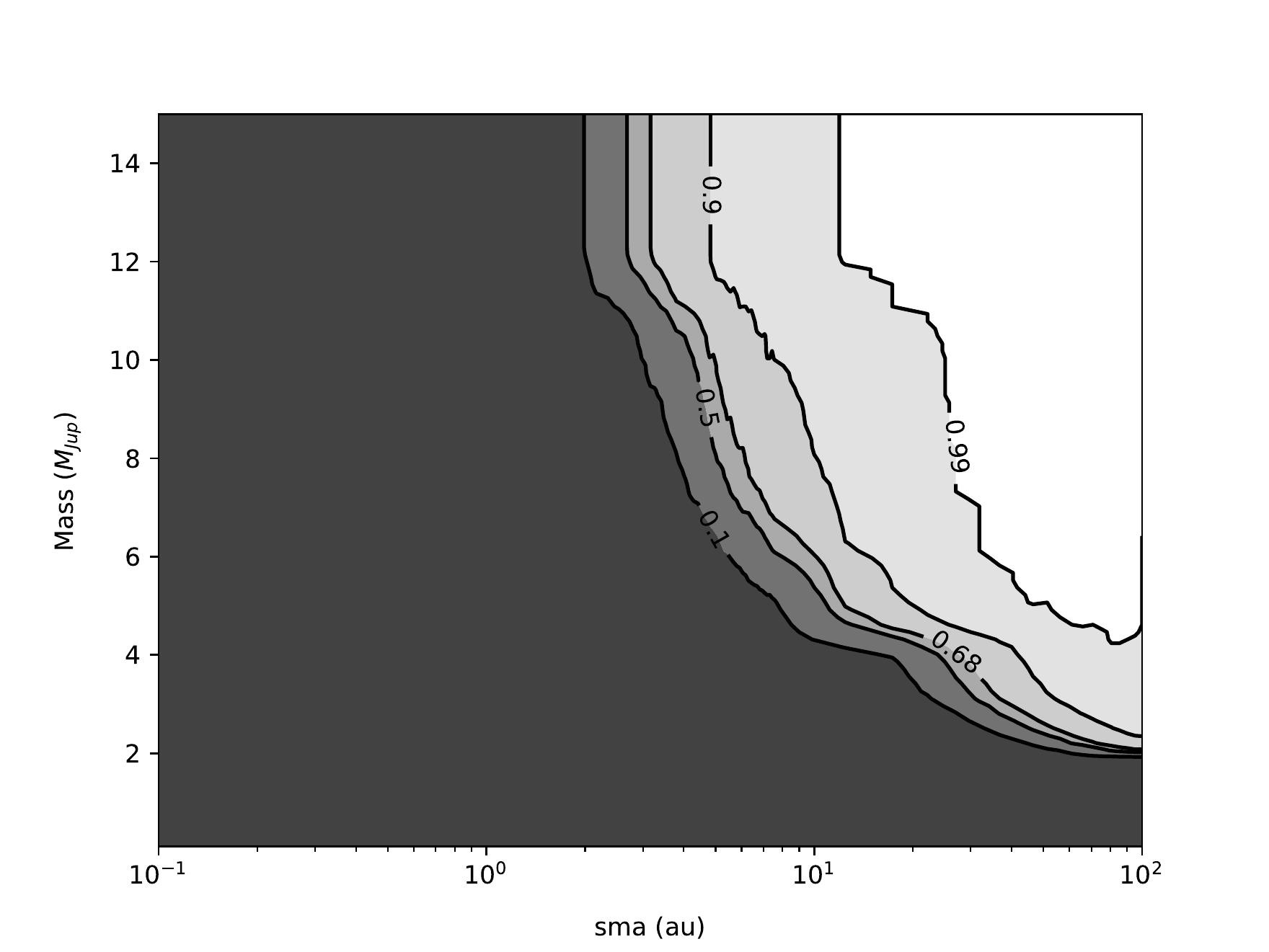}\quad
\includegraphics[width=.45\textwidth]{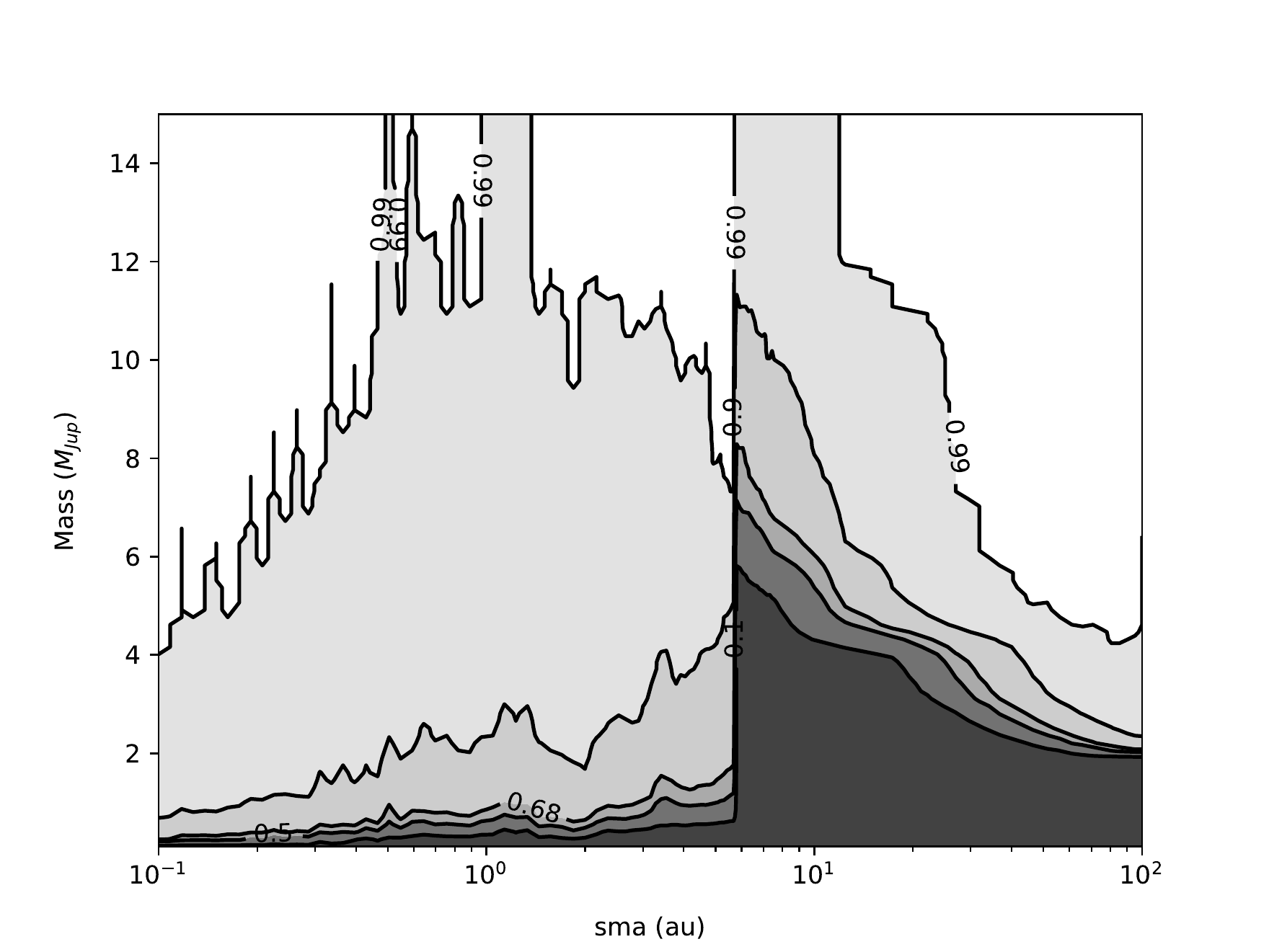}
\caption{Results for HD 224228, DI only (left), DI+RV (right). One epoch was available for DI.}
\end{figure}

\begin{figure}[htp]
\centering

\includegraphics[width=.45\textwidth]{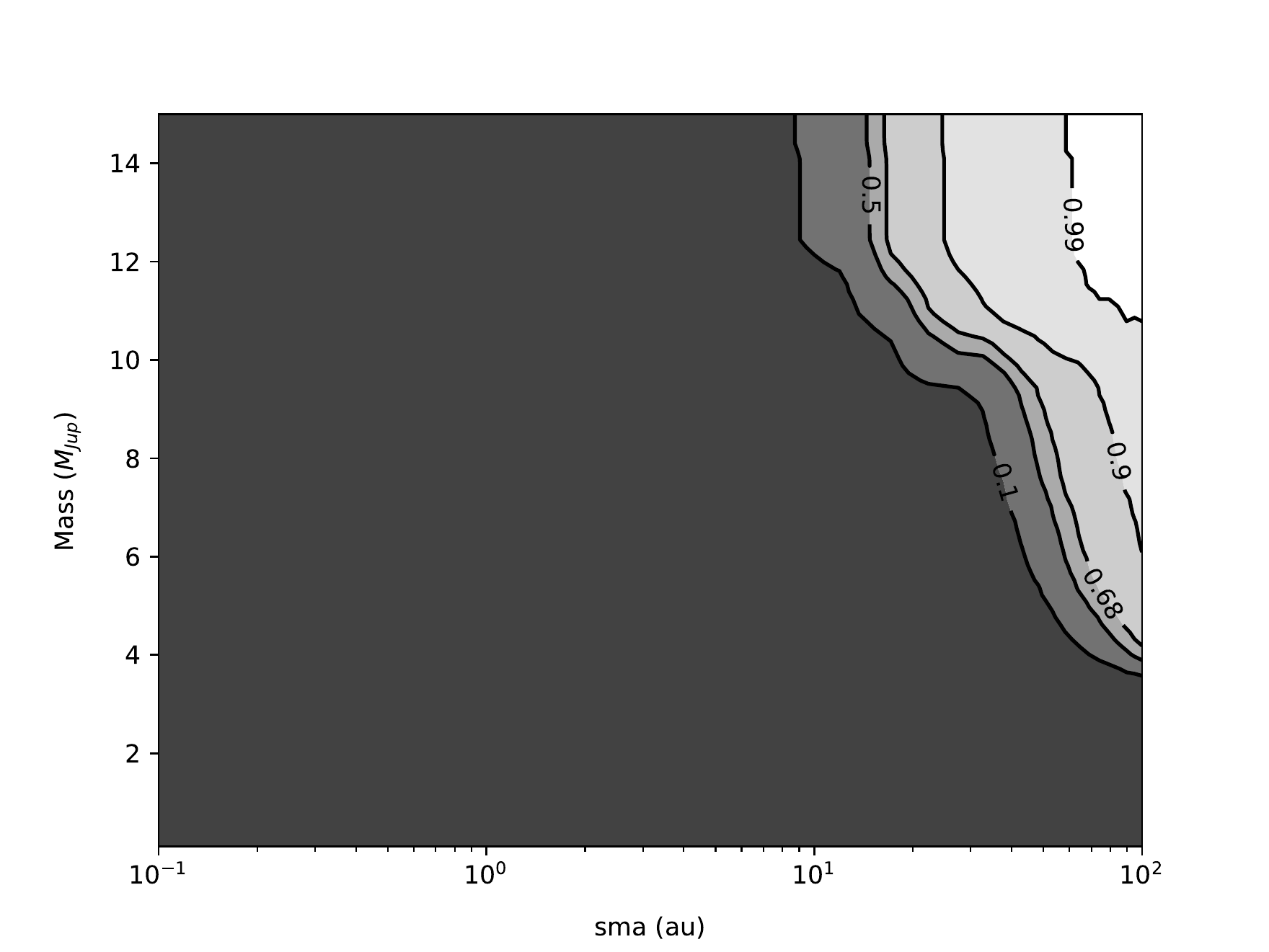}\quad
\includegraphics[width=.45\textwidth]{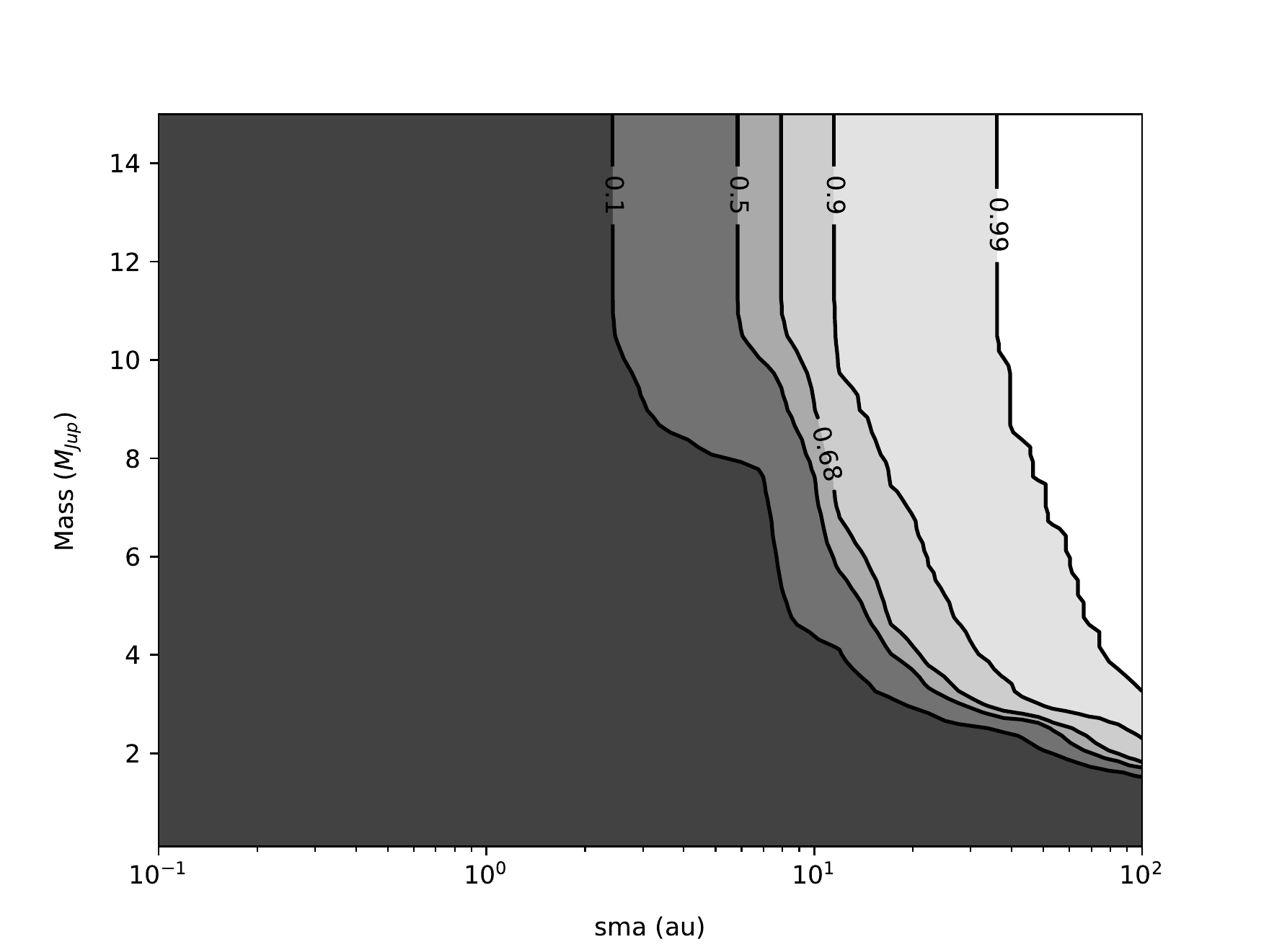}
\caption{Results for HD 90712 (left) and 
HD 197890 (right). No radial velocity data were available. One epoch was available for DI.}
\end{figure}

\begin{figure}[htp]
\centering

\includegraphics[width=.45\textwidth]{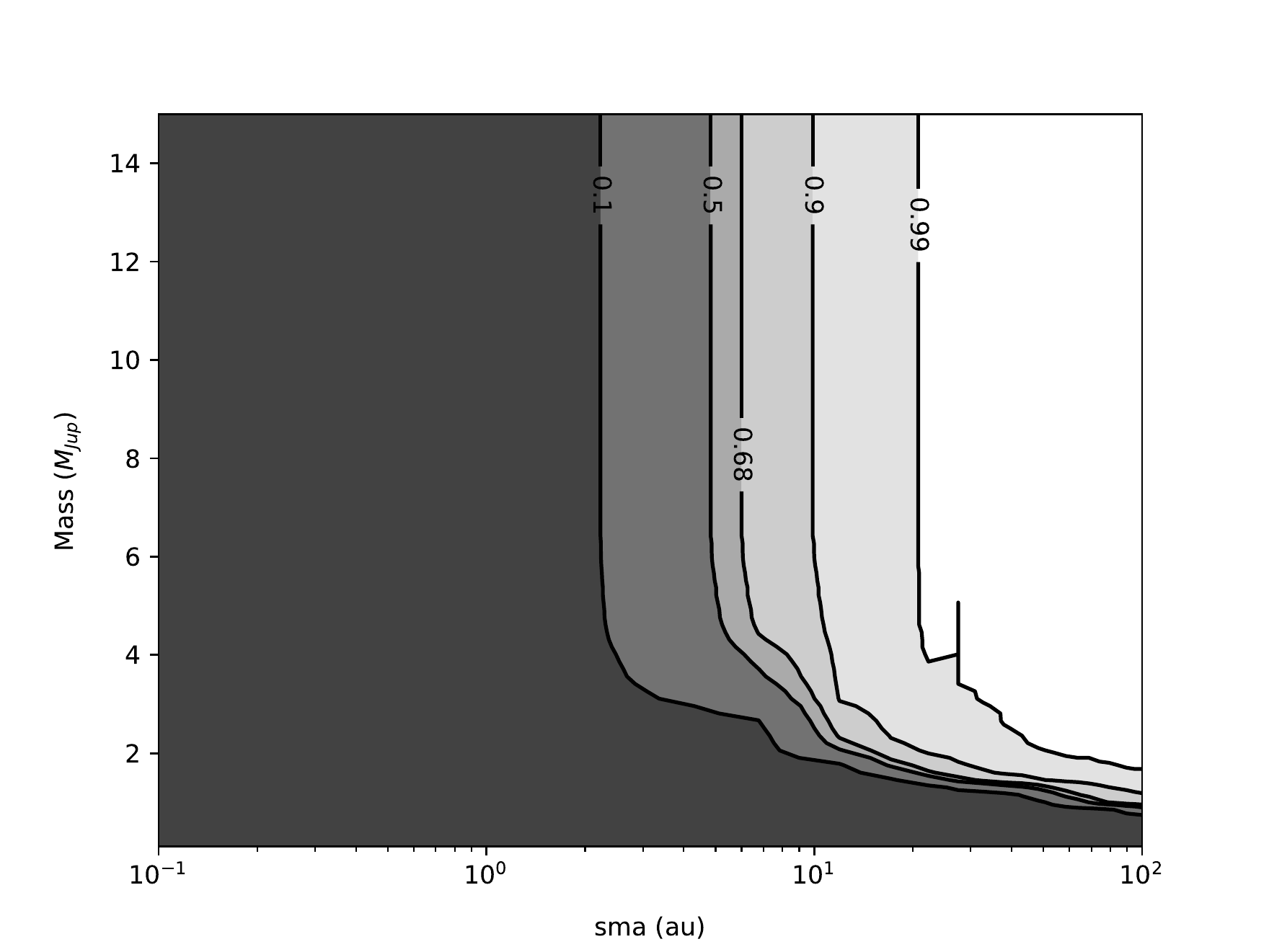}
\caption{Results for CD-31 16041. No radial velocity data were available. Two epochs were  available for DI.}
\end{figure}

\end{document}